\documentclass{aa}
\pdfoutput=1
\usepackage{txfonts}
\usepackage{natbib}
\usepackage{supertabular}
\usepackage{lscape}
\usepackage{indentfirst}
\usepackage{longtable}
\usepackage{graphicx}

\bibpunct{(}{)}{;}{a}{}{,}

\begin{document}

\title{Mid-infrared spectroscopy of \textit{Spitzer}-selected ultra-luminous 
starbursts at z\,$\sim$\,2 \thanks{Colour figures and the Appendix are only 
available in the electronic form via http://www.edpsciences.org}}
  
\author{N. Fiolet\inst{1,2}
  \and A. Omont\inst{1,2}
  \and G. Lagache\inst{3,4}
  \and B. Bertincourt\inst{3,4}
  \and D. Fadda\inst{5}
  \and A.~J. Baker\inst{6}
  \and A. Beelen\inst{3,4}
  \and S. Berta\inst{7}
  \and F. Boulanger\inst{3,4} 
  \and D. Farrah\inst{8}
  \and A. Kov\'acs\inst{9}
  \and C. Lonsdale\inst{10}
  \and F. Owen\inst{11}
  \and M. Polletta\inst{1,2,12}
  \and D. Shupe\inst{13}
  \and Lin Yan\inst{5}
   }


\institute{UPMC Univ Paris 06, UMR7095, Institut d'Astrophysique de Paris, 
F-75014, Paris, France 
\and CNRS, UMR7095, Institut d'Astrophysique de Paris, F-75014, Paris, France 
\and Univ Paris-Sud, Institut d'Astrophysique Spatiale, UMR8617, Orsay, F-91405, France
\and CNRS, Orsay, F-91405 
\and Spitzer Science Center, California Institute of Technology, MS 220-6, 
Pasadena, CA 91125, USA 
\and Department of Physics and Astronomy, Rutgers, the State University of New 
Jersey, 136 Frelinghuysen Road, Piscataway, NJ 08854, USA
\and Max-Planck Institut f\"ur extraterrestrische Physik, Postfach 1312, 85741 
Garching, Germany 
\and Department of Physics \& Astronomy, University of Sussex, Falmer, 
Brighton, BN1 9RH, UK
\and University of Minnesota, 116 Church St SE, Minneapolis, MN 55414, USA
\and North American ALMA Science Center, NRAO, Charlottesville, USA
\and National Radio Astronomy Observatory, P. O. Box O, Socorro, NM 87801, USA 
\and INAF-IASF Milano, via E. Bassini 15, 20133, Italy 
\and Herschel Science Center, California Institute of Technology, 100-22, 
Pasadena, CA 91125, USA 
}

\abstract{\textit{Spitzer}'s wide-field surveys and followup capabilities have 
allowed a new breakthrough in mid-IR spectroscopy up to redshifts $\geq 2$, 
 especially for 24\,$\mu$m detected sources.}
 {We want to study the mid-infrared properties and 
the starburst and AGN contributions, of 24\,$\mu$m sources at $z \sim 2$, through 
analysis of mid-infrared spectra combined with millimeter, radio, and infrared 
photometry. Mid-infrared spectroscopy allows us to recover accurate redshifts.}
 {A complete sample of 16 {\it Spitzer}-selected sources (ULIRGs) believed to 
be starbursts at $z \sim 2$ (``5.8\,$\mu$m-peakers") was selected in the 
(0.5\,deg$^{2}$) J1064+56 SWIRE Lockman Hole field (``Lockman-North'').  
 These sources have 
$S_{\rm 24\,\mu m} > 0.5$\,mJy, a stellar emission peak redshifted to 
5.8\,$\mu$m, and $r'_{\rm Vega} > 23$. The entire sample was observed with the 
low resolution units of the \textit{Spitzer}/IRS infrared spectrograph.  These 
sources have 1.2\,mm observations with IRAM 30\,m/MAMBO and very deep 20\,cm 
observations from the VLA. Nine of our sources also benefit from 350\,$\mu$m 
observation and detection from CSO/SHARC-II. All these data were jointly analyzed.}
 {The entire sample shows good quality IRS spectra dominated by strong PAH 
features. The main PAH features at 6.2, 7.7, 8.6, and 11.3\,$\mu$m have high S/N
average luminosities of $2.90 \pm 0.31$, $10.38 \pm 1.09$, $3.62 \pm 0.27$, 
and $2.29 \pm 0.26 \times 10^{10}\,L_{\odot}$, respectively. Thanks to their PAH spectra,
  we derived accurate redshifts spanning from 1.750 to 2.284.
 The average of these redshifts is 2.017$\pm$0.038. 
 This result confirms that the selection criteria of ``5.8\,$\mu$m-peakers" associated with 
 a strong detection at 24\,$\mu$m are reliable to select sources at $z \sim 2$. 
  We have analyzed the different 
correlations between PAH emission and infrared, millimeter, and radio 
emission. Practically all our sources are strongly dominated by starburst 
emission, with only one source showing an important AGN contribution.  We have 
also defined two subsamples based on the equivalent width at 7.7\,$\mu$m to 
investigate AGN contributions.}{Our sample contains strong starbursts and represents a particularly 
24\,$\mu$m-bright class of SMGs. The very good correlation between PAH and 
far-IR luminosities is now confirmed in high-$z$ starburst ULIRGs.  These 
sources show a small AGN contribution to the mid-IR, around $\sim$20\% or less 
in most cases.}


\maketitle

\section{Introduction}

The mid-infrared (MIR) regime is known to provide very rich diagnostics of the 
interstellar medium (ISM) of galaxies. Hot dust from starbursts and especially 
AGN contributes to continuum emission. Dust also produces various absorption 
features, most spectacularly from silicates. However, the MIR spectra of 
star-forming galaxies are dominated by strong emission features attributed to 
polycyclic aromatic hydrocarbons (PAHs). These large molecules, typically a 
few hundred carbon atoms in the ISM, are the carriers of the series of 
prominent emission bands observed between 3 and 19\,$\mu$m, especially between 
6 and 12\,$\mu$m \citep{Lege84,Alla85}. PAHs efficiently absorb UV and optical 
photons from young stars and reemit the energy mostly in the mid-infrared.

After their discovery in the local ISM and nearby star-forming regions, PAH 
features have proved to be common, not only in various regions of the Milky 
Way, but in other galaxies ranging from local systems to high-$z$ starbursts 
\citep[e.g., ][]{Tiel08}. PAH features were the focus of a good fraction of the
MIR spectroscopy programs of the infrared space observatories {\it ISO} and 
{\it Spitzer}. Important key projects of both missions were devoted to MIR 
spectroscopy of large samples of galaxies. The {\it ISO} Key Project on Nearby 
Galaxies \citep{Helo00} showed an amazing similarity of PAH spectra in various 
galaxy environments. More detailed studies with the SINGS spectral mapping 
program \citep{Kenn03} using the {\it Spitzer} Infrared Spectrograph 
\citep[IRS: ][]{Houc04} provided further evidence of interband PAH feature 
strength variations, especially in the presence of weak AGN \citep{Smit07}.

The high sensitivity of {\it Spitzer}/IRS allowed the extension of PAH 
spectroscopy to $z \ga 2$ for samples of hundreds of galaxies that could be 
observed in the $\sim 5$-35\,$\mu$m spectral range. This led to the 
measurement of a large number of redshifts of dust-enshrouded {\it Spitzer} 
sources at $1.5 < z < 2.5$ in the so-called ``redshift desert'' of optical 
spectroscopy. These MIR spectra allow one to distinguish power-law 
AGN-dominated spectra from PAH starburst-dominated spectra and composite 
systems. This long series of high-$z$ IRS studies has thus established the 
criteria for the identification of various types of sources from {\it Spitzer} 
3.5-24\,$\mu$m photometry 
\citep[][]{Fadd10,Houc05,Yan05,Yan07,Saji07,Weed06,Dasy09,Hern09,Bert09}.
Particular types of sources can be selected using various criteria: 
dust-enshrouded AGN from strong 24\,$\mu$m fluxes and/or power-law IRAC 
3.6-8.0$\mu$m spectra \citep[][]{Dey08,Dey09}, starbursts from the presence of 
a visible redshifted 1.6\,$\mu$m stellar bump 
\citep[][]{Farr06,Farr08,Lons09,Huan09,Desa09}, Type 1 or Type 2 QSOs 
\citep[][]{Lutz08,Mart08}, and submillimeter-selected galaxies 
\citep[SMGs: ][]{Vali07,Mene07,Mene09,Pope08a,Copp10}.

Indeed, strong PAH emission is so characteristic of starbursts that the 
strength of PAH features may be used as a star formation rate indicator. This 
is justified by the good correlation that has been found between the 
luminosity of PAH features (mostly at 6.2 and 7.7\,$\mu$m) and the total IR or 
far-IR luminosity, which in the absence of an AGN is taken to be a good tracer 
of star-forming activity \citep[][]{Kenn98}. Such a correlation has been 
checked for starburst galaxies at both low and high redshift 
\citep[e.g., ][]{Bran06,Pope08a,Mene09}. It is the origin of the correlation 
found between the {\it ISO} 7.0\,$\mu$m- or {\it Spitzer} 8.0$\,\mu$m-band 
flux and the star formation rate \citep[e.g., ][]{Elba02,Wu06,Alon06,Farr07} 
in low-$z$ luminous infrared galaxies. It is similarly related to the use of 
the single-band 24\,$\mu$m luminosity for determining star formation rates of 
galaxies at high redshift, especially at $z \sim 2$ 
\citep[e.g., ][ Fiolet et al. in prep.]{Bavo08,Riek08,Elba10}.

Such rich diagnostics of the properties of the ISM and star formation at high 
redshift provided by mid-IR PAH spectral features have justified the many 
studies performed with the unique capabilities of {\it Spitzer}/IRS. We have 
used this experience to focus one of the last observations of ``cold'' 
{\it Spitzer} to obtain high-quality mid-IR spectra of $z \sim 2$ starbursts. 
Our sample was {\it Spitzer}-selected in a field with very rich multiwavelength
data.  We aimed to select starburst-dominated ultra-luminous IR galaxies at 
$z \sim 2$ based on the presence of a rest-frame 1.6\,$\mu$m stellar bump--  
caused by the photospheric opacity minimum in giant or supergiant stars 
\citep[e.g. ][]{John88,Simp99,Weed06}-- in the 5.8\,$\mu$m IRAC band, together with very 
strong PAH emission implied by a high 24\,$\mu$m flux density.  Using these 
``5.8\,$\mu$m-peakers,'' we sought to address in the best conditions the 
various goals allowed by such studies: determination of redshifts from PAH 
features, accurate enough for allowing CO line observations with current 
millimeter facilities; detailed examination of the properties of PAH features 
in such high-$z$ sources, especially their luminosity ratios, and comparison 
to local systems; estimation of the ratio between PAH bands and the weak 
underlying continuum that may constrain the modest AGN contribution; and 
accurate assessment of the correlation of PAH luminosities with total and 
far-infrared luminosities (and with the radio luminosity). Silicate 
absorption at 9.7\,$\mu$m may provide additional information about 
 the distribution of interstellar dust. 

In this paper, we present the observations and the results from IRS 
spectroscopy of a complete sample of 16 sources selected by their IRAC and 
MIPS fluxes. The sample selection and the observations are described in 
Section~\ref{sample}. The method adopted to reduce these observations is 
presented in Section~\ref{datareduction}. The results and their analysis are 
discussed in Sections~\ref{resultats} and \ref{Analysis}. In 
Section~\ref{Discussion}, we discuss the discrimination between starburst- and 
AGN-dominated sources, and we make comparisons with the other samples of SMGs. 
Throughout the paper, we adopt a standard flat cosmology with $H_{0} = 71\,{\rm
km\,s^{-1}\,Mpc^{-1}}$, $\Omega_{M} = 0.27$, and $\Omega_{\Lambda} = 0.73$ 
\citep{Sper03}. 

\section{Sample selection and observations}\label{sample}

\subsection{Sample selection}\label{selection}\label{ancillary}

Our sample is based on the \textit{Spitzer}/SWIRE ``5.8\,$\mu$m-peaker'' 
sample of \citet{Fiol09} (hereafter F09). Using the 2007 SWIRE internal 
catalogue, we have selected the sixteen 24$\,\mu$m-brightest sources of this 
sample with $S_{\rm 24\,\mu m} > 0.5$\,mJy. These sources obey the 
``5.8\,$\mu$m-peaker'' criteria with $S_{\rm 3.6\,\mu m} < S_{\rm 4.5\,\mu m} <
S_{\rm 5.8\,\mu m} > S_{\rm 8.0\,\mu m}$. Some of our sources have no 
detection in the 8.0\,$\mu$m band, but we assume that these sources are also  
``5.8\,$\mu$m-peakers'' if their fluxes at 5.8\,$\mu$m are greater than the 
detection limits at 8.0\,$\mu$m ($< 40$\,mJy). To remove low-redshift 
interlopers, we also require that the sources be optically faint, i.e., 
$r'_{\rm Vega} > 23$ \citep{Lons06}. Our sample is complete for these criteria 
over 0.5\,deg$^{2}$. This area includes the ``Lockman-North'' field observed 
with the VLA \citep{Owen08,Owen09} and {\it Herschel} \citep{Oliv10}. For this 
sample from F09, we have 1.2\,mm flux densities ($S_{\rm 1.2\,mm}$) obtained with 
IRAM/MAMBO [Institut de Radioastronomie Millim\'etrique/Max-Planck Millimeter 
Bolometer] array \citep[]{Krey98}. 62\% (10/16) of our sources are detected with S/N\,$> 3$ at 1.2\,mm (Table~\ref{new})
 and are thus 
submillimeter galaxies (SMGs) with $S_{\rm 1.2\,mm} > 2\,{\rm mJy}$ 
($\langle S_{\rm 1.2\,mm}\rangle = 2.72 \pm 0.54$\,mJy).  The six other 
sources have an average $\langle S_{\rm 1.2\,mm}\rangle = 1.01 \pm 0.68$\,mJy 
(see Table~\ref{new} and F09). This sample is thus representative of a 
 \textit{Spitzer}-selected subclass of powerful $z \sim 2$ ULIRGs \citep[SMGs 
and OFRGs, Chapman et al. in prep;][]{Magd10}, 
 of which there are $\sim 2000$ in all SWIRE 
fields that peak at 4.5 or 5.8\,$\mu$m and have $S_{\rm24\,\mu m} > 
500\,\mu$Jy.

Nine of our sources detected at 1.2\,mm were observed at 350\,$\mu$m by 
\citet{Kova10}.  All these sources have been detected, and our 16 sources are 
also detected at 250\,--\,350\,$\mu$m by \textit{Herschel} \citep[Fiolet et al. in prep; ][]{Magd10}.
  The ``5.8\,$\mu$m-peaker'' criterium also implies 
that our sample is a mass selected-sample and has $z \sim 2$.

All of our sources benefit from very deep radio data at 1400, 610, and 
324\,MHz from the VLA and GMRT \citep[][ Owen et al. in prep.]{Owen08,Owen09}, 
with observed flux densities $S_{\rm 20\,cm} > 50\,\mu$Jy (Table~\ref{new} and F09).

Our sample differs from the other samples of ULIRGs essentially in its 
selection criteria and redshift range.  Indeed, \citet{Farr08} and 
\citet{Saji07} have selected bright sources at 24\,$\mu$m. However, the sample 
of \citet{Saji07} is built to select the reddest objects in the
[$24\,{\rm \mu m} - 8\,{\rm \mu m}$] vs. [$24\,{\rm \mu m} - R$] color-color 
diagram and is composed of PAH-weak and PAH-strong sources. In the comparisons
that follow, we consider only the PAH-strong sources of this sample. 
\citet{Pope08a}, \citet{Mene09}, and \citet{Vali07} have built samples of SMGs detected 
at 850\,$\mu$m. Nevertheless, the sources from \citet{Pope08a} selected with 
$S_{\rm 24\,\mu m} > 200\,\mu$ Jy are fainter than our sources ($S_{\rm 
24\,\mu m} > 500\, \mu$Jy). \citet{Mene09} and \citet{Vali07} have also 
selected their sample based on 1.4\,GHz flux density.  The sample of 
\citet{Desa09} is relatively similar to ours, with sources selected by a 
redshifted 1.6\,$\mu$m stellar bump in the IRAC bands and $S_{\rm 24\,\mu 
m}$\,$>$\,500\,$\mu$Jy.  These sources are also optically faint and satisfy 
the definition of Dust-Obscured Galaxies \citep[DOGs: ][]{Dey08}. 
The sample of \citet{Huan09} is also selected with $S_{\rm 24\,\mu m} > 
500\,\mu$Jy and with color-color criteria yielding star-forming galaxies. 
\citet{Copp10} have built a sample of SMGs selected to be AGN-dominated.
While our sample has redshifts from $\sim 1.7$ to $\sim 2.3$, the other 
samples span larger redshift ranges. \citet{Shi09} have selected sources at 
intermediate redshift around $z \sim 1$. \citet{Farr08} observed sources with 
$z = 1.3 - 2.2$ centered at $z \sim 1.7$ (``4.5\,$\mu$m-peakers''). The 
samples of SMGs from \citet{Pope08a,Mene09} and \citet{Vali07}, and the DOG 
sample of \citet{Desa09}, span redshifts from $\sim 1$ to $\sim 3$ 
\citep[0.6 to 3.6 for ][]{Mene09}. The sample of star-forming galaxies of 
\citet{Huan09} has redshifts between 1.7 and 2.1 centered at $z \sim 1.9$.
Finally, the sample of \citet{Copp10} ranges from $z \sim 2.45$ to $\sim 3.4$. 

\subsection{Observations}\label{observation}

The photometric redshifts of our targets, based on the IRAC bands, lie in the 
range $z \sim 1.5-2.7$ with uncertainties of $\pm 0.5$ (see F09). In this 
redshift range, the low-resolution ($R\sim60-127$) Long-Low observing modules 
of IRS, LL1 (19.5$-$38.0\,$\mu$m) and LL2 (14.0$-$21.3\,$\mu$m), can cover 
most of the PAH emission features from 6.2 to 11.3\,$\mu$m in the rest frame. 
Indeed, the actual redshifts have all proved to lie in the range $1.7 < z < 
2.3$ (see Sec.~\ref{redshift}). 

This project (GO50119, PI G. Lagache) was observed for a total time of 53 hours (41 hours 
on-source) during the last cold campaign of \textit{Spitzer}, on 
January 20-22, 2009.  We have divided our sample into four groups based on 
24$\,\mu$m flux density.  For each group, the on-target exposure time is 
summarized in Table~\ref{summary}. We chose these exposure times 
following the guidelines from the IRS IST report, based on IRS low-resolution 
observations of two faint sources by Teplitz et 
al.\footnote{http://ssc.spitzer.edu/irs/documents/irs\_ultradeep\_memo.pdf} 
and the Spectroscopy ``Performance Estimation Tool'' (SPEC-PET). We also used 
our own IRS observations of distant galaxies to check the expected S/N. The 
integration times have been fixed to achieve a S/N $\sim 4-5$ per wavelength 
element in LL1 and LL2. The observations were done in mapping mode to maximize 
the rejection of ``rogue'' pixels and (immediately after applying the 
``skydark'' calibrations) to limit contamination by latent charge from bright 
objects in our ``Long'' spectra of faint sources. 

\section{Data reduction}\label{datareduction}

The observed spectra were processed with version S18.5.0 of the 
\textit{Spitzer} Science Center pipeline. The pipeline produces the basic 
calibrated data (BCD), which are corrected for different effects (e.g., cosmic 
ray removal, masking of saturated pixels, and linearization of the science 
data signal ramps).

Applying the IDL code IrsLow, developed by \citet{Fadd10} and available 
on-line\footnote{http://web.ipac.caltech.edu/staff/fadda/IrsLow}, we have 
extracted the spectra from the BCD. IrsLow is a data reduction code built 
especially for low-resolution IRS observations. This code makes a background 
subtraction and extracts the spectra of the target. The background is 
estimated after the user has manually masked the regions of interest of the 
data. In practice, we mask the position of the sources on the spectra to estimate and subtract the background. 
The spectrum is then extracted using a PSF fitting. The code also allows the 
simultaneous extraction of two different spectra in the case of two nearby  
sources. The details of using IrsLow and some examples can be found in 
\citet{Fadd10}. The output of the program is an ASCII file containing the 
spectrum and its errors. To facilitate our analysis, we make some modifications to each of these 
files. First, we limit the wavelengths to 20.5\,$\mu$m through 35\,$\mu$m for 
the first order (i.e., LL1) -- the errors on the flux outside this range 
becoming quite large -- and to $<20.5$\,$\mu$m for the second order (i.e., 
LL2). Second, we sort the data by wavelength. Finally, we smooth the spectrum 
in IDL. This filter makes a running mean over five points providing thus an 
effective spectral channel width of 0.10\,$\mu$m. 
 To avoid artificial improvement of S/N, 
 the errors on the spectra are kept unchanged by the smooth filter. 

\section{Results: Spectra}\label{resultats}

The spectra of our 16 sources are shown in  Appendix~\ref{sed_spectra} (see 
also Figs.~\ref{decomposition} and \ref{Spectrald}). These spectra show 
high-S/N ($> 5\sigma$) detections of PAH features at 6.2, 7.7, 8.6, and (in 
many cases) 11.3\,$\mu$m for $z \sim 2$ (Table~\ref{few}).

Considering the uncertainties of our data, all our sources have MIPS $S_{\rm 
24\,\mu m}$ from the 2007 SWIRE internal catalogue that are compatible with 
the 24\,$\mu$m flux density ($S_{\rm24\,\mu m,spectra}$) that we have directly 
measured from the spectra after convolution with the MIPS 24\,$\mu$m filter 
profile (Fig.~\ref{sed_spectra}).  For 8/16, the difference is less than 10\%. 
These flux densities are reported in Table~\ref{new}. A similar result was 
obtained by \citet{Pope08a}. 

\begin{figure}[!htbp]
\resizebox{\hsize}{!}{\includegraphics{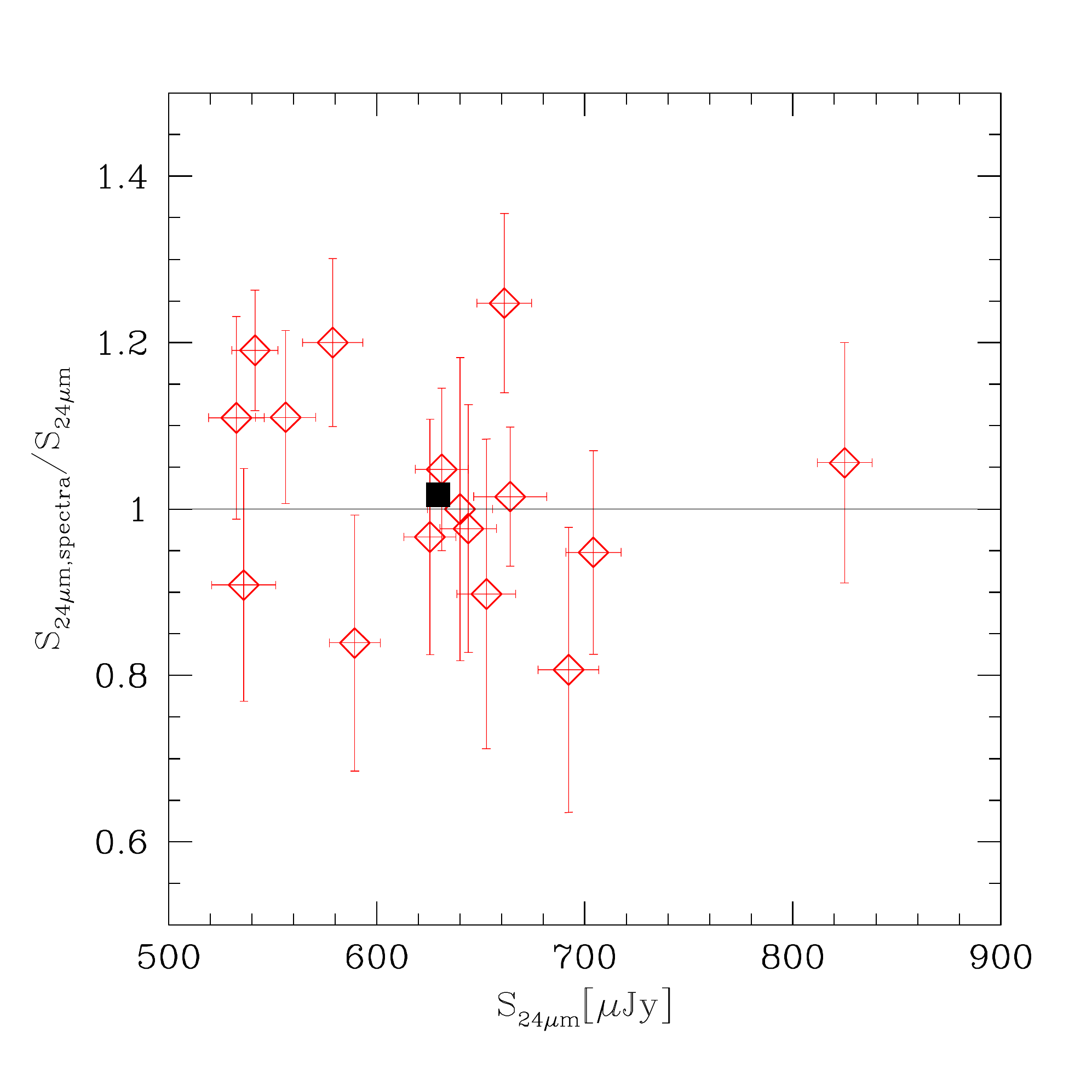}}
\caption{Ratio of $S_{\rm24\,\mu m,spectra}$ deduced from our spectra using 
the MIPS 24\,$\mu$m filter, to $S_{\rm24\,\mu m}$ from the 2007 SWIRE internal 
catalogue, as a function of $S_{\rm24\,\mu m}$.  The black square is the mean
for our full sample. The error bars correspond to 1\,$\sigma$.}
\label{2424}
\end{figure}

\subsection{PAH redshifts}\label{redshift}

\begin{figure}[!htbp]
\resizebox{\hsize}{!}{\includegraphics{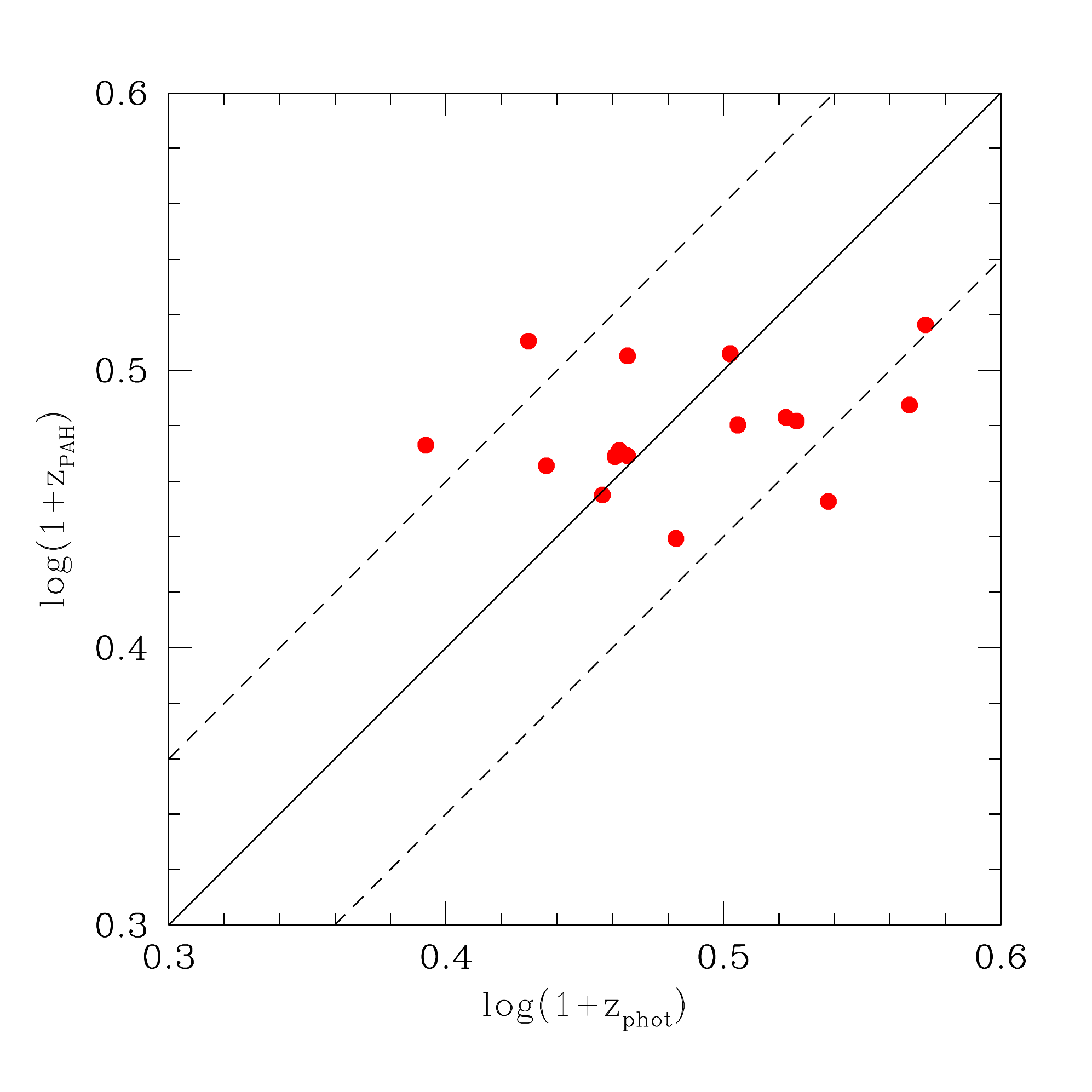}}
\caption{PAH spectroscopic redshift $z_{\rm PAH}$ as a function of photometric 
redshift from F09. The dashed lines correspond to $\Delta\,{\rm log}\,(1+z)= 
\pm 0.06$ or $\pm 15\%$.}
\label{z_zphot}
\end{figure}

Applying the method described in \citet{Bert09}, we have determined redshifts 
from PAH features ($z_{\rm PAH}$).  This method is based on the 
cross-correlation of 21 templates with our IRS spectra. These templates are 
spectra of different kinds of sources. Five templates are dominated by PAHs 
\citep{Smit07}. 13 are less PAH-dominated local ULIRGs from \citet{Armu07},  
including among others Arp\,220 and Mrk\,231. The remaining three are a radio 
galaxy, a quasar, and a Wolf-Rayet galaxy.  The best estimate of the redshift, 
$z_{\rm PAH}$, is the average of redshifts found for all templates after 
rejection in cases of very high $\chi^{2}$ ($\chi^{2}$/SNR\,$> 8$) or too low 
SNR ($< 1$). These redshifts are reported in Table~\ref{redshifts}. The 
estimated rms error of $z_{\rm PAH}$ is on average $\langle \Delta z_{\rm 
PAH}\rangle = 0.010$ (median\,=\,0.007; min\,=\,0.004; max\,=\,0.064). We also 
report in Table~\ref{redshifts} the redshift, $z_{\rm best}$, of the best-matching template. 
The small differences between $z_{\rm PAH}$ and the redshift for the 
best-matching template are very consistent with the estimated rms error 
$\langle \Delta z_{\rm PAH}\rangle$. We use $z_{\rm PAH}$ in the following 
analysis because this determination is less biased by template selection 
than the redshift from the best-matching template ($\langle z_{\rm best} 
\rangle = 2.011$; median\,=\,1.994; scatter\,=\,0.023).  We note that $z_{\rm 
PAH}$ and $z_{\rm best}$ are very similar for all of our sample, and that our 
sources are best matched by a specific ``PAH'' template from \citet{Smit07}.
The average of $z_{\rm PAH}$ is $2.017 \pm 0.38$ (median\,=\,1.997).  
 The IRS redshifts of our sample span a narrower range, 
from 1.7 to 2.3, than the photometric redshifts maybe
 due to the strong $S_{\rm 24\,\mu m}$.
  This result is compatible 
with the fact that ``5.8\,$\mu$m-peakers'' have a high probability of being 
dominated by a strong starburst at $z \sim 2$ because the stellar bump at 
1.6\,$\mu$m rest-frame is redshifted to around 5.8\,$\mu$m at redshift $\sim 2$
\citep[see, e.g., ][]{Lons09}. Compared to the photometric redshifts reported 
in F09 found using the Hyper-z code \citep[][]{Bolz00}, the difference is 
fairly significant (see Fig.~\ref{z_zphot}). Nevertheless, except for four 
sources, the photometric redshifts differ from $z_{\rm PAH}$ by less than 
15\%, the boundary for catastrophic outliers defined by \citet{Rowa08}. 
 
\subsection{Spectral decomposition}\label{thespectra}

\begin{figure}[!htbp]
\resizebox{\hsize}{!}{\includegraphics{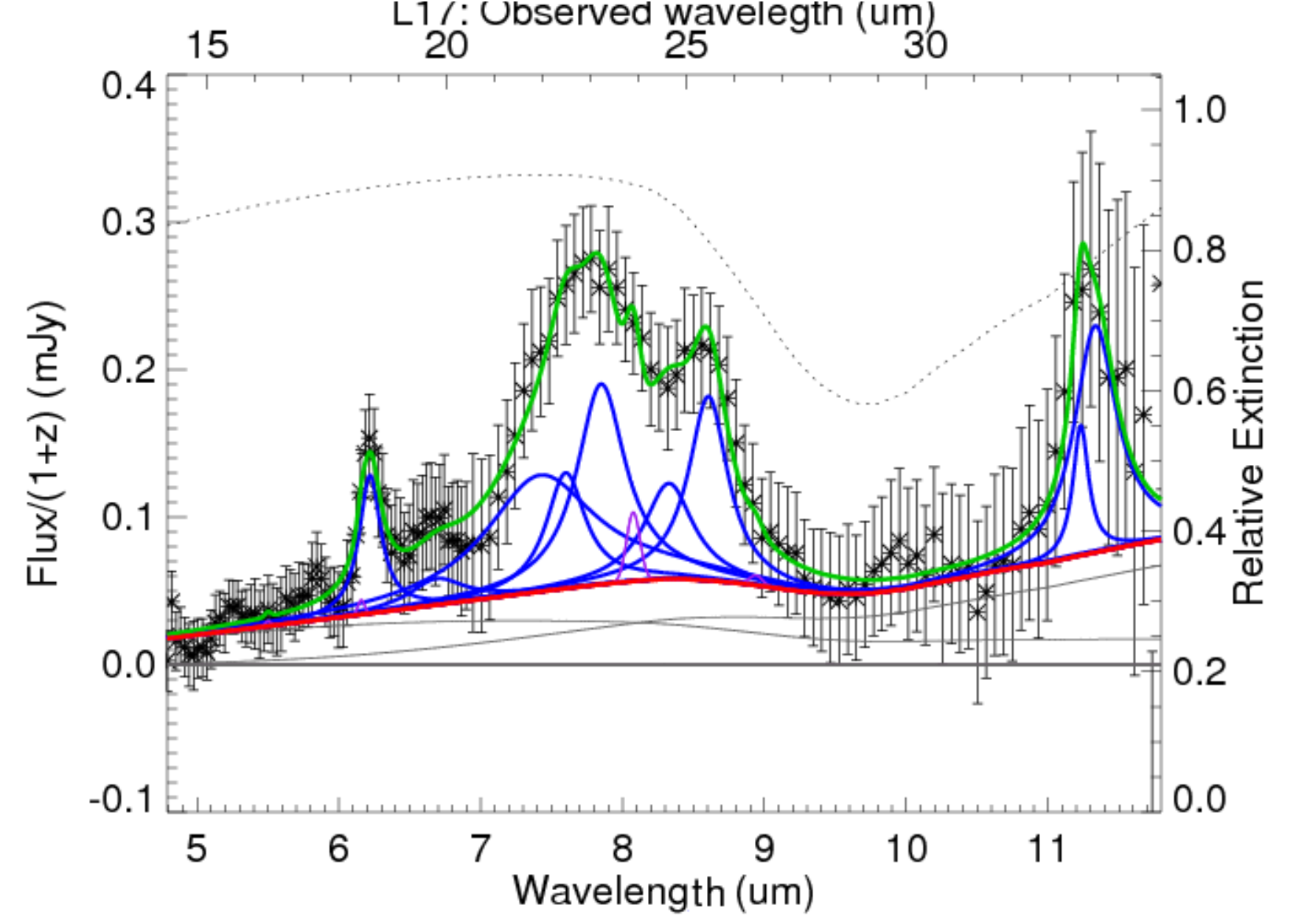}}
\caption{Spectral decomposition for source L17. The solid green line is the 
fitted model. Blue lines above the continuum are PAH features; violet lines 
are spectral lines. Thin grey lines represent thermal dust continuum 
components. The thick red line shows the total continuum (stars+dust). The 
dotted black line shows the extinction ($e^{-\tau_{\lambda}}$, $=1$ for no 
extinction). In this example, the top axis is labelled in observed wavelength. 
The redshift is $z_{\rm PAH} = 1.959$.  Spectral decompositions for the other 
sources are presented in Appendix B.}
\label{decomposition}
\end{figure}

We perform spectral decompositions using the PAHFIT code developed by 
\citet{Smit07}.  PAHFIT adjusts some PAH features and spectral lines as a 
function of rest wavelength (Fig.~\ref{decomposition}). The PAH features are 
represented by Drude profiles and the atomic and molecular spectral lines by 
Gaussians. The decomposition also tries to fit a stellar continuum component 
($T_{\star} = 5000\,{\rm K}$) and dust components for different temperatures 
(35, 50, 135, 200, 400, 600, 800, 1000, and 1200\,K) represented by modified 
blackbodies.  Only the intermediate dust temperatures make significant 
contributions to the mid-IR continuum. The dust extinction, mostly due to 
silicate absorption, is also taken into account. In our decompositions, we 
have adopted a screen geometry, corresponding to a uniform mask of dust 
absorbing the emission, to model the silicate absorption \citep{Smit07}.    
Figure~\ref{decomposition} presents an example of the result of the spectral 
decomposition. The PAH complexes at 6.2, 7.7, 8.6, and 11.3\,$\mu$m are 
clearly visible. As proposed by \citet{Smit07}, the 7.7, 8.6, and 11.3\,$\mu$m 
features are better fit by combinations of several features at 7.42, 7.60, and 
7.85\,$\mu$m for the 7.7\,$\mu$m feature, at 8.33 and 8.61\,$\mu$m for the 
8.6\,$\mu$m feature, and at 11.23 and 11.33\,$\mu$m for the 11.3\,$\mu$m 
feature. The results of our decompositions are somewhat uncertain, due to the 
 difficulty of estimating the silicate absorption. 
 However, this absorption is known to be weak to moderate in such starburst sources
  \citep[e.g. ][]{Spoo07,Farr08,Weed08,Li01,Yan05}. 
 Our targets appear to have especially weak absorption making them a good sample
  for analyzing with this decomposition approach.
 The spectral 
decompositions for the entire sample are reported in Appendix B for individual sources,
 and in Fig.~\ref{stackall} for the stacked spectrum of all 16 sources 
 (see also Fig.~\ref{stackew}). It is seen from individual spectra and the stack that the 
 level of the continuum between 5 and 11\,$\mu$m remains modest compared to the PAH 
 features (except for L10, Sec.~\ref{agn}).

In the stacked spectrum (Fig.~\ref{stackall}), a weak line is visible at $\sim$9.7\,$\mu$m
  and identified as the 0-0\,S(3) pure rotational 
transition of molecular hydrogen at 9.665\,$\mu$m ($J=5-3$) in the PAHFIT spectral decomposition. 
This line is also visible in the partial stacks of Fig.~\ref{stackew} and may be 
 present in the spectral decompositions of 13/16 sources (Fig.~\ref{Spectrald}). The detection
 of this line is confirmed by the detailed data analysis that 
 we perform in a separate paper (Fiolet et al. in prep.). The very large luminosity of this line,
  stronger than the CO luminosity, probably implies very strong shocks to excite its upper
  level $\sim$2500\,K above the ground state of H$_2$,
 although ${\rm H_2}$ excitation could also be
influenced by X-rays or UV from AGN.  We also see hints of other lines,
although not at significant levels: the S(5) and S(4) transitions of
molecular hydrogen at $6.86\,{\rm \mu m}$ and $8.08\,{\rm \mu m}$,
respectively, as well as the ionized species [Ar\,III] at
$8.98\,{\rm \mu m}$ and [S\,IV] at $10.52\,{\rm \mu m}$.\\

\begin{figure}[!htbp]
\resizebox{\hsize}{!}{\includegraphics{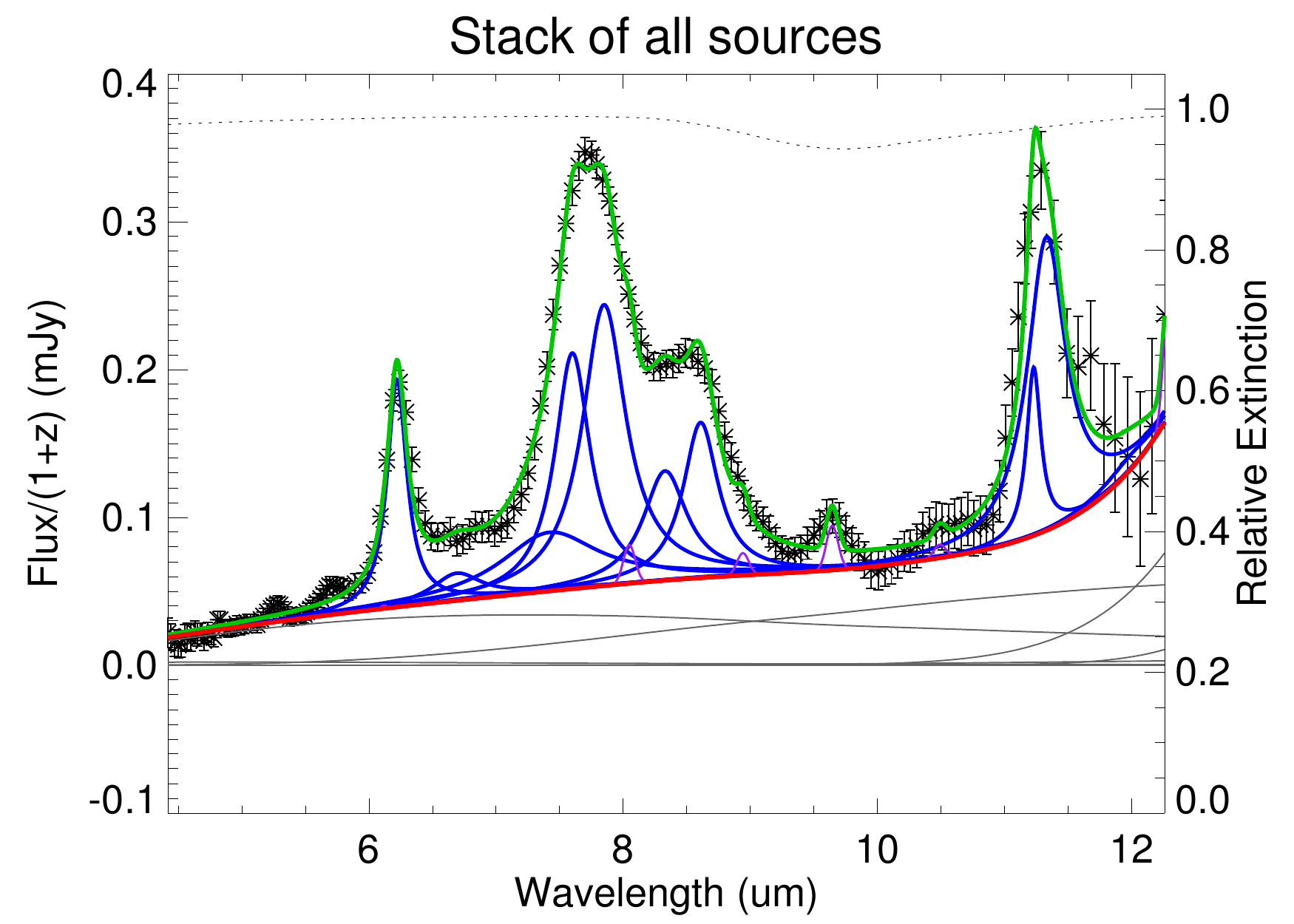}}
\caption{Spectral decomposition for the stacked spectrum of all 16 sources. 
The solid green line is the fitted model.  The blue lines above the continuum 
are the PAH features. The narrow violet lines are the spectral lines. The grey 
lines represent the thermal dust continuum components. The thick red line 
shows the total continuum (stars+dust). The dotted black line shows the 
extinction ($e^{-\tau_{\lambda}}$, $=1$ if no extinction).}
\label{stackall}
\end{figure}

\begin{figure}[!htbp]
\resizebox{\hsize}{!}{\includegraphics{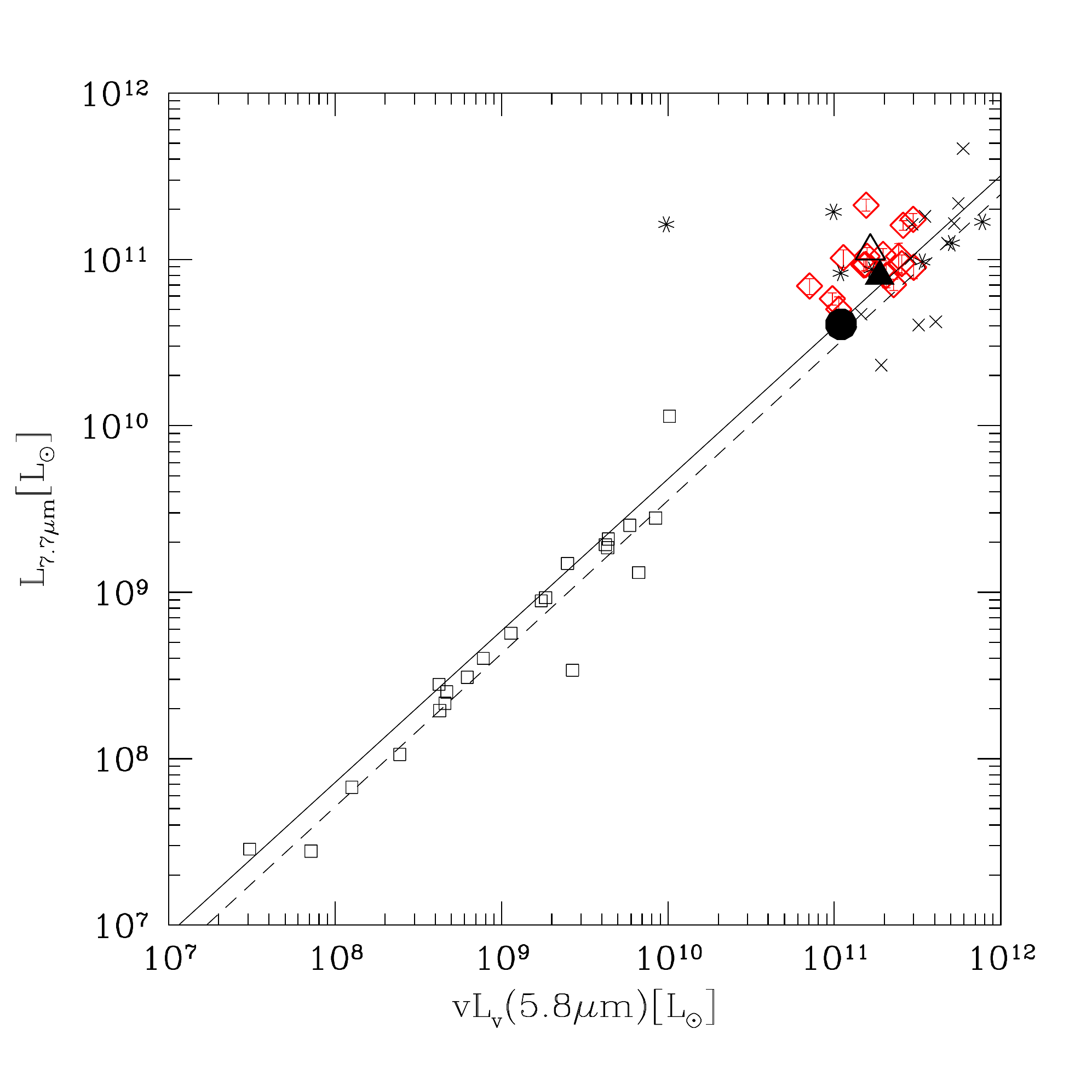}}
\caption{$L_{\rm 7.7\,\mu m}$ as a function of the continuum luminosity 
$\nu\,L_{\nu}$(5.8\,$\mu$m). The red diamonds are the sources of our sample. 
The open (full) black triangle is the stack of our sources with EW$_{\rm 
7.7\,\mu m}>\,(<)\,6\,{\rm \mu m}$. The full black circle is the composite 
spectrum of the SMG sample from \citet{Mene09}. The black crosses are the 
PAH-rich sources from \citet{Saji07}. The black asterisks are the 
AGN-dominated sources from \citet{Copp10}. The black line is the best fit of 
all these samples. The dashed black line is the best fit to the local 
starbursts from \citet{Bran06} (black squares).}
\label{58_77}
\end{figure}

For each spectrum, we evaluate the continuum flux density and luminosity 
density at 5.8\,$\mu$m ($\nu\,L_{\nu}$(5.8\,$\mu$m)). These values are reported
in Table~\ref{lum}. There is only a small dispersion of 
$\nu\,L_{\nu}$(5.8\,$\mu$m) for our sources (average\,=\,$1.87 \times 
10^{11}\,L_{\odot}$, median\,=\,$1.77 \times 10^{11}\,L_{\odot}$, 
min\,=\,$0.71 \times 10^{11}\,L_{\odot}$, max\,=\,$2.99 \times 
10^{11}\,L_{\odot}$). In Fig.~\ref{58_77}, we plot PAH luminosity at 
7.7\,$\mu$m ($L_{\rm 7.7\,\mu m}$) as a function of 
$\nu\,L_{\nu}$(5.8\,$\mu$m). The PAH luminosity at 7.7\,$\mu$m is calculated 
as the integral of the total flux between 7.3\,$\mu$m and 7.9\,$\mu$m after 
subtraction of the continuum contribution (see Eq.~\ref{eq1}). We have also 
plotted the sample of \citet{Saji07}. To make a true comparison, we have 
applied a similar reduction and spectral decomposition to the PAH-rich sources 
from \citet{Saji07}. The values of $\nu\,L_{\nu}$(5.8\,$\mu$m) found in this 
case show no significant difference from the published continuum luminosities.
We see, in Fig.~\ref{58_77}, that our sample shows somewhat weaker 
$\nu\,L_{\nu}$(5.8\,$\mu$m) than the PAH-rich sample of \citet{Saji07} or the 
AGN-dominated sample of \citet{Copp10}, while the values of $L_{\rm 7.7\,\mu 
m}$ stay in the same range. This is likely due to a stronger AGN contribution 
in the sample of \citet{Saji07}. Note also that the ratio $L_{\rm 7.7\,\mu 
m}/\nu\,L_{\nu}$(5.8\,$\mu$m) is very consistent with that found for the local 
starbursts of \citet{Bran06}. 

\subsection{PAH features}\label{features}

\begin{figure}[!htbp]
\resizebox{\hsize}{!}{\includegraphics{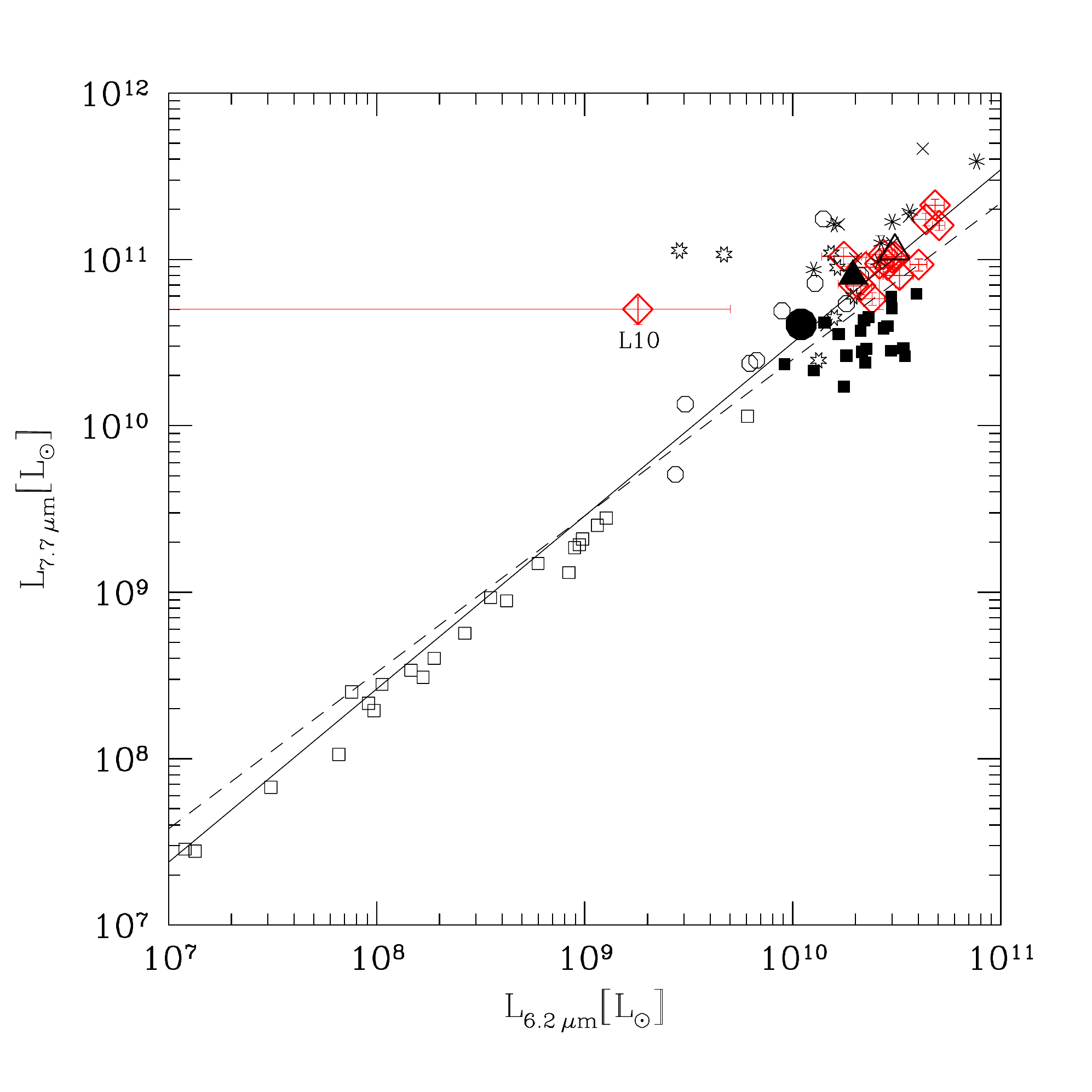}}
\caption{$L_{\rm 7.7\,\mu m}$ as a function of $L_{\rm 6.2\,\mu m}$. The red diamonds are the sources of 
our sample. The open (full) black triangle is the stack of our subsample with 
EW$_{\rm 7.7\,\mu m}>~(<)~6~{\rm \mu m}$.  The full black circle is the 
stacked spectrum of the SMG sample from \citet{Mene09}. The black circles 
are the intermediate-redshift star-forming galaxies from \citet{Shi09}. The 
black stars are the sample of SMGs from \citet{Pope08a}. The black crosses are 
the PAH-rich sources from \citet{Saji07}. The black asterisks are the 
AGN-dominated sources from \citet{Copp10}. The full black squares are the 
sources from \citet{Farr08}. The open black squares are the local star-forming 
galaxies from \citet{Bran06}. The solid line is our best fit to all these 
samples. The dashed line is the best fit found by \citet{Pope08a} for their 
SMGs and the local starbursts.}
\label{77_62}
\end{figure}  

\begin{figure}[!htbp]
\resizebox{\hsize}{!}{\includegraphics{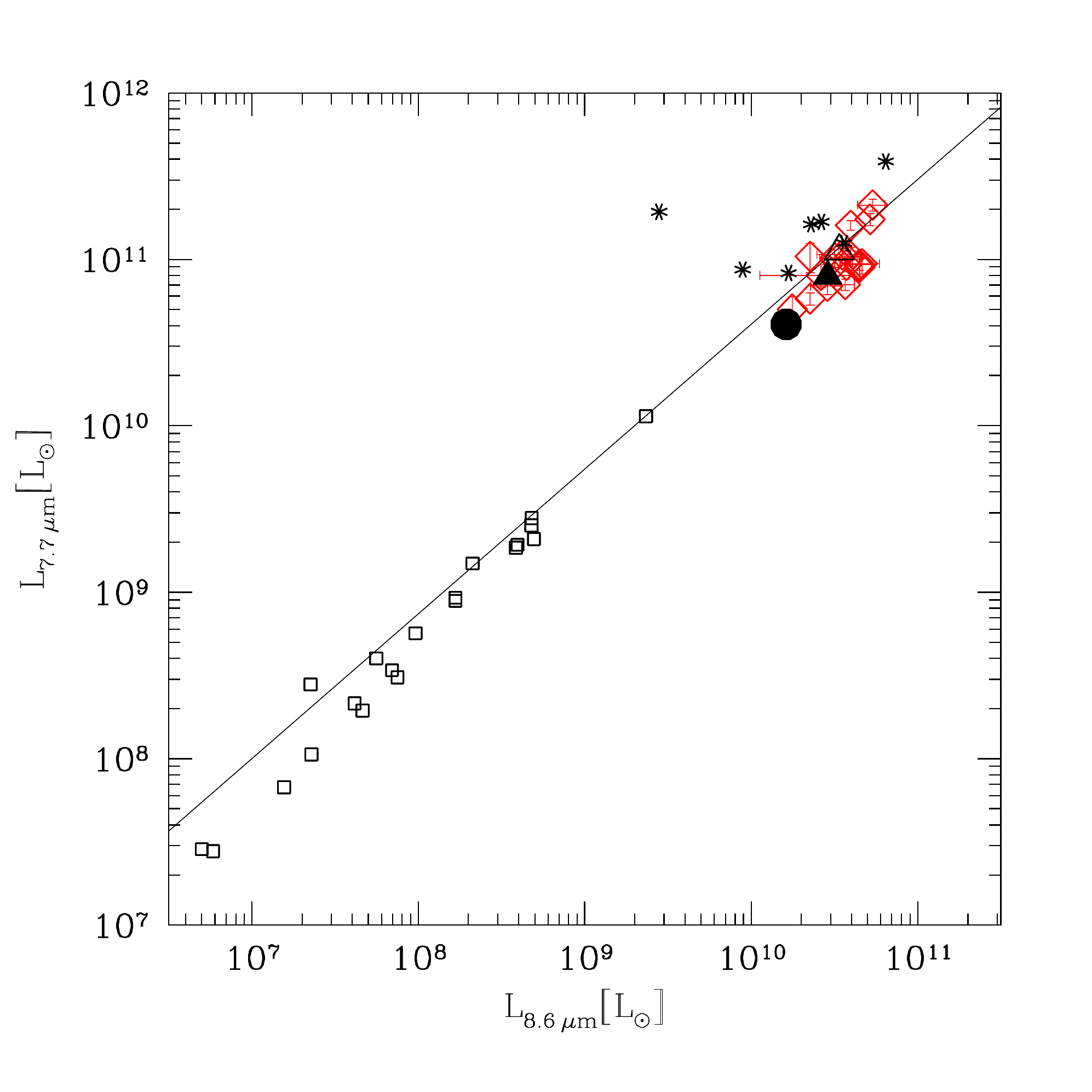}}
\caption{$L_{\rm 7.7\,\mu m}$ as a function of $L_{\rm 8.6\,\mu m}$. The 
symbols and the line are the same as in Fig.~\ref{77_62}.}
\label{77_86}
\end{figure}  

\begin{figure}[!htbp]
\resizebox{\hsize}{!}{\includegraphics{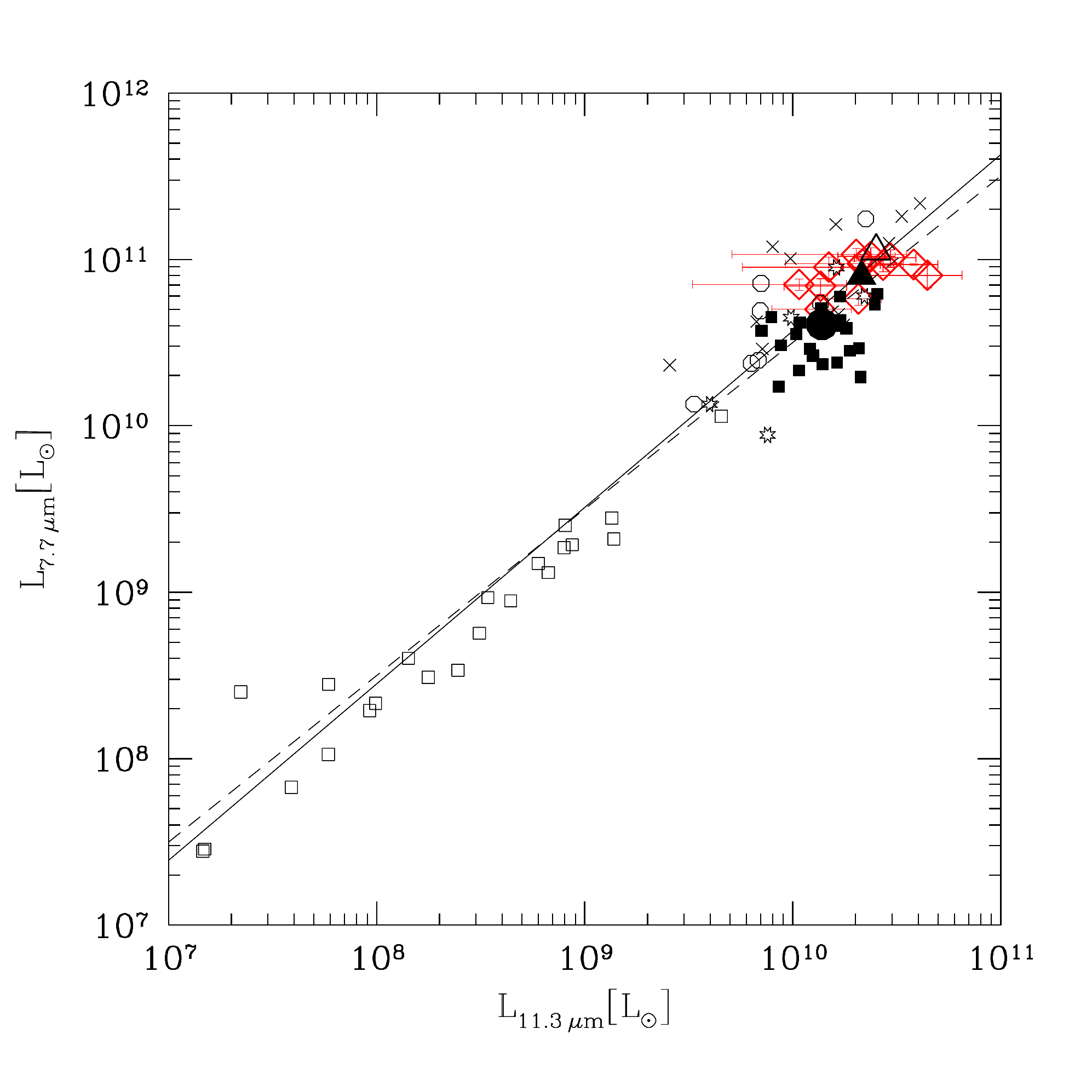}}
\caption{$L_{\rm 7.7\,\mu m}$ as a function of $L_{\rm 11.3\,\mu m}$. The 
symbols and the lines are the same as in Fig.~\ref{77_62}.}
\label{77_113}
\end{figure}

Thanks to the spectral decomposition, we have determined the fluxes ($F$) and 
equivalent widths (EW) for the PAH complexes at 6.2, 7.7, 8.6, and 
11.3\,$\mu$m (Table~\ref{few}). The flux is calculated by integrating the 
spectra between 6.2 and 6.3\,$\mu$m, 7.3 and 7.9\,$\mu$m, 8.3 and 8.7\,$\mu$m, 
and 11.2 and 11.4\,$\mu$m, respectively, after subtracting the continuum flux. 
We calculate the associated luminosities $L_{\rm 6.2\,\mu m}$, $L_{\rm 
7.7\,\mu m}$, $L_{\rm 8.6\,\mu m}$, and $L_{\rm 11.3\,\mu m}$, reported in 
Table~\ref{lum}, using:
\begin{equation}
L_{i}=4\pi D_{L}^{2} F_{i}
\label{eq1}
\end{equation}

\noindent where $D_L$ is the luminosity distance and 
$F_{i}=\int_{\lambda_{1}}^{\lambda_{2}}(S_{\lambda}-S_{c})d\lambda$ with 
$S_{\lambda}$ and $S_{c}$  the total and the continuum flux densities  at the 
considered wavelengths.

The equivalent width is calculated following:
\begin{equation}
EW=\int_{\lambda_{1}}^{\lambda_{2}}\frac{S_{\lambda}-S_{c}}{S_{c}}d\lambda
\label{eq2}
\end{equation}
All our sources show  $L_{\rm 7.7\,\mu m} \ga 5 \times 10^{10}\,L_{\odot}$ and 
EW$_{\rm 7.7\,\mu m} > 2\,\mu$m, i.e., well above the EW$_{\rm 7.7\,\mu m} > 
0.8\,\mu$m limit for ``strong'' PAH sources defined by \citet{Saji07}. All our 
sources except L10 also show EW$_{\rm 6.2\,\mu m} \gg 0.2\,\mu$m and 
$L_{\rm 6.2\,\mu m} \gg 2\times 10^{10}\,L_{\odot}$. These values are usually 
used to separate starburst- from AGN-dominated sources \citep{Armu07}. 

Our sample has PAH luminosities on average comparable
 to those of the PAH-rich sources of \citet{Saji07} and 
 slightly larger than those of the SMGs observed 
by \citet{Pope08a} and \citet{Mene09}, and of the ``4.5\,$\mu$m-peakers'' of 
\citet{Farr08}. Such a behavior might be naturally explained by stronger
 $S_{\rm 24\,\mu m}$ than SMG samples samples and higher redshift than \citet{Farr08}'s sample.
  The S/N of our average luminosities are higher than the other samples
  (Table~\ref{lumsam}). Indeed, the mean values of $L_{\rm 
6.2\,\mu m}$, $L_{\rm 7.7\,\mu m}$, $L_{\rm 8.6\,\mu m}$, and $L_{\rm 
11.3\,\mu m}$ are $2.90 \pm 0.31$, $10.38 \pm 1.09$, $3.62 \pm 0.27$, and 
$2.29 \pm 0.26 \times 10^{10}\,L_{\odot}$, respectively, for our sample. The 
stated uncertainties are the standard deviations of the means. These values of 
luminosities are simple averages. The weighted averages and the measurements 
from the stacked spectra of our sources give luminosities consistent with the 
simple averages within the uncertainties. For example, the weighted average of 
$L_{\rm 6.2\,\mu m}$ is ($2.70 \pm 0.09) \times 10^{10}\,L_{\odot}$, and the 
value derived from the stacked spectrum is ($2.81 \pm 0.94) \times 
10^{10}\,L_{\odot}$. 

As seen in Figures~\ref{77_62}, \ref{77_86}, and \ref{77_113}, and in 
Table~\ref{ratio}, our PAH luminosities $L_{\rm 6.2\,\mu m}$, $L_{\rm 7.7\,\mu 
m}$, $L_{\rm 8.6\,\mu m}$, and $L_{\rm 11.3\,\mu m}$ have ratios similar to 
those of SMGs, as well as local starbursts and star-forming galaxies at 
$z \sim 1$. The AGN-dominated sample from \citet{Copp10} seems to have 
slightly different ratios, with a higher $L_{\rm 7.7\,\mu m}/L_{\rm 8.6\,\mu 
m}$.

The samples plotted in Figs.~\ref{77_62}, \ref{77_86}, and \ref{77_113} follow 
the relations
\begin{eqnarray}
\rm{log}\,({\it L}_{\rm 7.7\,\mu m}) & = & (1.04 \pm 0.03)\,\rm{log}\,({\it L}_{\rm 
6.2\,\mu m}) + (0.10 \pm 0.32) \label{eq3} \\
\rm{log}\,({\it L}_{\rm 7.7\,\mu m}) & = & (0.87 \pm 0.04)\,\rm{log}\,({\it L}_{\rm 
8.6\,\mu m}) + (1.91 \pm 0.34) \label{eq3b} \\
\rm{log}\,({\it L}_{\rm 7.7\,\mu m}) & = & (1.06 \pm 0.03)\,\rm{log}\,({\it L}_{\rm 
11.3\,\mu m}) - (0.03 \pm 0.30) \label{eq4}
\end{eqnarray}
Note that these relations are just plain fits of the displayed samples of 
 Figs.~\ref{77_62}--\ref{77_113}. As these samples may be biased in various ways, 
 and the properties of the galaxies may vary with redshift, 
 one should not expect a strong statistical consistency for these relations. 
The relations for $L_{\rm 6.2\,\mu m}$ and $L_{\rm 11.3\,\mu m}$ are 
consistent with the relations found by \citet{Pope08a} for local starbursts 
and SMGs.

These relations show that both $L_{\rm 7.7\,\mu m}/L_{\rm 6.2\,\mu m}$ and 
$L_{\rm 7.7\,\mu m}/L_{\rm 11.3\,\mu m}$ are $\sim 4$ for our sources at 
$z \sim 2$ (see also Figs.~\ref{figra} and \ref{ioni} and Table~\ref{ratio}),
i.e., about twice as large as for the local starbursts of \citet{Bran06} and 
the sources of \citet{Farr08} at $z \sim 1.7$. On the other hand, the ratio 
$L_{\rm 7.7\,\mu m}/L_{\rm 8.6\,\mu m}$ is about twice as small in these luminous galaxies 
  at $z \sim 2$  
as in fainter local starbursts (Table~\ref{ratio}). 
 This behavior is similar to that of SMGs (Table~\ref{ratio} and Sec.~\ref{studies}). 
  The difference with respect to local starbursts could be explained by 
 a difference in the size or ionization of PAH grains 
 (Sec.~\ref{agn}, \ref{studies} and Fig.~\ref{ioni}), or by a different extinction
 due to silicates around 10\,$\mu$m (affecting the 8.6 and 11.3 \,$\mu$m bands)
  or ice at $\sim$6\,$\mu$m (absorbing the 6.2\,$\mu$m band) \citep{Mene09,Spoo02}.
  However, the effects of extinction seem somewhat contradictory for the different
  band ratios. The finding of a lower 7.7/8.6 ratio compared to normal, local starbursts,
  is consistent with a low silicate extinction in our sample, as does the generally low 
 level of extinction in the spectral fits. However, the 'larger'  7.7/11.3 ratio
  that we report is going in the other direction; similarly, our 'larger'  7.7/6.2
 ratio would imply a larger ice absorption than in local starbursts.

L10 is the only source that seems not to follow the relation between 
$L_{\rm 7.7\,\mu m}$ and $L_{\rm 6.2\,\mu m}$, showing too weak a feature at 
6.2\,$\mu$m. However, Table~\ref{lum} and Fig.~\ref{77_62} show that the 
deviation of L10 from the correlation remains within the statistical errors of 
this noisy spectrum. The offset is probably due to higher extinction or a more 
energetically important AGN \citep{Rigo99}; see also Sec.~\ref{agn}.
  
\subsection{Silicate absorption}\label{absorption}

The determination of the optical depth of the silicate absorption at 
9.7\,$\mu$m ($\tau_{9.7}$), reported in Table~\ref{few}, is a product of the 
general fit of the spectra with PAHFIT. It is based on interstellar extinction 
similar to that of  the Milky Way \citep{Smit07}. We note that the silicate 
absorption is weak on average, $\langle \tau_{9.7} \rangle = 0.28$ (0.50 if we 
consider only the sources with $\tau_{9.7} \neq 0$) vs. 1.09 for the PAH-rich 
sample of \citet{Saji07}. Indeed, for 44\% (7/16) of our sources, PAHFIT 
provides $\tau_{9.7} = 0$. It is clear that all values of extinction 
($\tau_{9.7}$) are very uncertain because of the weakness of the 
continuum and its uncertainty, and because of the lack of measurements at long 
wavelengths. We thus state only the qualitative conclusion that the average 
extinction is likely weak in most of our sources.  Considering the values of
EW$_{6.2}$ and $\tau_{9.7}$, most of our sources belong to class 1C of the 
diagram of \citet[][their Figs. 1 \& 3]{Spoo07}. This class is defined by 
 spectra dominated by PAH emission with weak silicate absorption. 

\section{PAH emission and star formation}\label{Analysis}

As noted above, it is well known that infrared PAH emission is strongly 
correlated with star formation through excitation and IR fluorescence of PAHs 
induced by the UV radiation of young stars. The exceptional quality of our PAH 
spectra at $z \sim 2$ allows us to further assess the correlation between PAH 
luminosity and total IR luminosity, and to compare with other starburst 
samples at $z \sim 2$ and at lower redshift. The deep radio data allow a 
precise comparison with the radio luminosity which is also known to trace the 
starburst IR luminosity well.

\subsection{The correlation between IR and PAH luminosities}\label{ir}

\begin{figure}[!htbp]
\resizebox{\hsize}{!}{\includegraphics{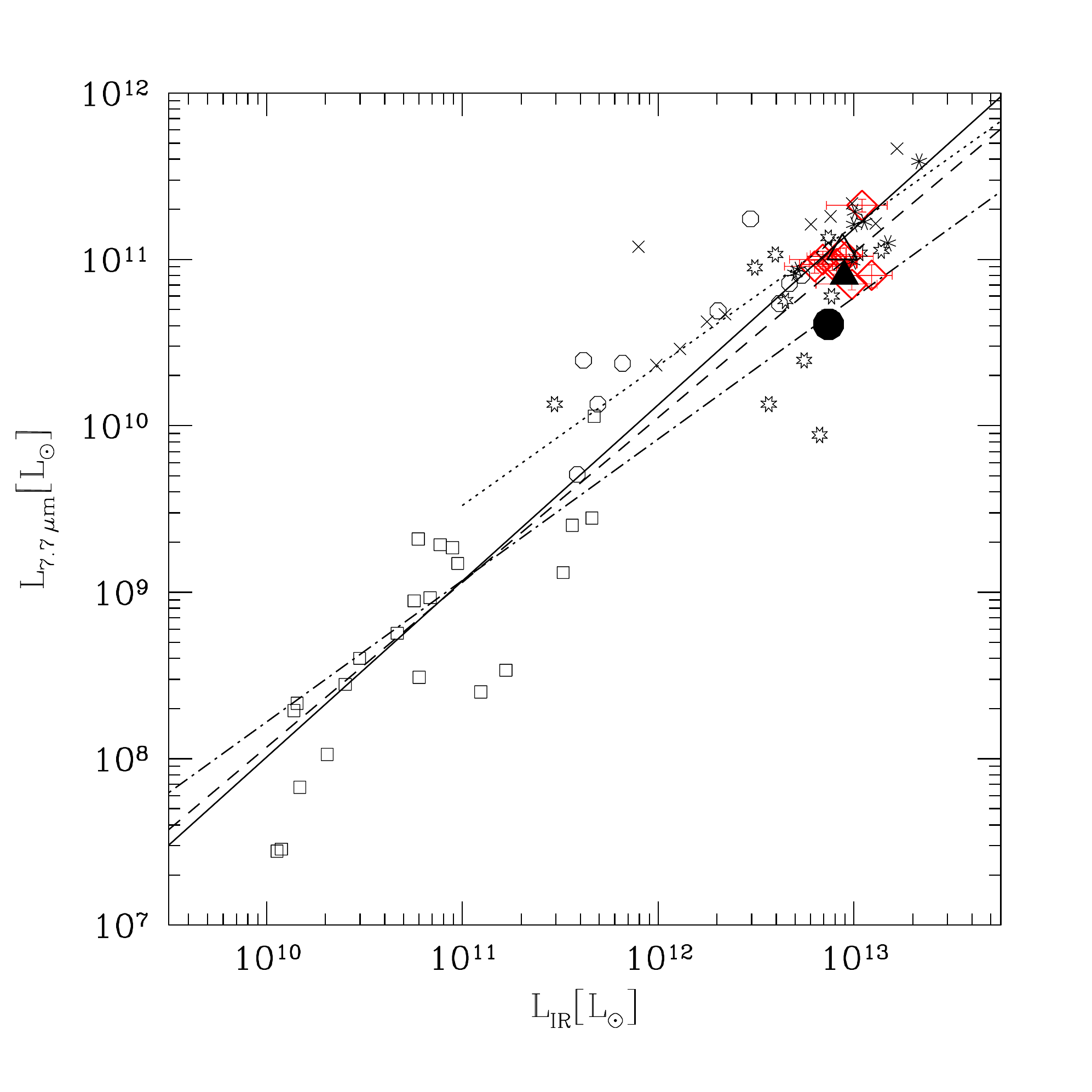}}
\caption{$L_{\rm 7.7\,\mu m}$ as a function of $L_{\rm IR}$. The symbols are 
the same as in Fig.~\ref{77_62}. The solid line is our best fit to all samples.
The dashed line is the best fit found by \citet{Pope08a} from fitting both 
their SMGs and the local starbursts. The dotted line is the best fit found by 
\citet{Shi09}. The dot-dashed line is the best fit found by \citet{Mene09}.}
\label{77_ir}
\end{figure}

The correlation between the strength of the PAH emission and the far-IR 
luminosity in starbursts is well established although not understood in detail.
Remarkably, it has been verified among local starburst galaxies over two 
orders of magnitude in $L_{\rm IR}$, from 10$^{10}$ to 10$^{12}\,L_{\odot}$, 
by \citet{Bran06}. The approximate proportionality between $L_{\rm 7.7\,\mu 
m}$ and $L_{\rm IR}$ has been extended through the ULIRG range up to 
10$^{13}\,L_{\odot}$ and $z \ge 2$ by several studies 
\citep{Shi09,Pope08a,Mene09}. Our sample gives us an opportunity to check how 
tight the relation is among the strongest starbursts at $z \sim 2$.

For the nine sources of our sample observed at 350\,$\mu$m by \citet{Kova10} --
all SMGs (F09) -- we have an accurate estimate of the infrared luminosity 
$L_{\rm IR}$ computed by \citet{Kova10}, who fit a distribution of grey-body 
models with different dust temperatures (multi-temperature model). All these 
sources are powerful ULIRGs with $L_{\rm IR}$ close to 10$^{13}\,L_{\odot}$ 
(average\,=\,$8.87 \times 10^{12}\,L_{\odot}$, $\langle\Delta L_{\rm 
IR}\rangle = 2.81 \times 10^{12}\,L_{\odot}$; see Table~\ref{lum}).
The average value of the ratio of IR and PAH luminosities is
\begin{equation}
\label{eq52}
\langle L_{\rm 7.7\,\mu m}/L_{\rm IR}\rangle = 11.7 \times 10^{-3}
\end{equation}
It is seen in Table~\ref{ratio} and Fig.~\ref{77_ir} that the $L_{\rm 
7.7\,\mu m}/L_{\rm IR}$ ratio is within a factor $\sim 2$ of those for the 
other high-$z$ samples. However, it is a factor $\sim 3.5$ larger than for 
less luminous local starbursts as yielded by Eq.~\ref{eq52}. \citet{Schw06} have also 
 demonstrated the existence of a correlation between the strength
 of the PAH emission and the far-IR 
luminosity  for local starburst-dominated ULIRGs and QSO hosts. 
 Nevertheless, the $\langle L_{\rm 7.7\,\mu m}/L_{\rm IR}\rangle$ ratio that they found 
 is probably lower because of the more compact nature of those objects. 

In Fig.~\ref{77_ir}, we plot $L_{\rm 7.7\,\mu m}$ as a function of $L_{\rm 
IR}$ for different samples. We can see that our sample has $L_{\rm IR}$ 
comparable to those of the other SMG samples \citep{Pope08a,Mene09}, and the 
PAH-rich sources from \citet{Saji07}. The star-forming galaxies of 
\citet{Shi09} have $L_{\rm IR}$ values slightly lower than our sample. We also 
note that the best fit of all these samples is consistent with the relations 
found by \citet{Pope08a} and \citet{Mene09} for SMGs and local starbursts 
\citep{Bran06}. The AGN-dominated sample of \citet{Copp10} shows 
$L_{\rm IR}$ and $L_{\rm 7.7\,\mu m}$ greater than our sources, but their 
ratio is not very different from our sample. The sources of \citet{Shi09} 
seem to follow a slightly different relation but with the same slope as the 
fit found by \citet{Mene09}. The best fit of all these samples, from local 
starbursts to SMGs, is
\begin{equation}
{\rm log}\,(L_{\rm 7.7\,\mu m}) = (1.06 \pm 0.04)\,{\rm log}\,(L_{\rm IR}) - 
(2.57\pm0.52)\\  
\label{eq6}
\end{equation}

This relation is consistent with $L_{\rm 7.7\,\mu m}$  
 being a good tracer of high-z starbursts: a strong $L_{\rm 7.7\,\mu m}$ should be the sign of a high SFR.
 If we use our relation between $L_{\rm 7.7\,\mu m}$ and $L_{\rm IR}$ 
 and assume the relation from \citet{Kenn98} between SFR and 
$L_{\rm IR}$ based on a 0.1--100\,$M_{\odot}$ Salpeter-like IMF, we obtain the following relation
\begin{equation}
{\rm log}\,({\rm SFR}) = 0.94\,{\rm log}\,(L_{\rm 7.7\,\mu m}) - 7.27
\label{eq5b}
\end{equation}
where SFR is in units of $M_{\odot}$\,yr$^{-1}$.

This relation is slightly different from that found by \citet{Mene09}.
  The latter is based on 
 the luminosity at 7.7\,$\mu$m calculated from the total flux between 7.1\,$\mu$m and 8.3\,$\mu$m after continuum substraction,
  instead 
 of between 7.3\,$\mu$m and 7.9\,$\mu$m as in our case. 
 However, the values of $L_{\rm 7.7\,\mu m}$ computed with the two definitions are very comparable. 
 Applying 
Eq.~\ref{eq5b}, we find on average $\langle {\rm SFR} \rangle = 1260 \pm 
470\,M_{\odot}$\,yr$^{-1}$. This value is consistent with the average SFR of 
1060\,M$_{\odot}$\,yr$^{-1}$ computed from the radio luminosity in F09, and 
close to the median SFR of $\sim$ 1200\,M$_{\odot}$\,yr$^{-1}$ found for SMGs 
by \citet[][]{Mene09}.

\subsection{The correlation between 1.4\,GHz and PAH luminosities}

The correlation between radio luminosity ($L_{\rm 1.4\,GHz}$) and 
far-infrared luminosity in star-forming regions and in local starbursts is 
well known \citep[e.g., ][]{Helo85,Cond92,Craw96,Sand96}. This correlation has 
been confirmed at $z \sim 2$ by \citet{Chap05} \citep[see also F09; ]
[]{Cond92,Smai02,Yun02,Ivis02,Kova06,Saji08,Youn09,Ivis10,Magn10}. 

\begin{figure}[!htbp]
\resizebox{\hsize}{!}{\includegraphics{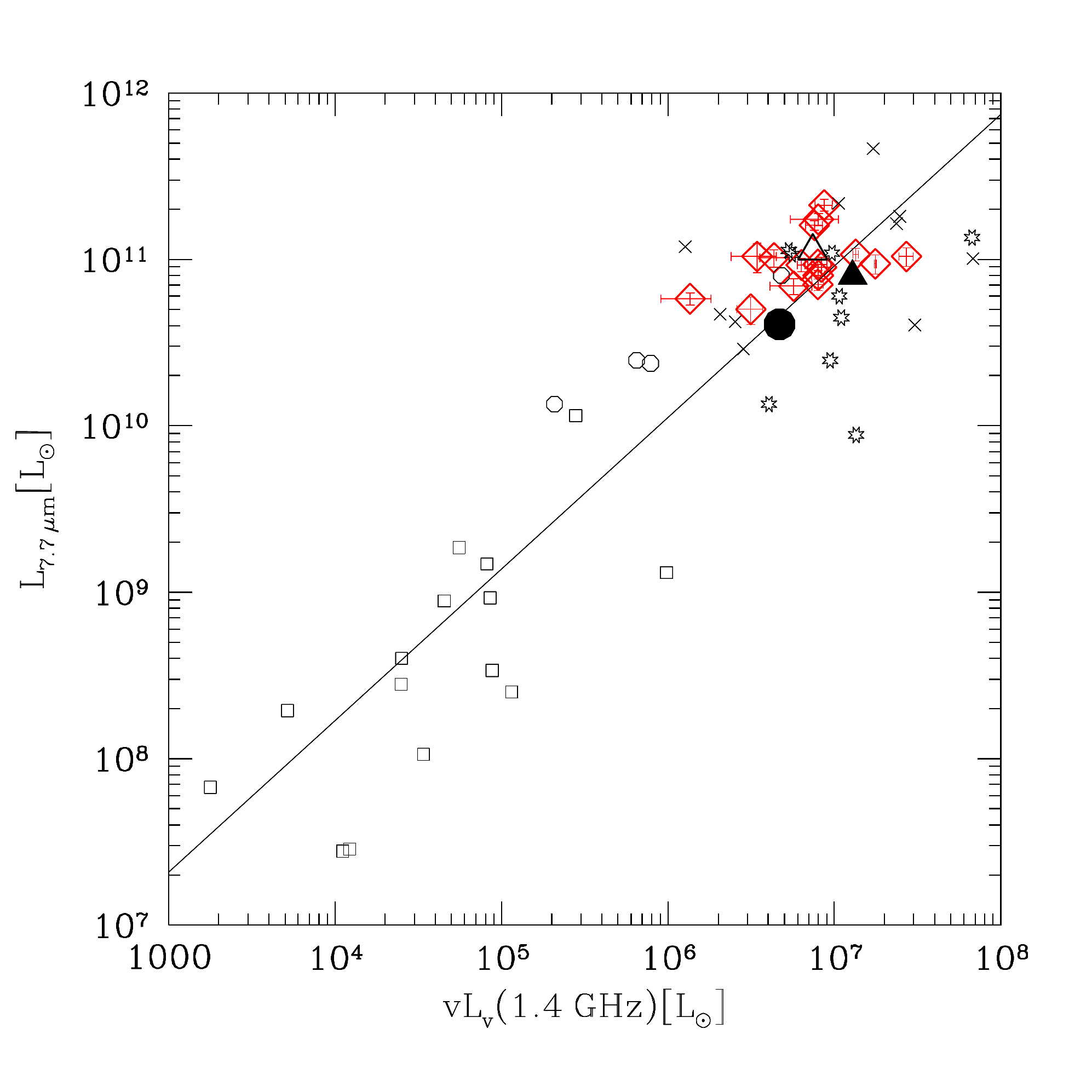}}
\caption{$L_{\rm 7.7\,\mu m}$ as a function of rest-frame 
$\nu\,L_{\nu}$(1.4\,GHz). The symbols are the same as in Fig.~\ref{77_62}.
 The black line is our best fit to all these samples.}
\label{77_14}
\end{figure}

From the correlation between $L_{\rm IR}$ and $L_{\rm 7.7\,\mu m}$ 
(Fig.~\ref{77_ir}, Eq.~\ref{eq6}), we might expect a very good correlation 
between PAH and radio luminosities, as previously observed by \citet{Pope08a}. 
Since our sources benefit from exceptionally good radio data at 20\,cm, 50\,cm,
and 90\,cm (Sec.~\ref{selection}), it is worthwhile to directly check the 
tightness of this relation.

It is straightforward to compute the rest-frame luminosity at 1.4\,GHz 
($L_{\rm 1.4\,GHz} = \nu\,L_{\nu}$(1.4\,GHz)) using, e.g., the formula given 
by  \citet{Kova06}:
\begin{equation}
\nu\,L_{\nu}({\rm 1.4\,GHz}) = 4\pi D_L^2\,S_{\rm 20\,cm}\,(1.4 \times 
10^{9})\,(1 + z)^{(-\alpha-1)}
\label{lum_radio}
\end{equation}
where $D_L$ is the luminosity distance and $\alpha$ is the radio spectral 
index defined  by the best power law fit, $S_{\nu}\propto \nu^{\alpha}$, 
between $S_{\rm{20\,cm}}$, $S_{\rm{50\,cm}}$, and $S_{\rm{90\,cm}}$. We have 
computed $\nu\,L_{\nu}$(1.4\,GHz) for our sample and those of  
\citet{Saji07}, \citet{Pope08a}, \citet{Shi09}, and \citet{Bran06}. For our sample, we have used 
$\alpha$ as computed in F09 and reported in Table~\ref{new}. For the PAH-rich 
sample of \citet{Saji07}, we have used the values published in \citet{Saji08}.
 For the other samples, we use the average value of $\alpha$, 
$\langle\alpha\rangle = -0.64$ found in F09, which is consistent with the 
typical radio spectral index for star-forming galaxies 
\citep[e.g., ][]{Cond92}. The values of $\nu\,L_{\nu}$(1.4\,GHz) are reported 
in Table~\ref{lum}. 

As expected, we observe a correlation between $L_{\rm 1.4\,GHz}$  and 
$L_{\rm 7.7\,\mu m}$. Considering our sample, but also the local star-forming 
galaxy sample \citep{Bran06}, the sample of SMGs \citep{Pope08a}, the PAH-rich 
sources from \citet{Saji07}, and the intermediate-redshift starbursts 
\citep{Shi09}, we obtain the following relation (Fig~\ref{77_14}): 
\begin{equation}
{\rm log}\,(L_{\rm 7.7\,\mu m}) = (0.91\pm0.06)\,{\rm log}\,(L_{\nu}(\rm 
1.4\,GHz))+(4.60\pm0.36)
\label{eq7}
\end{equation}
The effect of the positive $k$-correction and the value of $\alpha$ are 
particularly important for the determination of $L_{\rm 1.4\,GHz}$ and may 
significantly affect the reliability of the relation stated in Eq.~\ref{eq7}. 
 The correlation between  $L_{\rm 1.4\,GHz}$ and $L_{\rm 7.7\,\mu m}$ seems to be more scattered than the correlation between 
$L_{\rm IR}$ and $L_{\rm 7.7\,\mu m}$. It is not surprising considering the uncertainties on $L_{\rm 1.4\,GHz}$ and the scatter
  of the correlation between $L_{\rm IR}$ and $L_{\rm 1.4\,GHz}$ as shown by e.g.~\citet{Kova10}.

\subsection{PAH emission and stellar mass}

\begin{figure}[!htbp]
\resizebox{\hsize}{!}{\includegraphics{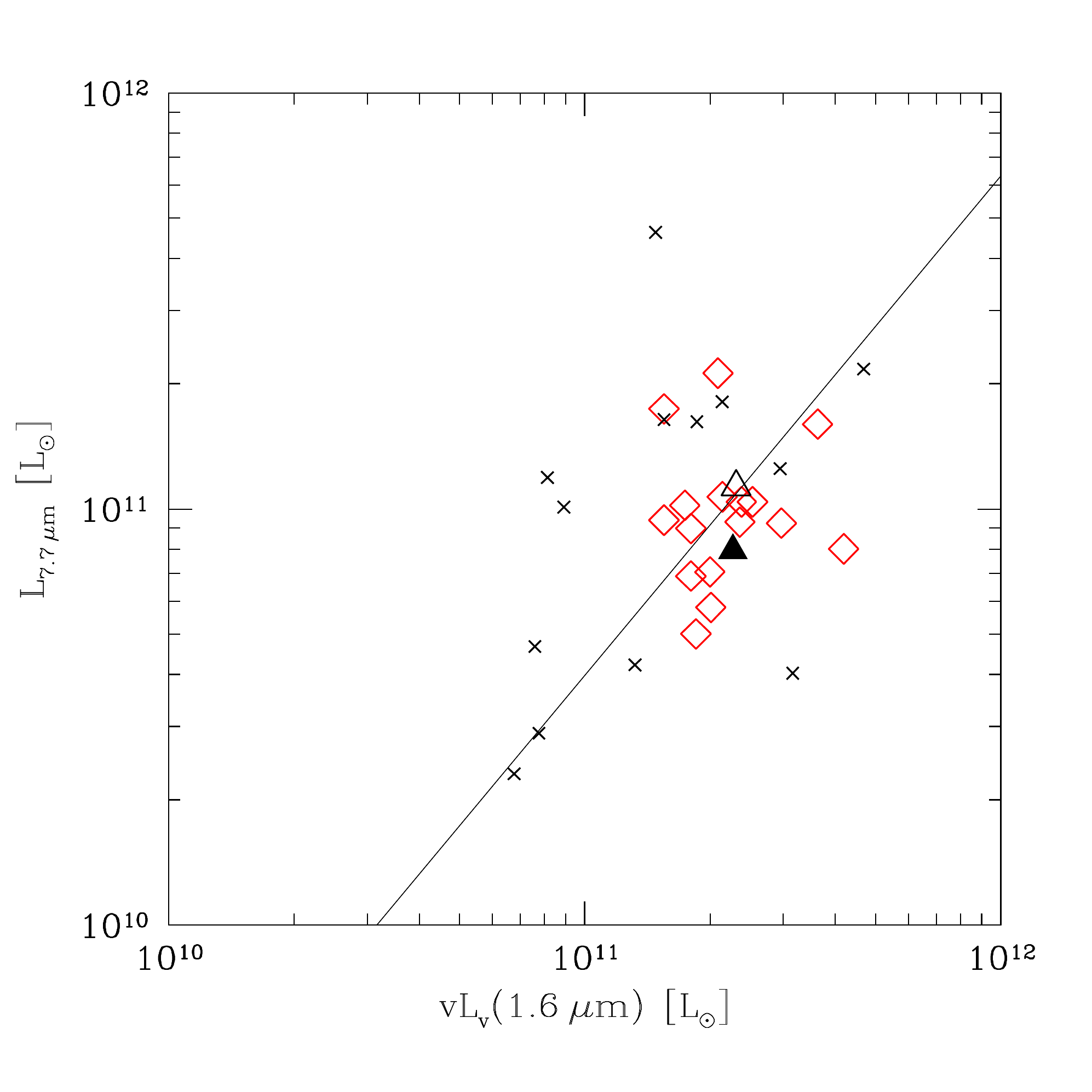}}
\caption{$L_{\rm 7.7\,\mu m}$ as a function of the stellar bump luminosity 
$L_{\rm 1.6\,\mu m}$. The symbols are the same as in Fig.~\ref{77_62}.  Only the samples at $z \sim 2$ are plotted. 
The  line is the best fit found for the ``normal'' star-forming galaxies of 
\citet{Lu03}.}
\label{16_77}
\end{figure}

The stellar bump luminosity at 1.6\,$\mu$m ($\nu\,L_{\nu}$(1.6\,$\mu$m), 
reported in Table~\ref{lum}, gives an estimate of the stellar mass 
($M_{\star}$) of our sources \citep[see, e.g., F09; ][]{Seym07,Lons09}.
Applying the same mass-to-light ratio as F09 
($M_{\star}/\nu\,L_{\nu}({\rm 1.6\,\mu m}) = 0.7\,M_{\odot}/L_{\odot}$), 
$M_{\star}$ derived from $\nu\,L_{\nu}$(1.6\,$\mu$m) ranges from 1.1 to 
$3.0 \times 10^{11}\,M_{\odot}$ (average\,=\,1.6$\times$10$^{11}\,M_{\odot}$, 
median\,=\,1.4$\times$10$^{11}\,M_{\odot}$). 

In Fig.~\ref{16_77}, we plot $L_{\rm 7.7\,\mu m}$ as a function of 
$\nu\,L_{\nu}$(1.6\,$\mu$m) for our sources and the PAH-rich sample from 
\citet{Saji07}. Our sample shows no clear difference from the sources of 
\citet{Saji07} in terms of $\nu\,L_{\nu}$(1.6\,$\mu$m). Both samples follow 
the relation found for ``normal'' star-forming galaxies 
\citep[][]{Saji07,Lu03}. This result implies that $L_{\rm 7.7\,\mu m}$ and 
$\nu\,L_{\nu}$(1.6\,$\mu$m) do not trace AGN 
\citep{Saji07}. Since the PAH luminosity is tightly correlated with the star 
formation rate, the relation between $\nu\,L_{\nu}$(1.6\,$\mu$m) and 
$L_{\rm 7.7\,\mu m}$ seems to confirm the existence of a correlation between star formation 
rate and stellar mass at $z \sim 2$ as previously observed by \citet{Pann09} and \citet{Moba09}. 
 However, note that the mass-to-light ratio remains uncertain because of uncertainty 
 in the age of the dominant stellar population.

\section{Discussion}\label{Discussion}

\subsection{Distinction between starburst and AGN}\label{agn}

\begin{figure}[!htbp]
\begin{minipage}{0.5\linewidth}
\includegraphics[scale=0.45]{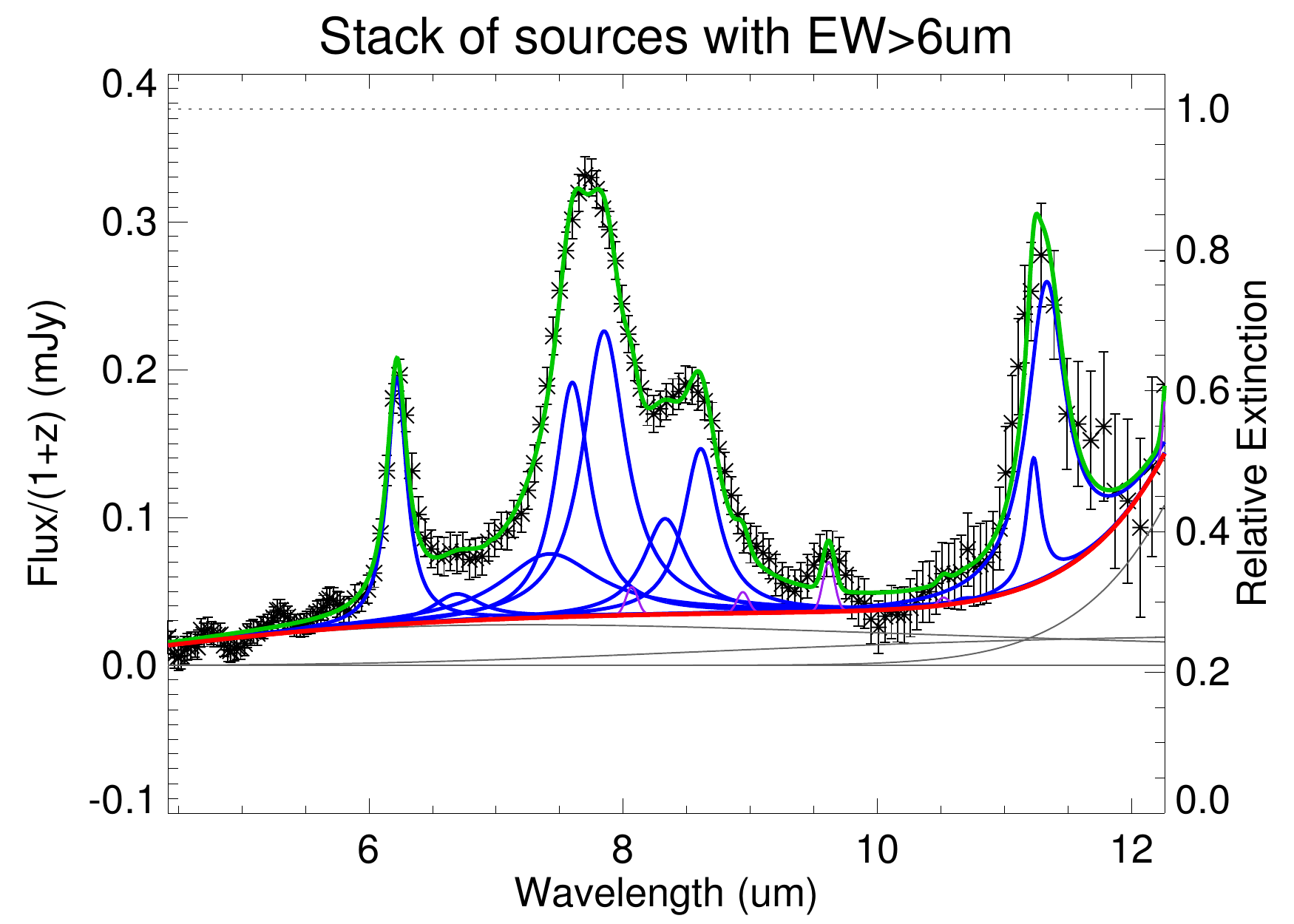}
\end{minipage}
\begin{minipage}{0.5\linewidth}
\includegraphics[scale=0.45]{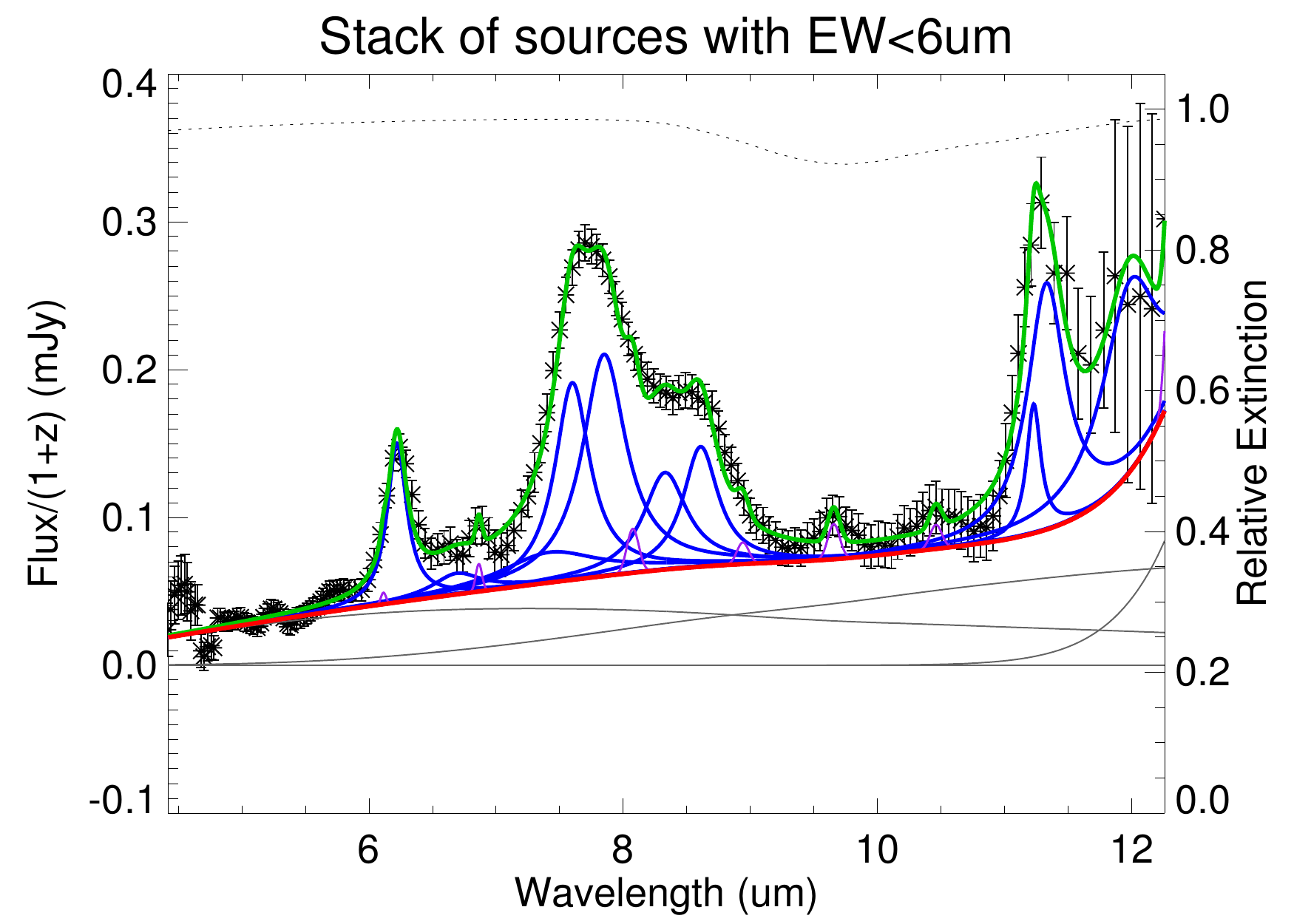}
\end{minipage}
\caption{PAHFIT spectral decomposition of stacked spectra. \textit{Upper 
panel:} 7 sources with EW$_{\rm 7.7\,\mu m} > 6\,\mu$m. \textit{Lower 
panel:} 9 sources with EW$_{\rm 7.7\,\mu m} < 6\,\mu$m. The solid green line 
is the fitted model. The blue lines above the continuum are the PAH features. 
The narrow violet lines are the spectral lines. The grey lines represent the 
thermal dust continuum components. The thick red line shows the total 
continuum (stars+dust). The dotted black line shows the extinction 
($e^{-\tau_{\lambda}}$\, $=1$ if no extinction).}
\label{stackew}
\end{figure}

As previously discussed, the relative strength of the PAH features and the 
mid-IR continuum is a very efficient criterion for distinguishing starburst- 
from AGN-dominated ULIRGs \citep[e.g., ][]{Yan07,Saji07,Weed06,Houc05}.
\citet{Genz98} and \citet{Rigo99} use the parameter 7.7\,$\mu$m-(L/C), defined 
as the ratio between the peak flux density at 7.7\,$\mu$m and the continuum 
flux density at the same wavelength, to distinguish between AGN and starburst.
AGN-dominated composite sources show values of 7.7\,$\mu$m-(L/C) weaker than 
starburst-dominated systems. An equivalent straightforward indicator is the 
equivalent width of the PAH features, especially EW$_{\rm 7.7\,\mu m}$, since 
it is proportional to the ratio of $L_{\rm 7.7\,\mu m}$ to the 7.7\,$\mu$m 
continuum. The equivalent widths of our sources span a broad range, more than 
a factor of 5. Despite the large uncertainty in the continuum, e.g., as 
estimated by PAHFIT, it is clear that EW$_{\rm 7.7\,\mu m}$ should be a 
relevant parameter for discussion of AGN contributions. For our entire sample, 
we find  $\langle {\rm EW}_{\rm 7.7\,\mu m}\rangle = 5.64\,{\rm \mu m}$ 
(median\,=\,4.88) vs. 3.04\,$\mu$m for the stacked spectrum of \citet{Mene09}.

Based on the values of EW$_{\rm 7.7\,\mu m}$ (see Table~\ref{few}), we have 
constructed two subsamples: the first has EW$_{\rm 7.7\,\mu m} > 6\,\mu$m 
(7 sources, $\langle {\rm EW}_{\rm 7.7\,\mu m}\rangle = 8.33\,\mu$m) and the 
second has EW$_{\rm 7.7\,\mu m} < 6\,\mu$m (9 sources, $\langle {\rm 
EW}_{\rm 7.7\,\mu m}\rangle = 3.55\,\mu$m). We have produced stacked spectra 
of these two subsamples, built from the weighted means of the individual 
spectra scaled to $\langle F_{\rm 7.7\,\mu m}\rangle$. They are plotted in 
Fig.~\ref{stackew}. It is obvious, as expected, that the second subsample has 
a significantly stronger continuum. 

Applying the same method as \citet{Mene09}, we estimate an X-ray luminosity 
($L_{X}$) from the 10\,$\mu$m continuum flux (Table~\ref{few}) via the \citet{Krab01} relation. 
Then assuming the $L_{X}/L_{\rm IR}$ found for AGN by \citet{Alex05}, 
we estimate the contribution of AGN to $L_{\rm IR}$. We find that the  
AGN contribution is $\sim 20\%$ for our whole sample and $\sim 10\%$ and 
$\sim 25\%$ for the subsamples with EW$_{\rm 7.7\,\mu m} > 6\,\mu$m and with 
EW$_{\rm 7.7\,\mu m} < 6\,\mu$m, respectively. \citet{Mene09} have found 
$\sim 32\%$ to be the AGN contribution in SMGs.  The difference can be 
interpreted as a difference in the continuum flux density, with continuum 
larger by a factor $\sim 2$ for the sample of \citet{Mene09}.

In Fig.~\ref{77_ir} (see also Table~\ref{lumsam}), it is seen that the two 
subsamples are not different in term of $L_{\rm IR}$ and follow the 
correlation between $L_{\rm 7.7\,\mu m}$ and $L_{\rm IR}$ found in 
Sec.~\ref{ir}. This result tends to argue that the majority of the IR 
luminosity is associated with the starburst and not with the AGN. This is not 
the case in the sample of \citet{Mene09}, which shows a smaller  
$L_{\rm 7.7\,\mu m}/L_{\rm IR}$ ratio, implying that the AGN contributes more 
to $L_{\rm IR}$ (Table~\ref{ratio}).

Another possible sign of a strong AGN contribution is a large  $L_{\rm 
7.7\,\mu m}/L_{\rm 6.2\,\mu m}$ ratio, as discussed by, e.g., \citet{Rigo99} 
and \citet{Pope08a}. They have observed such a rise in $L_{\rm 7.7\,\mu m}$ 
versus $L_{\rm 6.2\,\mu m}$ for some SMGs, which they attribute to greater 
silicate absorption in the presence of a more luminous AGN than is seen in  
most SMGs or starburst galaxies. Stronger silicate absorption significantly 
affects $L_{\rm 8.6\,\mu m}$ and $L_{\rm 11.3\,\mu m}$, while it may lead to 
an overestimate of $L_{\rm 7.7\,\mu m}$. However, as seen in Figs.~\ref{77_62} 
and \ref{77_113}, there is no significant difference in the value of 
$L_{\rm 7.7\,\mu m}/L_{\rm 6.2\,\mu m}$ between sources with large and small 
EWs, perhaps because the silicate absorption is small.

We see in Fig.~\ref{77_14} and Table~\ref{lumsam} that sources with 
EW$_{\rm 7.7\,\mu m} < 6\,\mu$m seem to have an average value of 
$\nu\,L_{\nu}$(1.4\,GHz) somewhat greater than that of sources with 
EW$_{\rm 7.7\,\mu m} > 6\,\mu$m. Radio excess could be a good indication of 
AGN strength \citep{Seym08,Arch01,Reul04}. However, we note that practically 
all of this difference is due to two sources in the small-EW subsample, L11 
and L17, which are close to the radio loudness limit at $z \sim 2$ as defined 
by \citet{Jian07} and \citet{Saji08}: $\nu$\,L$_{\nu}({\rm 1.4\,GHz}) = 3.66 \times 10^{7}\,L_{\odot}$. 
 They are thus good candidates for having 
significant AGN emission. Applying the same method as previously to determinate the AGN contribution, 
 we find 26\% and 19\% for L11 and L17, respectively. 

To summarize: although small, it seems that the AGN contribution is slightly 
greater in the sources with EW$_{\rm 7.7\,\mu m} < 6\,\mu$m than the systems 
with EW$_{\rm 7.7\,\mu m} > 6\,\mu$m. Nevertheless, all these sources show 
strong PAH emission and individually follow more or less the relations found for local 
starbursts \citep{Bran06}. They are probably all starburst-dominated or 
AGN-starburst composite sources, with at most a weak AGN contribution to the 
mid-infrared. 

Our sample is particularly homogeneous in term of PAH luminosity (see 
Table~\ref{lum}). Only one source, L10, is different. Indeed, this source
shows a spectrum slightly different from the others. Its continuum is 
stronger and has a larger slope (Fig.~\ref{Spectrald}). Moreover, L10 deviates 
from the correlation between $L_{\rm 6.2\,\mu m}$ and $L_{\rm 7.7\,\mu m}$ 
with a low $L_{\rm 6.2\,\mu m}$. L10 also presents the weakest EW$_{\rm 
6.2\,\mu m}$, EW$_{\rm 7.7\,\mu m}$, and EW$_{\rm 8.6\,\mu m}$ of our sample, 
with values of 0.10, 2.32, and 0.77\,$\mu$m, respectively. Such a difference 
can be explained by a larger AGN contribution to the mid-infrared emission, 
and also by a low signal-to-noise ratio (see Fig~\ref{sed_spectra}). 
Nevertheless, the fact that L10 follows the other correlations -- especially 
the correlation between $\nu\,L_{\nu}$(1.4\,GHz) and $L_{\rm 7.7\,\mu m}$ -- 
seems to prove that this source is starburst-dominated. The same kind of 
source was observed by \citet{Pope08a}.

\subsection{Comparison with other studies}\label{studies}

\begin{figure}[!hbtp]
\begin{minipage}{0.5\linewidth}
\includegraphics[scale=0.45]{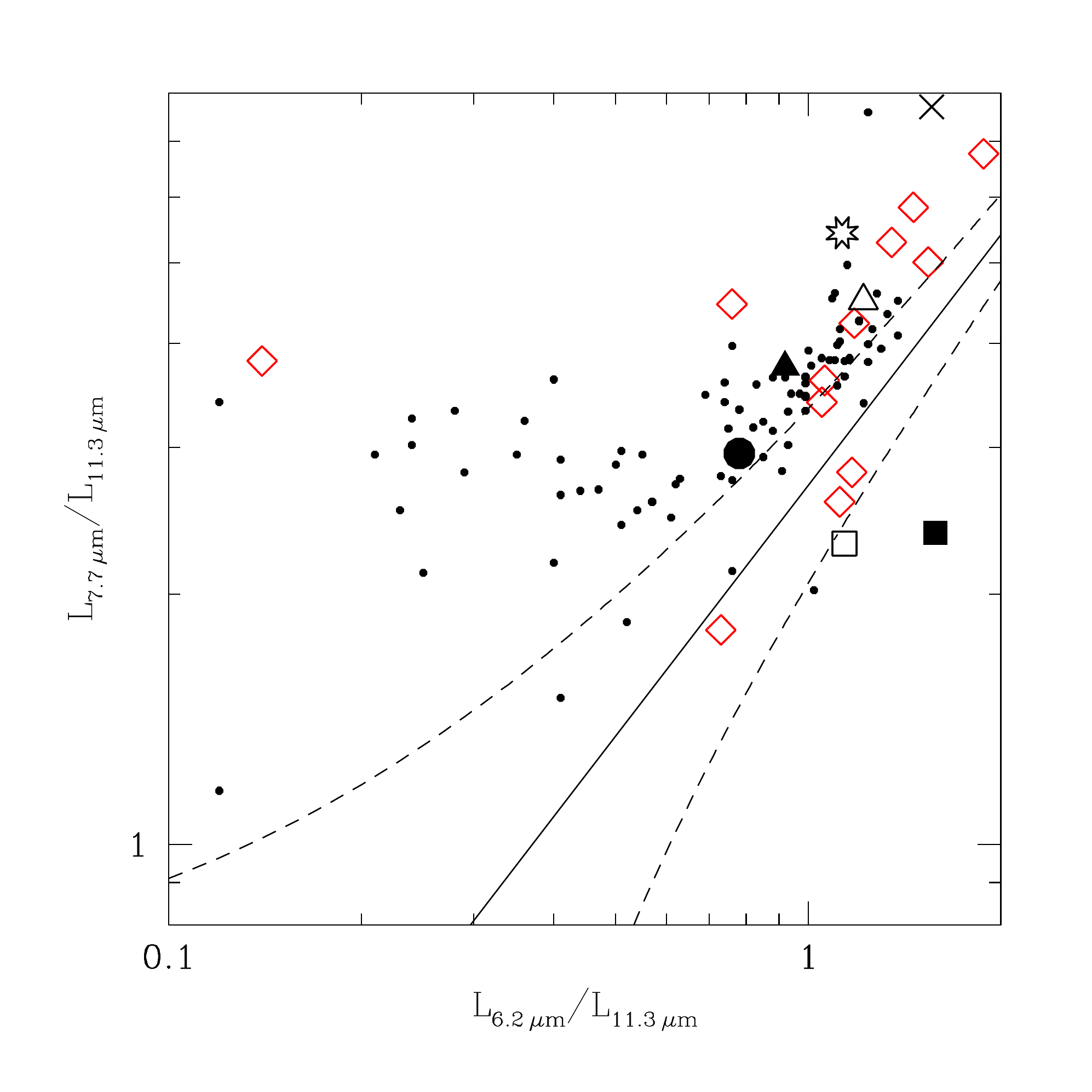}
\end{minipage}
\begin{minipage}{0.5\linewidth}
\includegraphics[scale=0.45]{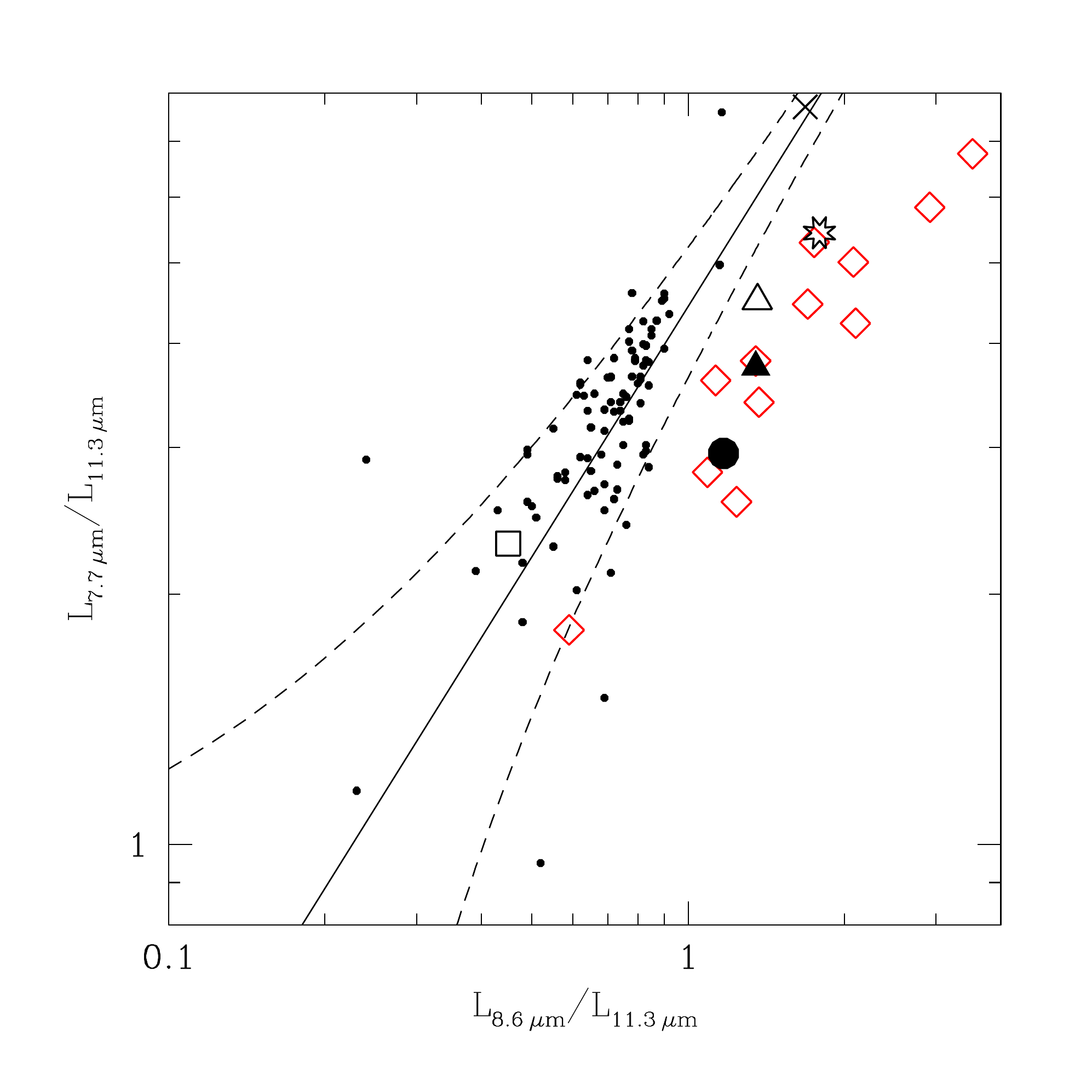}
\end{minipage}
\caption{{\it Upper panel:} $L_{\rm 7.7\,\mu m}/L_{\rm 11.3\,\mu m}$ as a 
function of $L_{\rm 6.2\,\mu m}/L_{\rm 11.3\,\mu m}$. {\it Lower panel:} 
$L_{\rm 7.7\,\mu m}/L_{\rm 11.3\,\mu m}$ as a function of $L_{\rm 8.6\,\mu 
m}/L_{\rm 11.3\,\mu m}$. In both panels, the red diamonds are the sources in  
our sample. The small black circles are the sample of \citet{Odow09}. The open 
(full) black triangle is the stack of our sources with EW$_{\rm 7.7\,\mu 
m}>\,(<)\,6\,\mu$m. The black cross shows the average for the PAH-rich sources 
from \citet{Saji07}. The black star shows the average for the sample of SMGs 
from \citet{Pope08a}. The open black square shows the average for the local 
star-forming galaxies from \citet{Bran06}. The full black square is the average for the sample of ``4.5\,$\mu$m-peakers'' from \citet{Farr08}.
 The large black circle is based on 
the composite spectrum of the SMG sample from \citet{Mene09}. The black line 
and the dotted lines are the best linear regression and the $1\sigma$ 
dispersion found by \citet{Gall08} for Galactic HII regions, dwarf spirals, 
and starburst galaxies.} 
\label{figra}
\end{figure}

The present work is based on a complete sample of 16 sources selected to be 
starburst ``5.8\,$\mu$m-peakers'' that are bright at 24\,$\mu$m, with $S_{\rm 
24\,\mu m} > 0.50$\,mJy. The average flux density of our sample is $\langle 
S_{\rm 24\,\mu m} \rangle = 0.63$\,mJy. This is a brighter sample of SMGs 
compared to that of \citet{Pope08a} ($\langle S_{\rm 24\,\mu m} \rangle = 
0.45$\,mJy) or \citet{Mene09} ($\langle S_{\rm 24\,\mu m} \rangle = 
0.33$\,mJy). On the other hand, the PAH-rich sample of \citet{Saji07} is 
biased toward the brightest $24\,{\rm \mu m}$ sources, with a mean 
$\langle S_{\rm 24\,\mu m} \rangle = 1.19$\,mJy. Our sample is also brighter than the ULIRGs at 
 $z \sim 2$ of \citet{Fadd10} ($\langle S_{\rm 24\,\mu m} \rangle = 
0.27$\,mJy). All these samples have an 
average redshift $z \sim 2$, but ours spans a narrower redshift range than the 
others. Indeed, our sample has $z_{\rm PAH}$ between 1.75 and 2.28, while the 
samples of \citet{Pope08a} and \citet{Mene09}, the ULIRGs from \citet{Fadd10}  and the PAH-rich sample of 
\citet{Saji07} span $z = 0.93 - 2.55$, 0.69--3.62, 1.62--2.44, and 0.82--2.47, 
respectively.

The differences between PAH luminosities for all the samples are small, but 
 our PAH luminosities show an higher S/N than the other high-redshift samples  
(Table~\ref{lumsam}). Our sample has PAH luminosities in all bands quite 
comparable to those of \citet{Saji07} and about twice as large as those of 
\citet{Pope08a} and \citet{Mene09}.

As regards the PAH-band luminosity ratios, we see in  Table~\ref{ratio} that 
our sample is not very different in term of ratios from the other SMG samples 
\citep{Pope08a,Mene09,Saji07}. Nevertheless, \citet{Mene09} find lower ratios 
for $L_{\rm 6.2\,\mu m}/L_{\rm IR}$ and  $L_{\rm 7.7\,\mu m}/L_{\rm IR}$. 
This is probably due to a greater AGN contribution to $L_{\rm IR}$ 
(Sec.~\ref{agn}). 

Our sample also shows a higher $L_{\rm 7.7\,\mu m}/L_{\rm IR}$ ratio than the 
lower-redshift sample of \citet{Bran06}. This difference could be explained by a 
more extended PAH distribution in our sources. Indeed, if the PAH distribution is extended, 
 a larger fraction of PAH can survive in strong UV radiation field from AGN or young stars. 
 Then $L_{\rm 7.7\,\mu m}/L_{\rm IR}$ becomes higher than in compact PAH distributions where 
 PAH are easier destroyed by UV radiation field \citep[e.g., ][]{Huan09}.

In Fig.~\ref{figra}, we plot the different PAH ratios of these samples 
following \citet{Odow09}. We find that our sample and the other SMG samples 
show $L_{\rm 7.7\,\mu m}/L_{\rm 11.3\,\mu m}$ vs. $L_{\rm 6.2\,\mu m}/L_{\rm 
11.3\,\mu m}$ ratios consistent with the sample of \citet{Odow09}. The sample 
of \citet{Bran06} falls in the border of the dispersion for the PAH ratios of 
Galactic HII regions, dwarf spirals, and starburst galaxies found by 
\citet{Gall08}. The plot of $L_{\rm 7.7\,\mu m}/L_{\rm 11.3\,\mu m}$ vs. 
$L_{\rm 8.6\,\mu m}/L_{\rm 11.3\,\mu m}$ shows that all these samples have 
ratios different from those of \citet{Odow09} because of stronger emission at 
8.6\,$\mu$m. Our sample does not follow the relation found by \citet{Gall08}.

\begin{figure}[!htbp]
\resizebox{\hsize}{!}{\includegraphics{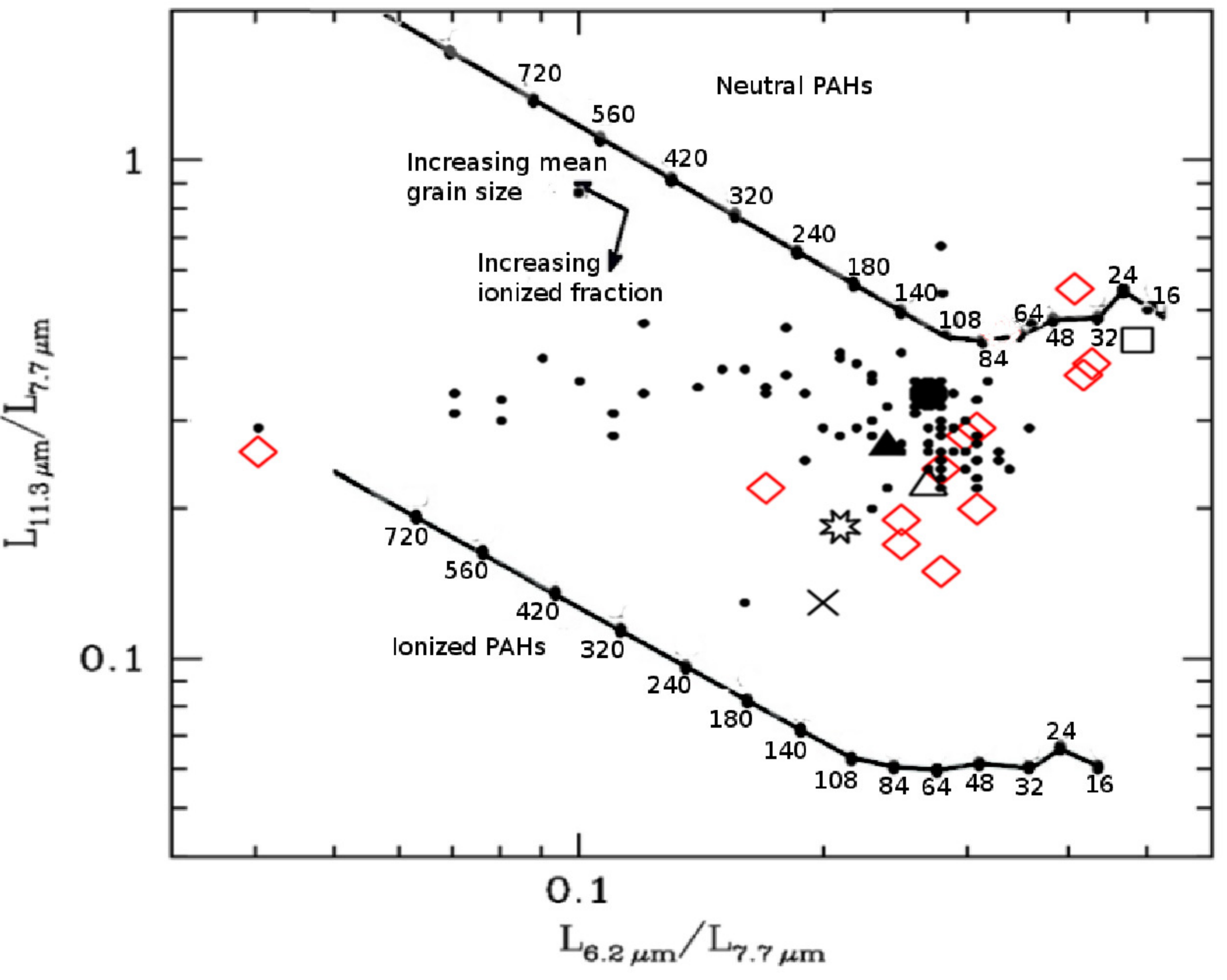}}
\caption{Adapted from \citet{Odow09}: $L_{\rm 11.3\,\mu m}/L_{\rm 
7.7\,\mu m}$ as a function of $L_{\rm 6.2\,\mu m}/L_{\rm 7.7\,\mu m}$. The 
symbols are the same as in Fig.~\ref{figra}. The black lines represent the 
expected ratios for fully neutral or fully ionized PAH molecules of a given 
number of carbon atoms from the models of \citet{Drai01}.}
\label{ioni}
\end{figure}

A likely explanation of these differences in PAH ratios is modification of the 
size or ionization of PAH grains. Fig.~\ref{ioni}, adapted from \citet{Odow09},
seems to prove that our sample has PAH grains with sizes comparable to those 
in the other samples of SMGs, but with a slightly lower fraction of ionized 
grains than the samples of \citet{Pope08a} and (especially) \citet{Saji07}. 
\citet{Bran06} find smaller and more neutral grains in local starbursts than our sample. The 
difference could arise due to the strongest ionizing UV radiation from young 
stars in the most powerful starbursts but less ionising than in the sample of 
 \citet{Pope08a} and  \citet{Mene09} probably because of lower AGN contamination. 
  
\section{Summary and conclusions}\label{conclusion}

In this paper, we have presented the results of {\it Spitzer}/IRS observations 
of a complete sample of 16 sources selected to be 24\,$\mu$m-bright and 
``5.8\,$\mu$m-peakers.'' The spectra obtained show very strong PAH features at 
6.2, 7.7, 8.6, and 11.3\,$\mu$m, along with a weak continuum. Thanks to these 
exceptionally strong features, we have estimated accurate PAH redshifts that 
span the range $z = 1.75-2.28$. Our average PAH redshift is $\langle z_{\rm 
PAH}\rangle = 2.02 \pm 0.15$, and the average error of the individual $z_{\rm 
PAH}$ is $\langle \Delta z_{\rm PAH}\rangle = 0.010$ (median\,=\,0.007; 
min\,=\,0.004; max\,=\,0.064). The stacked spectrum of our 16 sources displays evidence for the pure 
rotational 0-0\,S(3) molecular hydrogen line at $\lambda = 9.67\,\mu$m, 
 which is confirmed and analyzed in a separate paper.

Thanks to the very good quality of our IRS spectra, 
 we have calculated PAH luminosities $L_{\rm 6.2\,\mu m}$, $L_{\rm 7.7\,\mu m}$,
$L_{\rm 8.6\,\mu m}$, and $L_{\rm 11.3\,\mu m}$ to have average values $2.90 
\pm 0.31$, $10.38 \pm 1.09$, $3.62 \pm 0.27$, and $2.29 \pm 0.26 \times 
10^{10}\,L_{\odot}$, respectively. These luminosities have a S/N higher than the other SMGs samples. 
 We have studied the correlation between  
$L_{\rm 7.7\,\mu m}$ and the other PAH luminosities. All our sources, except 
perhaps L10, follow a correlation similar to that found for local starbursts, and they are 
not very different from other samples of SMGs at $z \sim 2$. We have also 
confirmed the very good correlation, previously observed, between $L_{\rm 
7.7\,\mu m}$ and $L_{\rm IR}$ at high redshift, and we have verified its 
extension to the radio luminosity  $\nu\,L_{\nu}$(1.4\,GHz). It also extends 
more loosely to stellar mass through $\nu\,L_{\nu}$(1.6\,$\mu$m).

All these relations, the luminosity ratios, and the equivalent widths allow us 
to estimate the AGN contribution to the mid-IR luminosity. We conclude that 
our sources are starburst-dominated and that the AGN contribution is 
$\sim$20\%. This sample is the most pure selection of 
 massive starbursts at $z \sim 2$ compared to the other $z \sim 2$
 \textit{Spitzer}-selected samples. The known fact that the equivalent width of PAH features
  is a good discriminant between starburst-dominated and AGN-dominated sources is reinforced 
 by the study of two subsamples 
of sources with EW$_{\rm 7.7\,\mu m}\,>6\,\mu$m and EW$_{\rm 7.7\,\mu m} < 
6\,\mu$m. The subsample with EW$_{\rm 7.7\,\mu m} < 6\,\mu$m shows a larger 
AGN contribution ($\sim 25\%$) than the EW$_{\rm 7.7\,\mu m} > 6\,\mu$m 
subsample ($\sim 10\%$), although the AGN contribution is small in both 
subsamples.

\begin{acknowledgements}
We thank Karin Men\'endez-Delmestre and H\'el\`ene Roussel for their helpful 
contributions. This work is based primarily on IRS observations and 
observations made within the context of the SWIRE survey with the {\it Spitzer 
Space Telescope}, which is operated by the Jet Propulsion Laboratory, 
California Institute of Technology under NASA contract. This work includes 
observations made with IRAM, which is supported by INSU/CNRS (France), MPG 
(Germany) and IGN (Spain). The VLA is operated by NRAO, the National Radio 
Astronomy Observatory, a facility of the National Science Foundation 
operated under cooperative agreement by Associated Universities, Inc. The CSO 
is funded by NSF Cooperative Agreement AST-083826. G.L., B.B., A.B. and F.B. 
 had support for this work provided by ANR ``D--SIGALE" ANR--06-- BLAN--0170 grant. 

\end{acknowledgements}

\bibliographystyle{aa}

\bibliography{15504bib}

\begin{appendix}

\section{Spectra and best fit-templates}\label{app}

In Fig. \ref{sed_spectra}, we present our spectra and the best-fit templates 
among the 21 templates described in Sec~\ref{redshift} using the method described in Bertincourt et al. (2009). 
 We also plot the 24\,$\mu$m flux densities from the SWIRE catalogue 
and the values determined from the IRS spectra convolved with the MIPS 
24\,$\mu$m filter.

\begin{figure*}
\label{sed_spectra}
\begin{minipage}{0.5\linewidth}
\includegraphics[height=6cm,width=9cm]{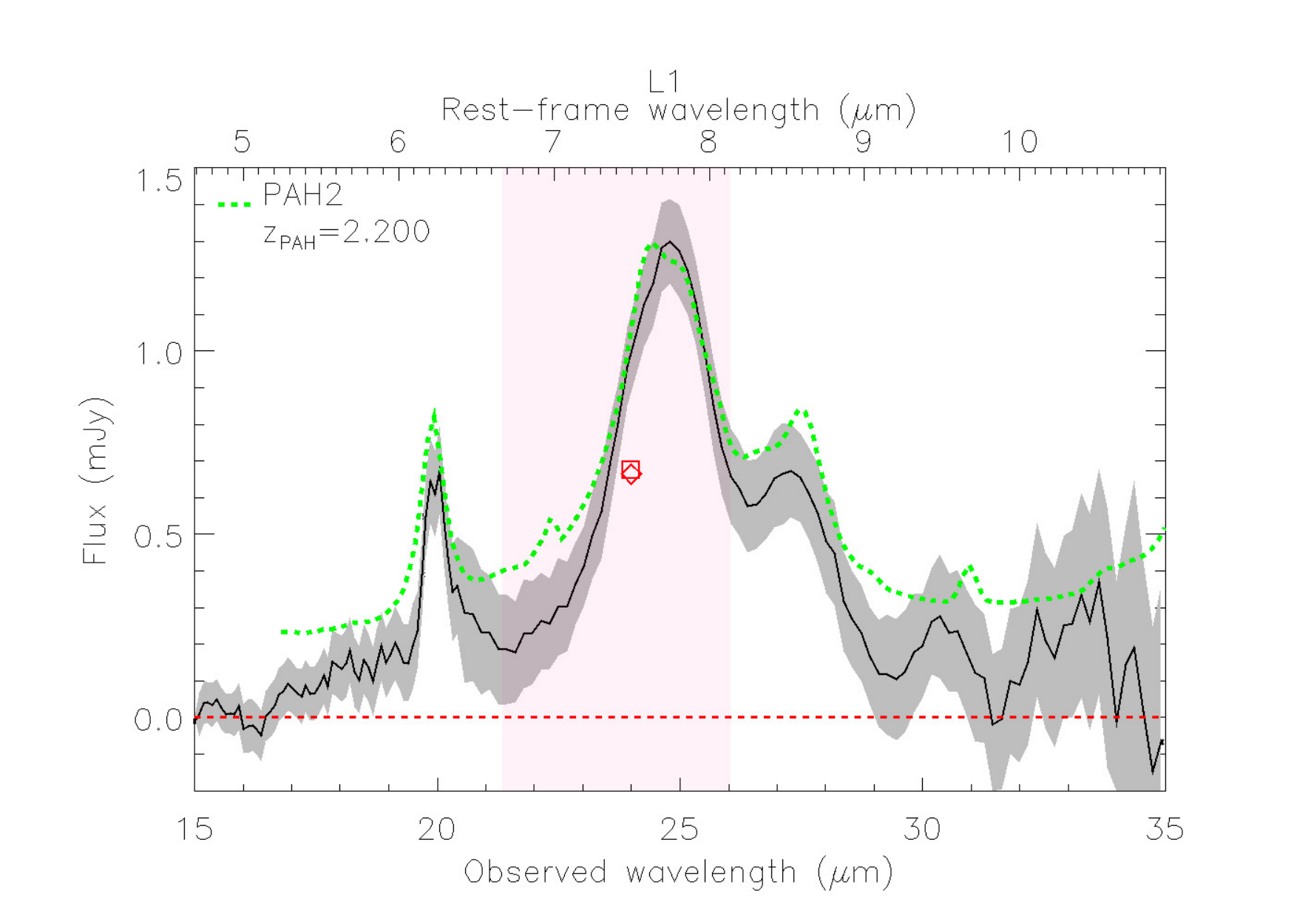}
\end{minipage}
\begin{minipage}{0.5\linewidth}
\includegraphics[height=6cm,width=9cm]{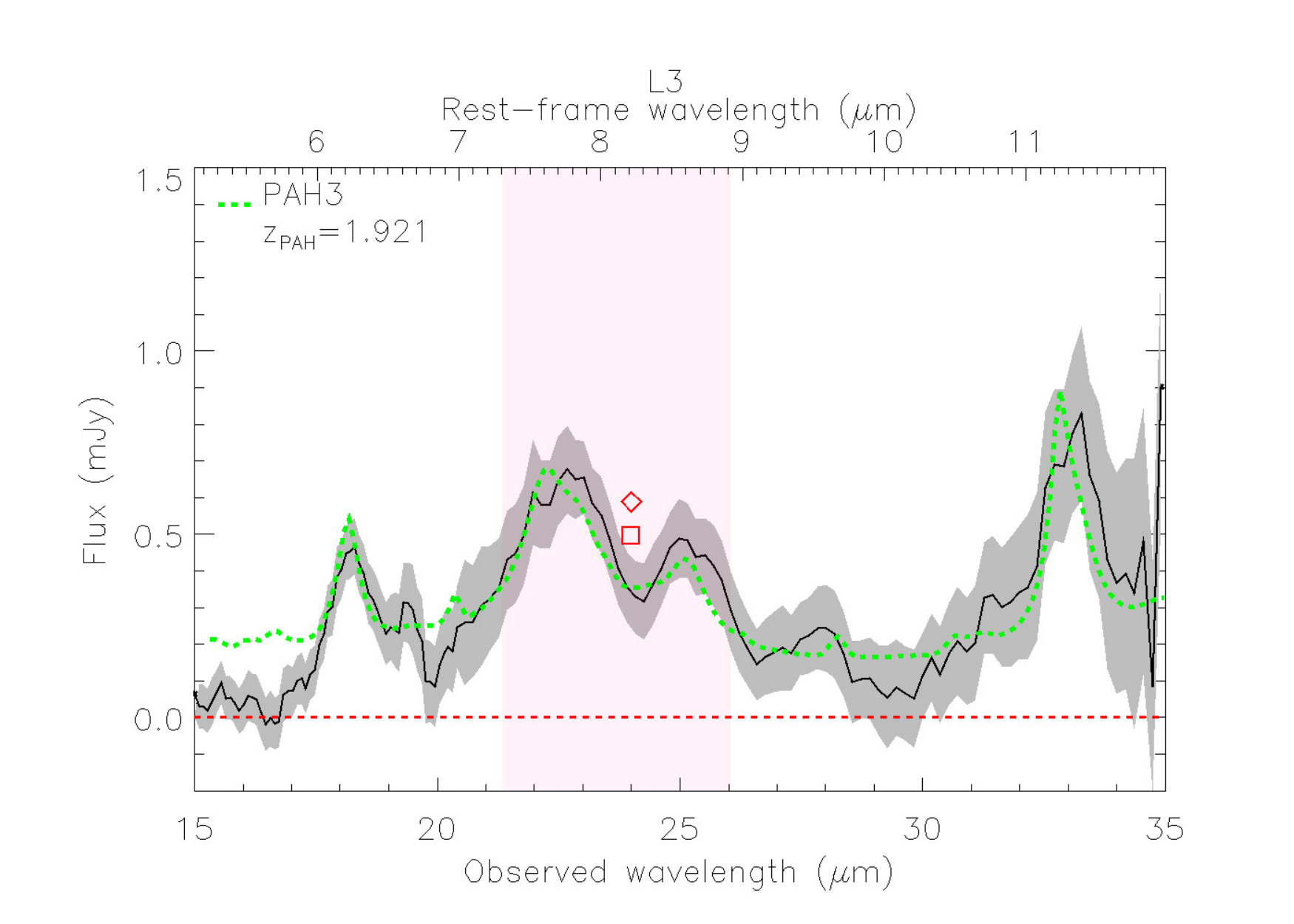}
\end{minipage}
\begin{minipage}{0.5\linewidth}
\includegraphics[height=6cm,width=9cm]{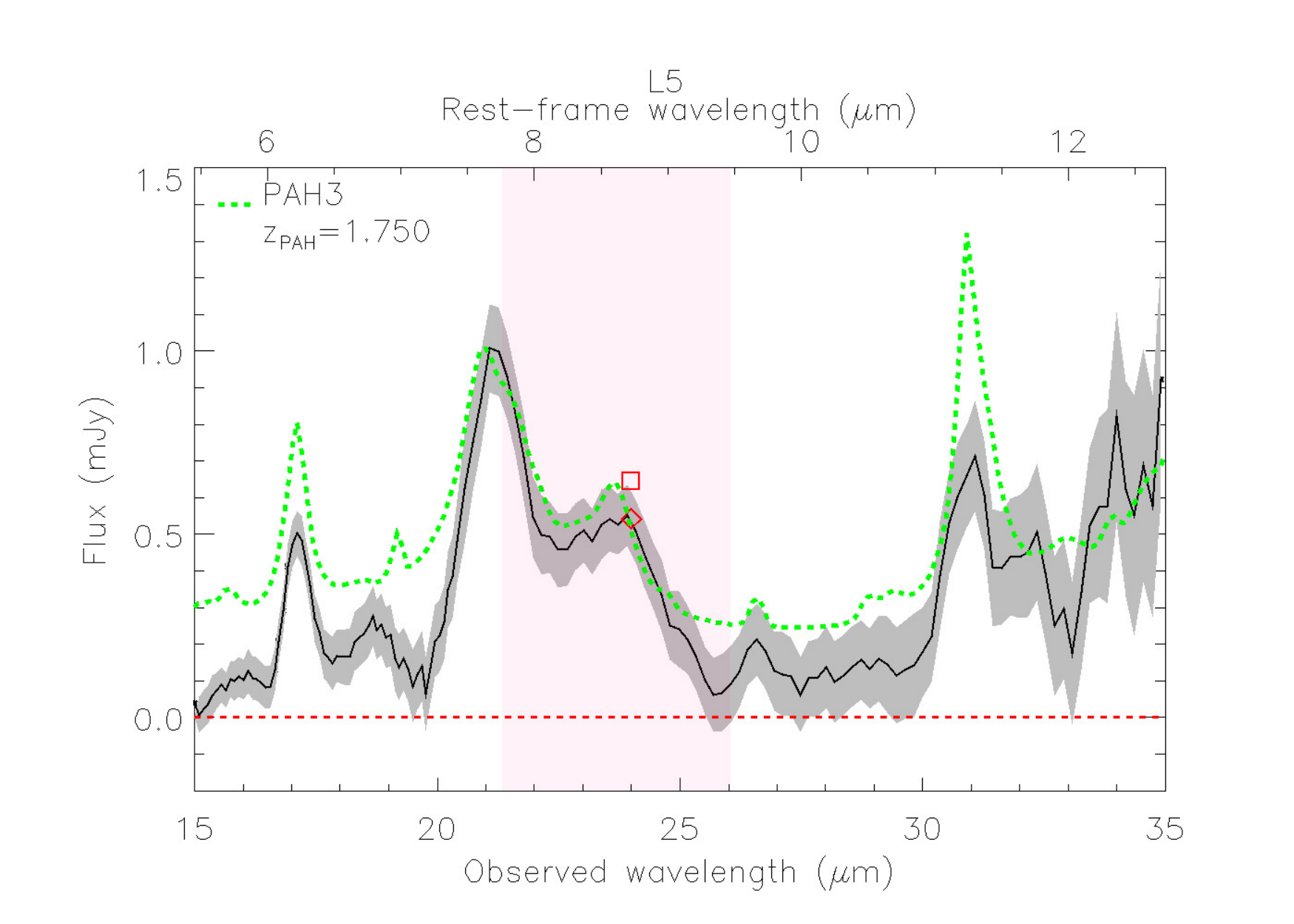}
\end{minipage}
\begin{minipage}{0.5\linewidth}
\includegraphics[height=6cm,width=9cm]{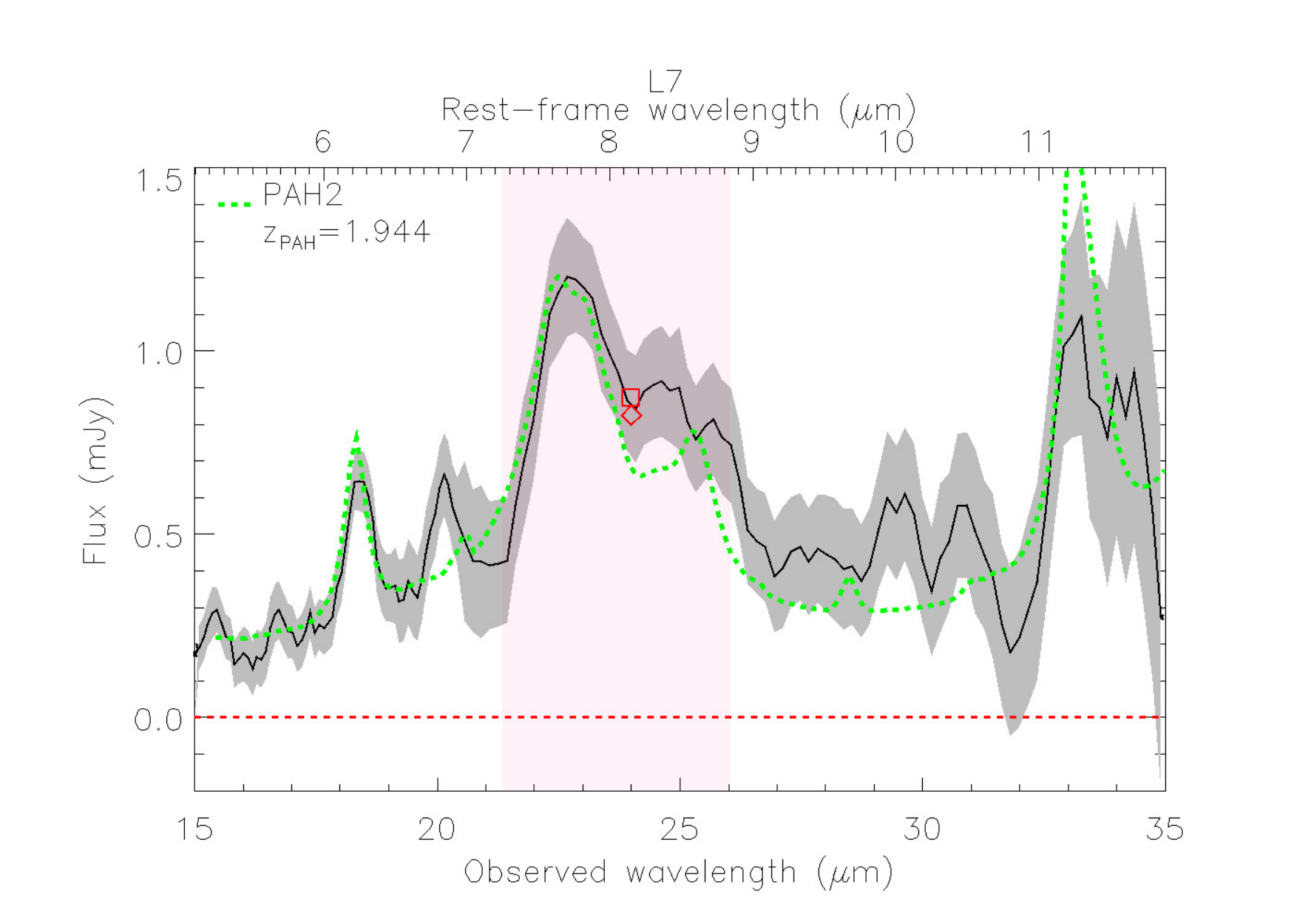}
\end{minipage}
\begin{minipage}{0.5\linewidth}
\includegraphics[height=6cm,width=9cm]{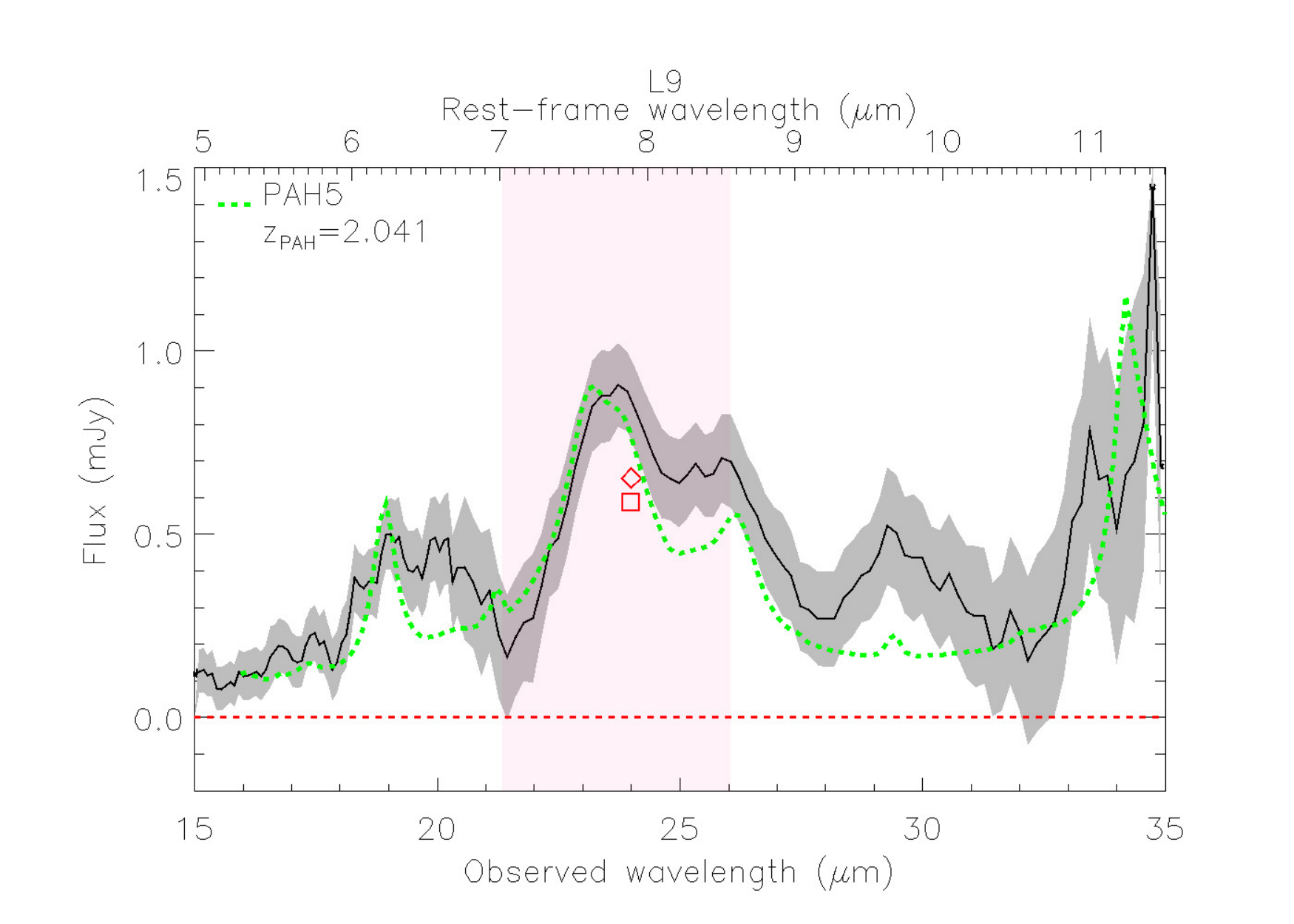}
\end{minipage}
\begin{minipage}{0.5\linewidth}
\includegraphics[height=6cm,width=9cm]{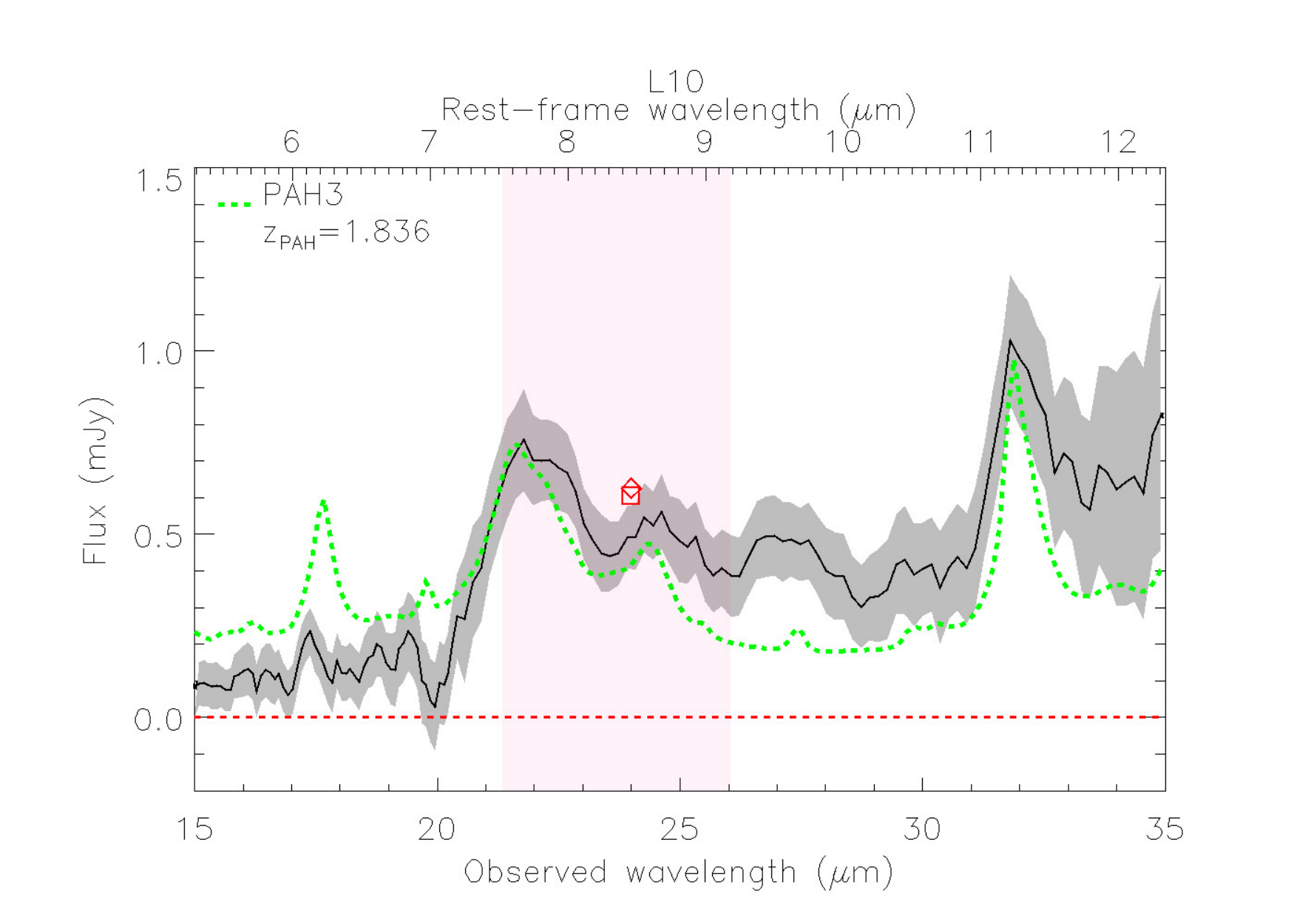}
\end{minipage}
\begin{minipage}{0.5\linewidth}
\includegraphics[height=6cm,width=9cm]{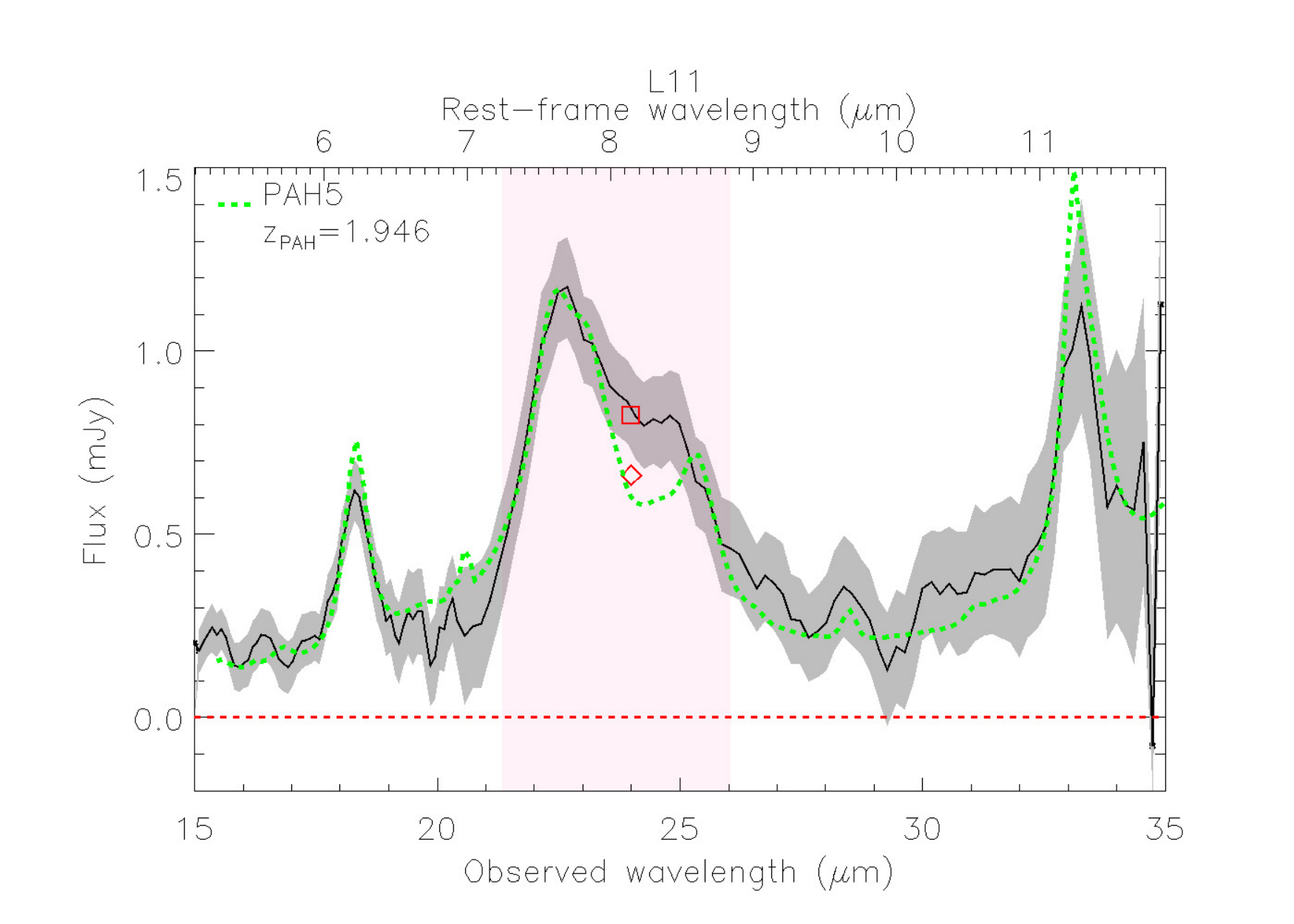}
\end{minipage}
\begin{minipage}{0.5\linewidth}
\includegraphics[height=6cm,width=9cm]{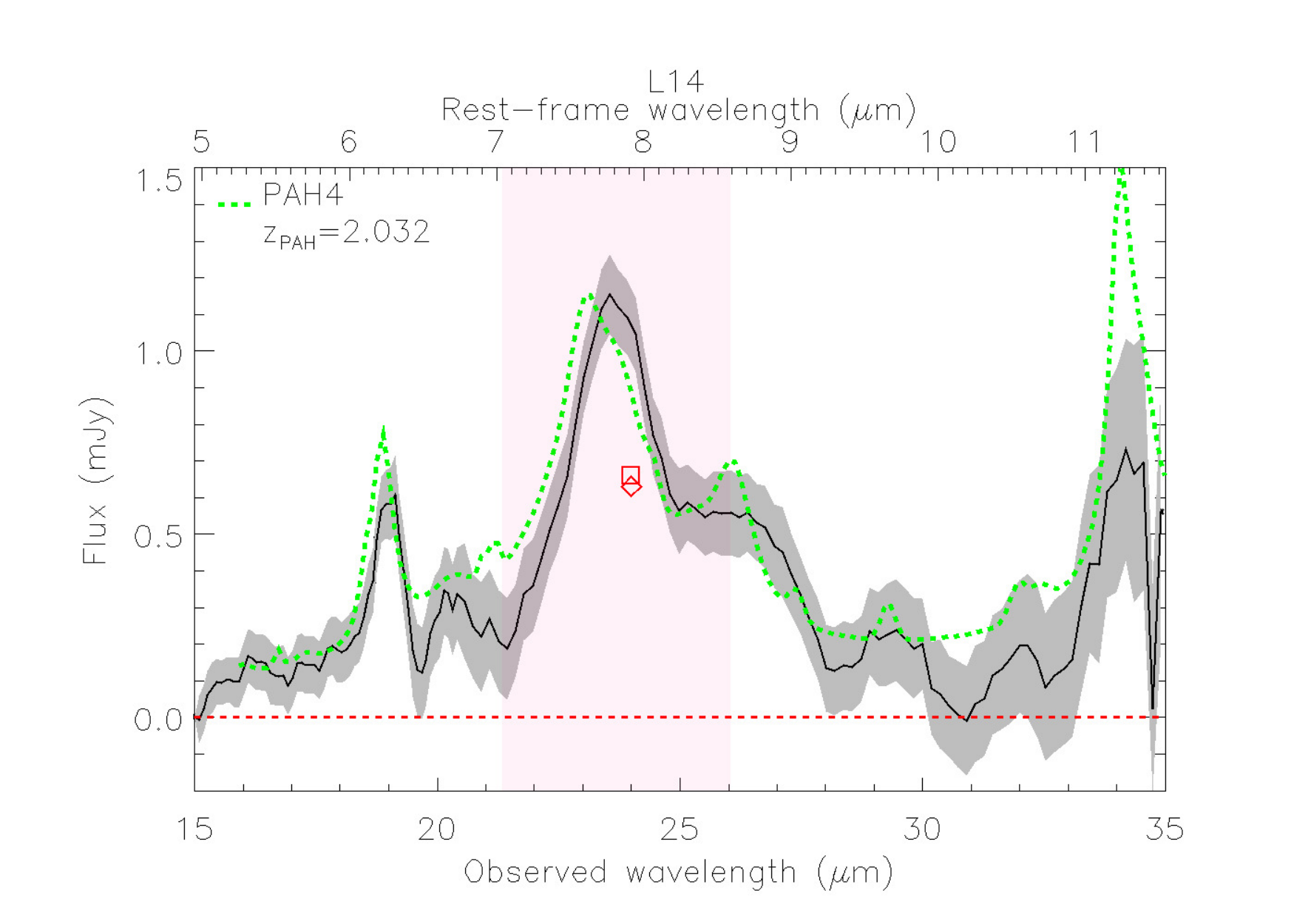}
\end{minipage}
 \caption{IRS spectra of our sources. The black line is our data. In shaded 
grey, we plot the 1$\sigma$ deviation. The green short-dashed line is the best-fit 
PAH template from \citet{Smit07}. The red diamond is the 24\,$\mu$m 
flux density from the SWIRE catalogue. The red square is the 24\,$\mu$m flux 
density extracted from the spectrum with the 24\,$\mu$m filter profile. The 
pink shape shows the region where the transmission of the 24\,$\mu$m is 
maximum. All these spectra are plotted in observed wavelength. The templates 
are scaled to the ''$(1+z) \times 7.7\,\mu$m'' flux. The source ID, the names 
of the best-fit templates, and $z_{\rm PAH}$ are reported in each
 panel.}
\end{figure*}
\addtocounter{figure}{-1}

\begin{figure*}
\begin{minipage}{0.5\linewidth}
\includegraphics[height=6cm,width=9cm]{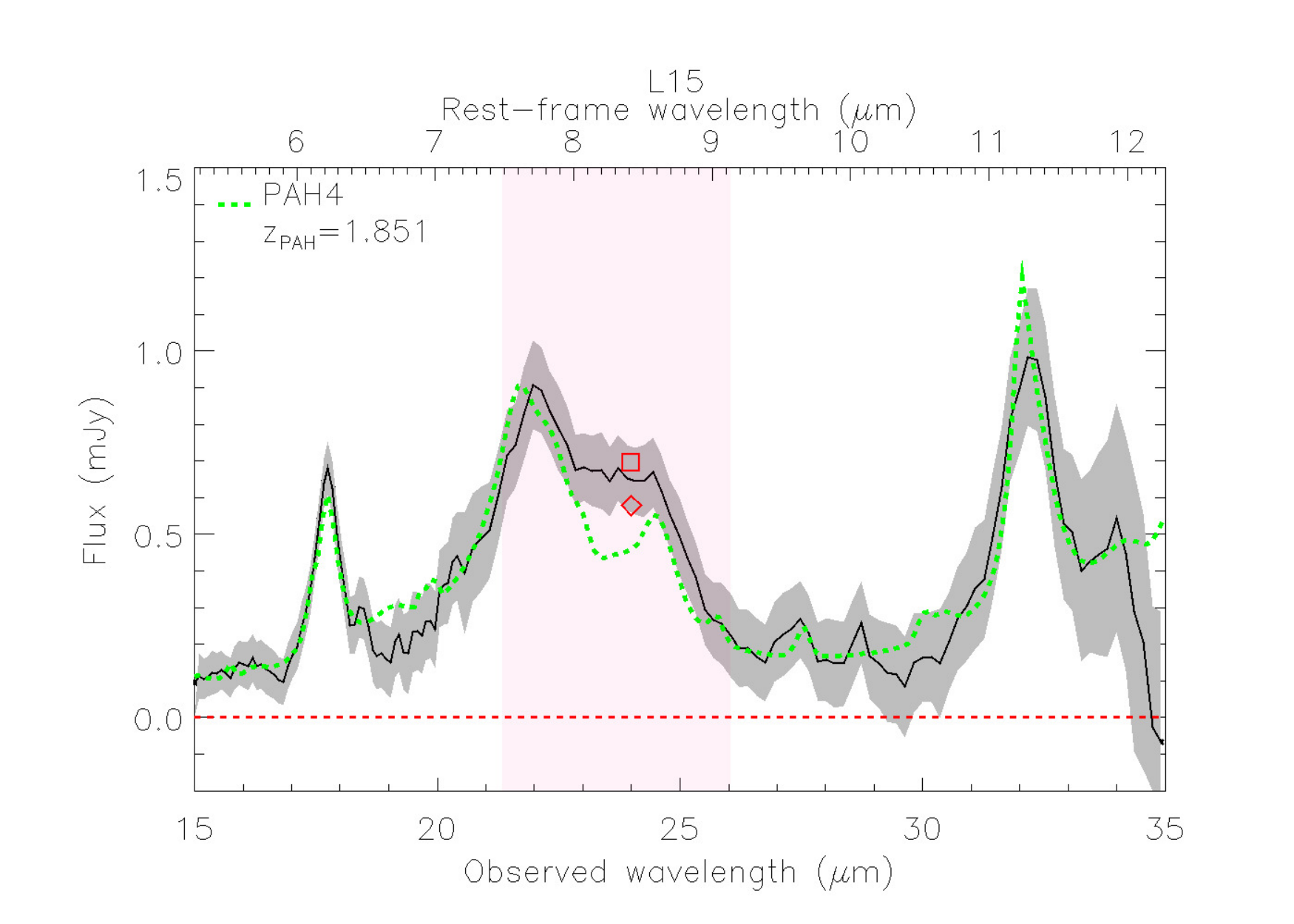}
\end{minipage}
\begin{minipage}{0.5\linewidth}
\includegraphics[height=6cm,width=9cm]{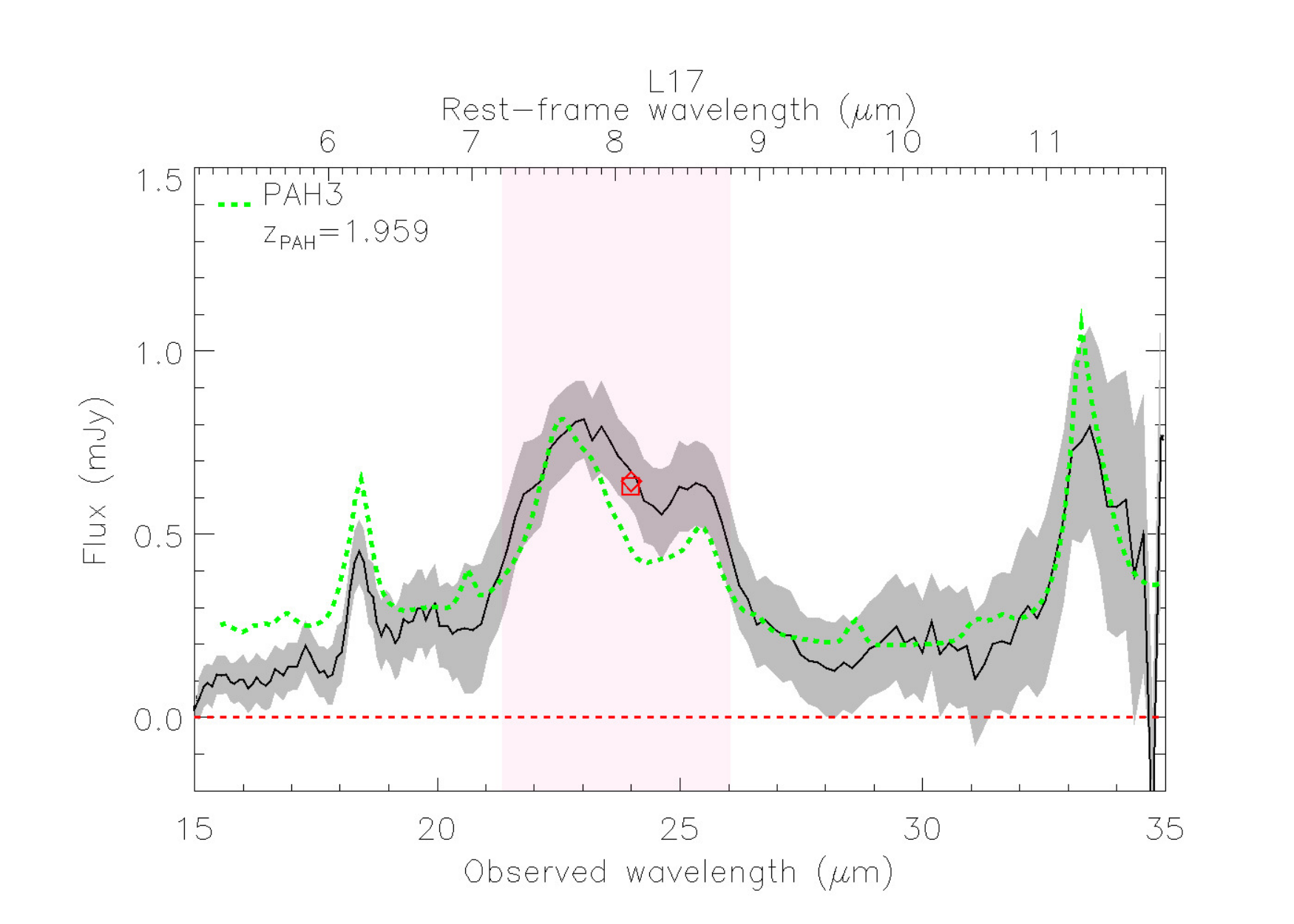}
\end{minipage}
\begin{minipage}{0.5\linewidth}
\includegraphics[height=6cm,width=9cm]{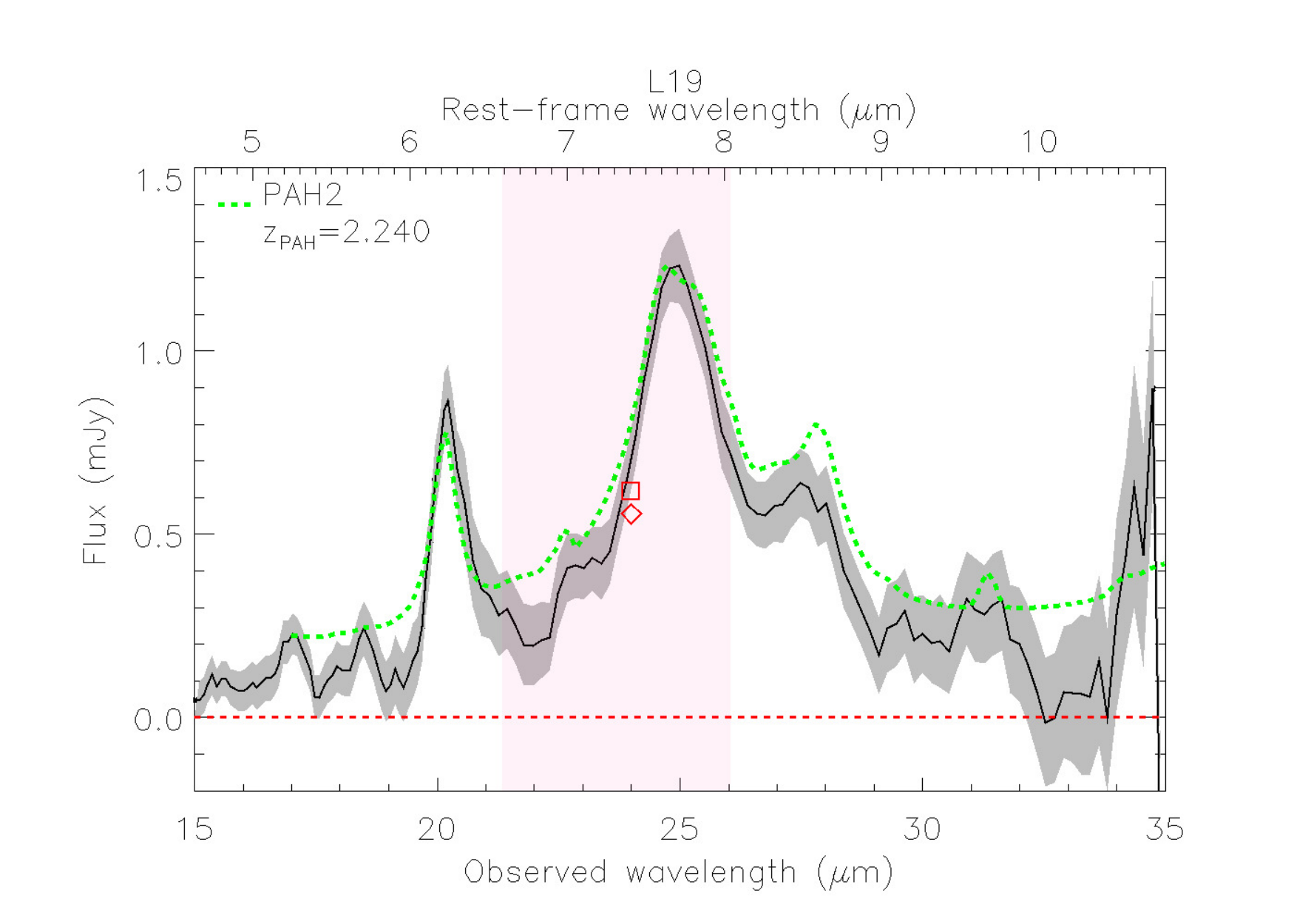}
\end{minipage}
\begin{minipage}{0.5\linewidth}
\includegraphics[height=6cm,width=9cm]{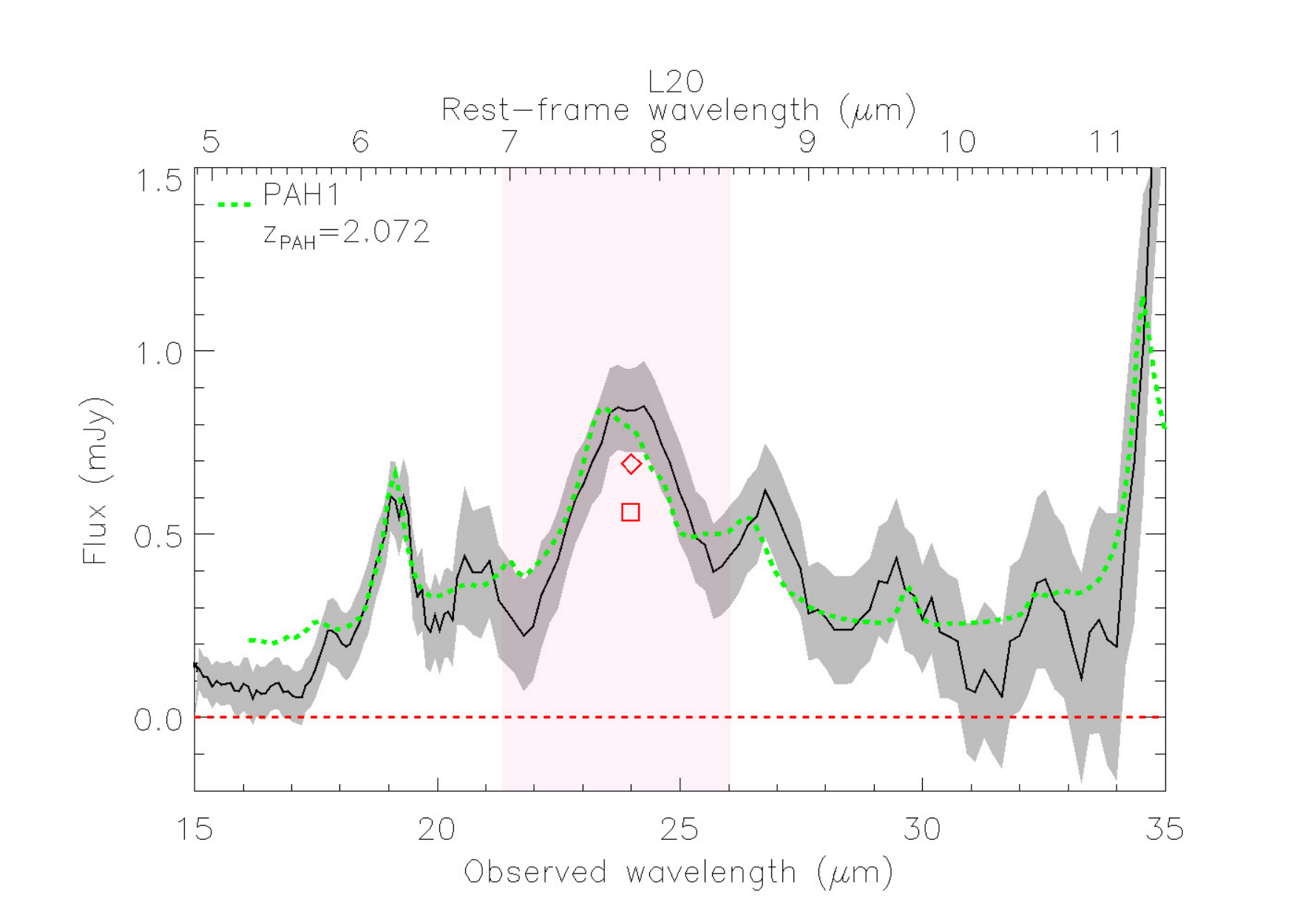}
\end{minipage}
\begin{minipage}{0.5\linewidth}
\includegraphics[height=6cm,width=9cm]{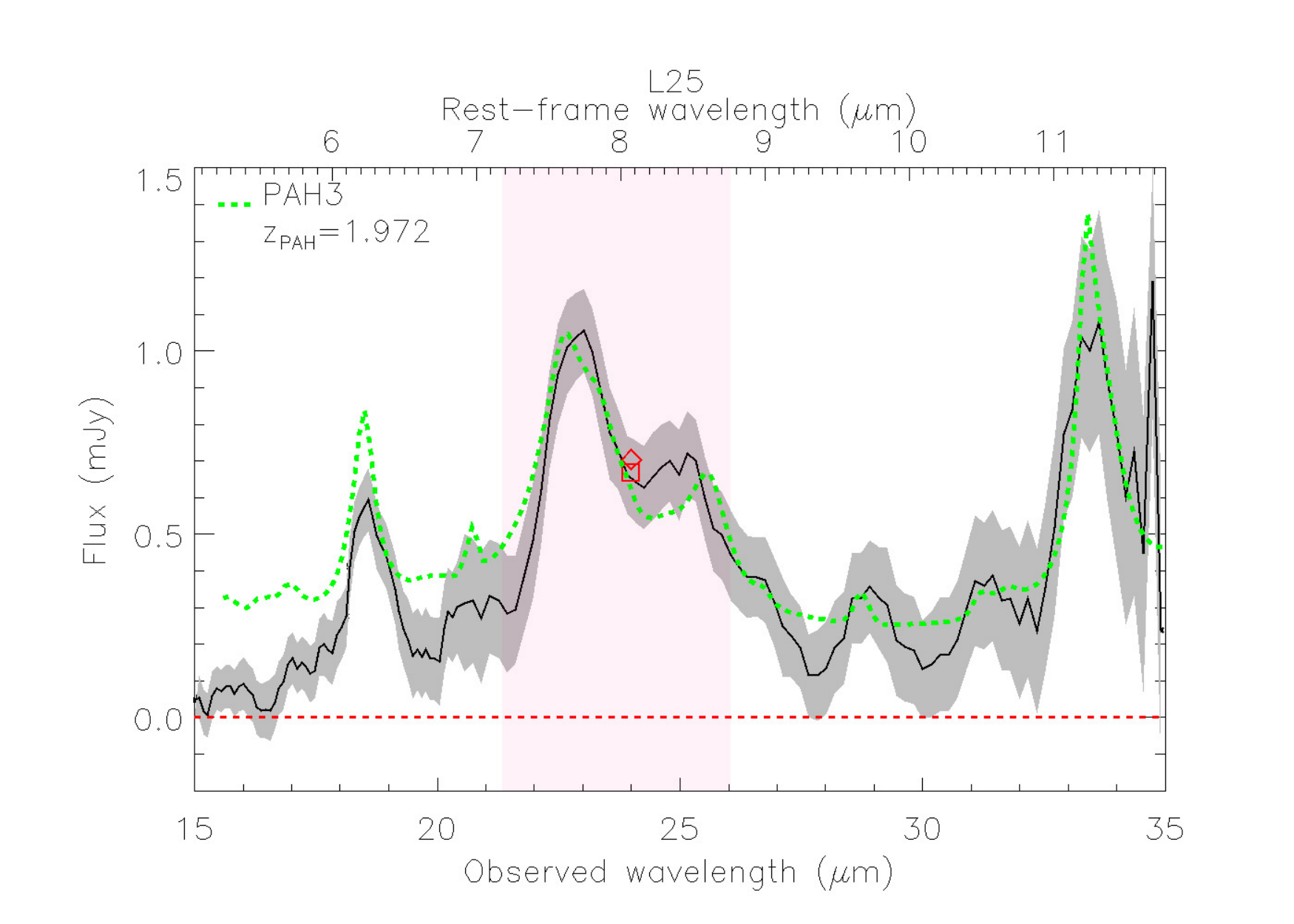}
\end{minipage}
\begin{minipage}{0.5\linewidth}
\includegraphics[height=6cm,width=9cm]{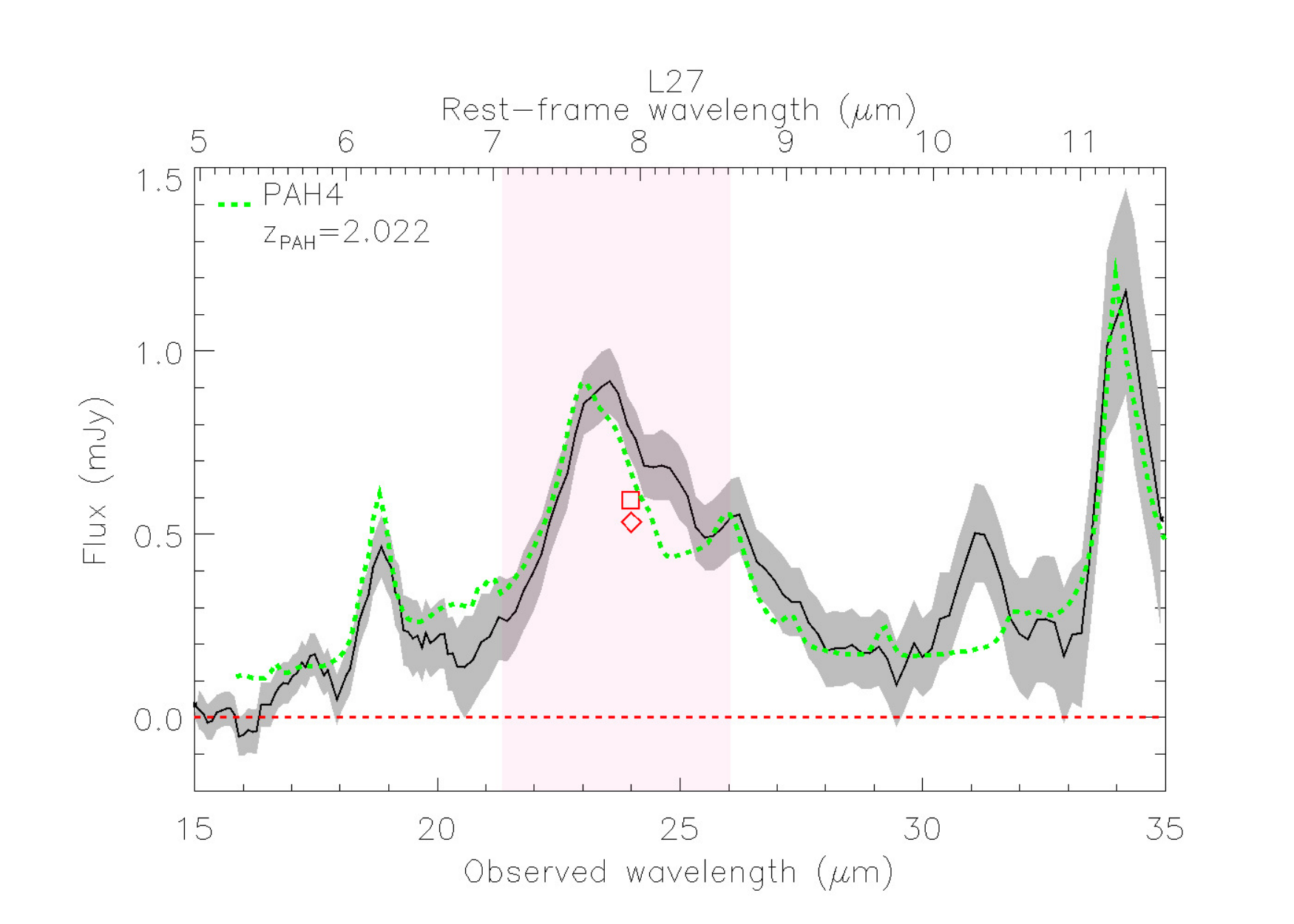}
\end{minipage}
\begin{minipage}{0.5\linewidth}
\includegraphics[height=6cm,width=9cm]{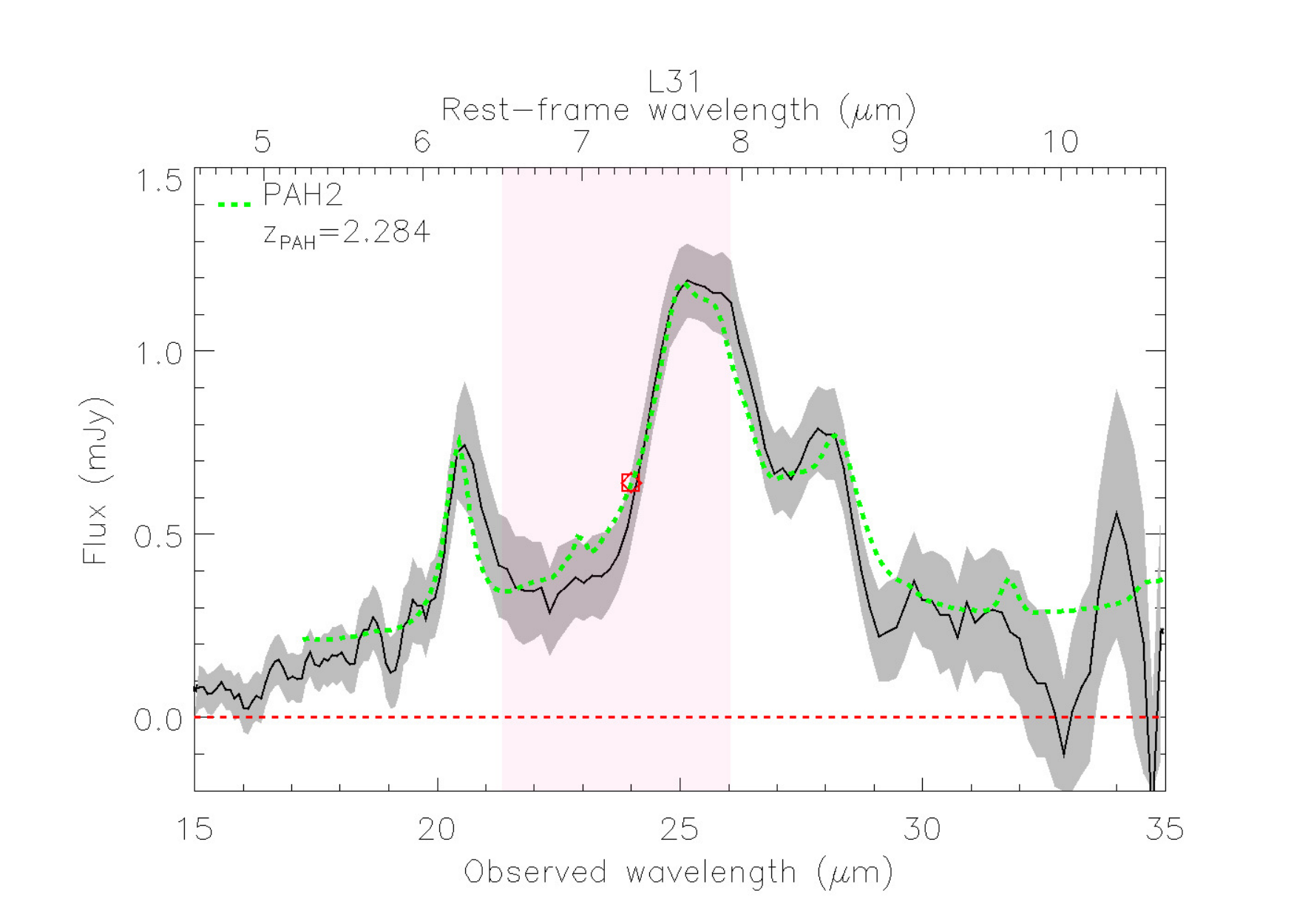}
\end{minipage}
\begin{minipage}{0.5\linewidth}
\includegraphics[height=6cm,width=9cm]{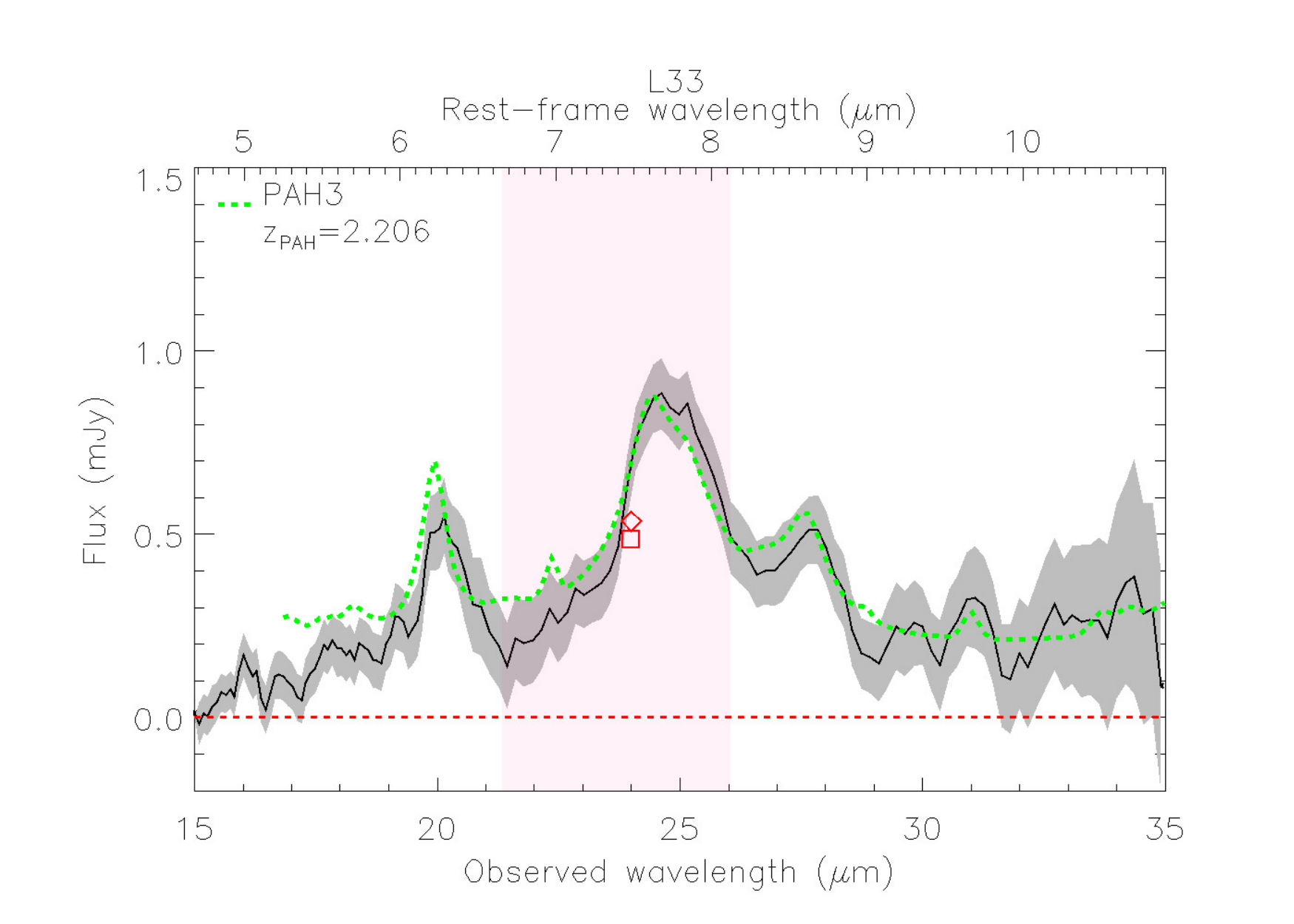}
\end{minipage}
\caption{\it{Continued}}
\end{figure*}

\end{appendix}

\begin{appendix}

\section{Spectral decomposition of the spectra}\label{app2}

In Fig. \ref{Spectrald}, we present the decompositions of our spectra made 
with the PAHFIT code developed by \citet{Smit07}.

\begin{figure*}
\label{Spectrald}
\begin{minipage}{0.5\linewidth}
\includegraphics[height=6cm,width=9cm]{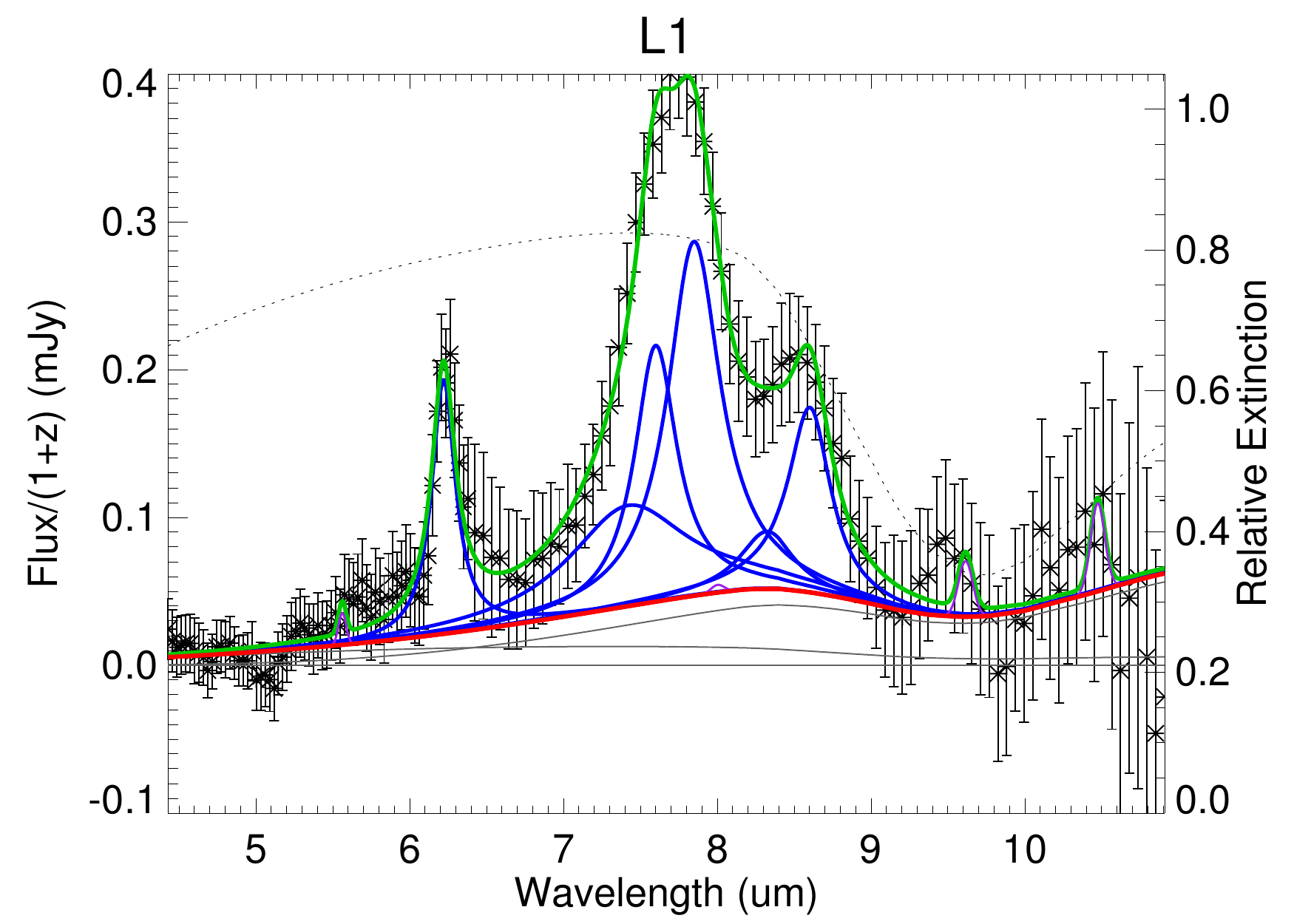}
\end{minipage}
\begin{minipage}{0.5\linewidth}
\includegraphics[height=6cm,width=9cm]{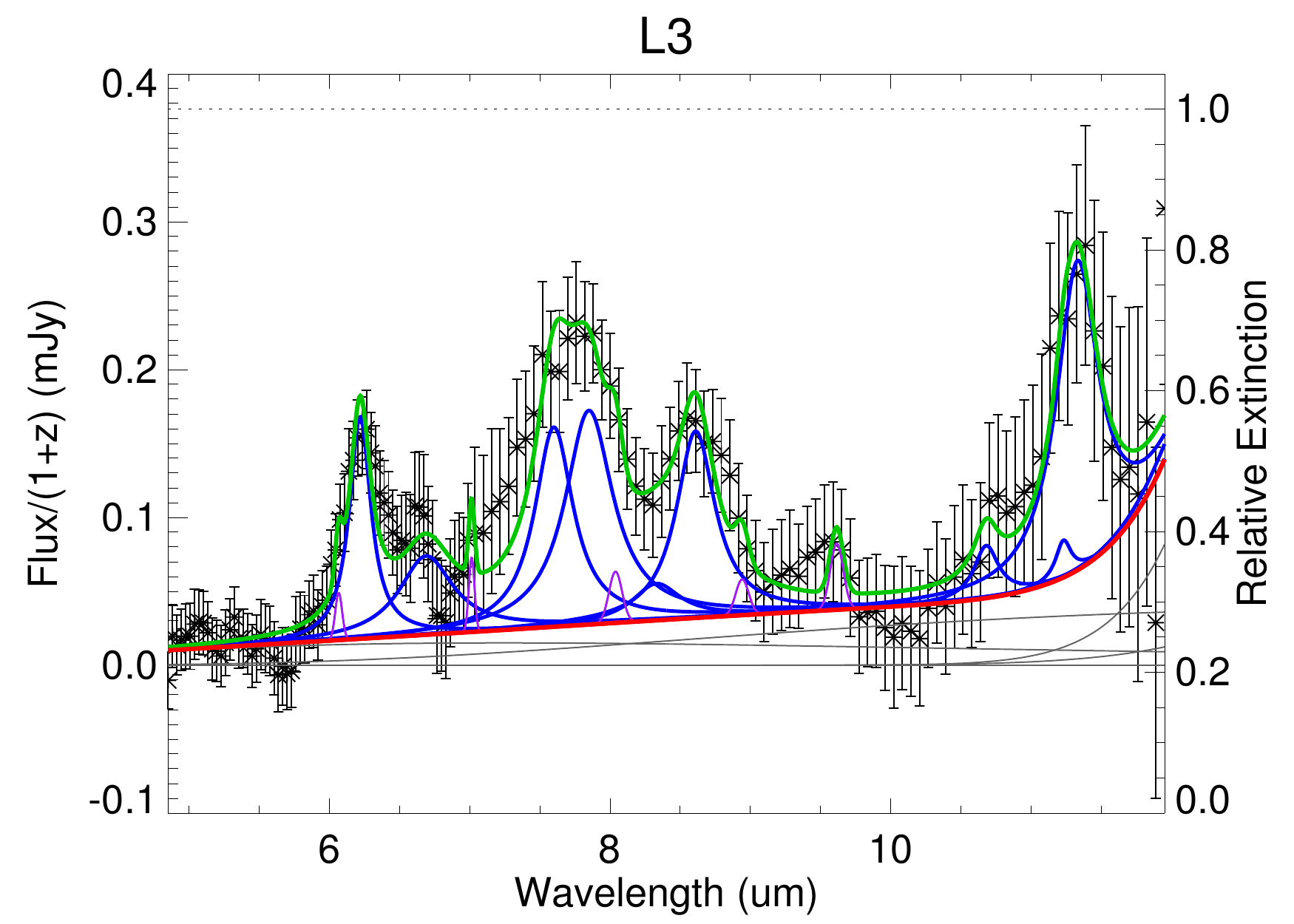}
\end{minipage}
\begin{minipage}{0.5\linewidth}
\includegraphics[height=6cm,width=9cm]{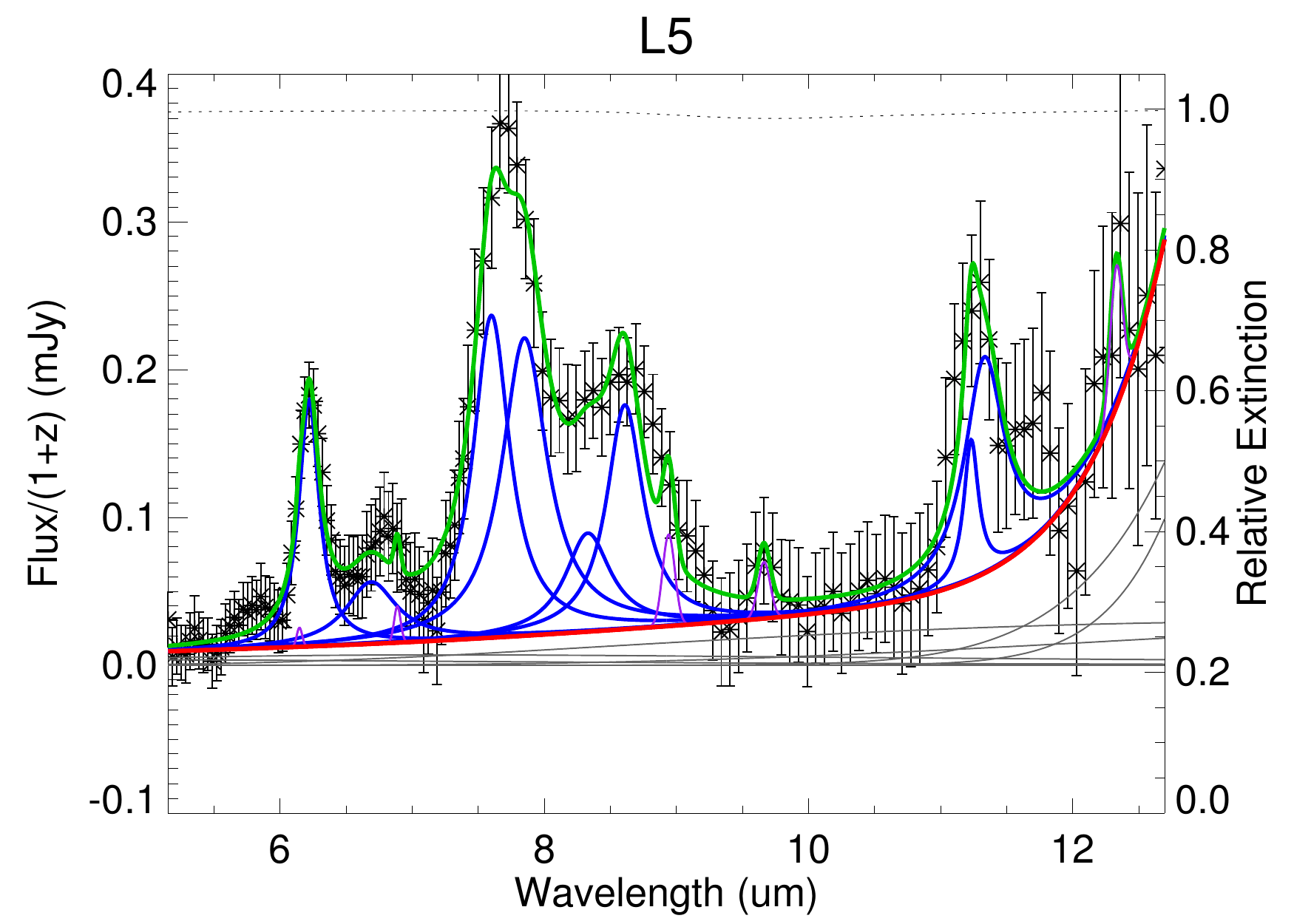}
\end{minipage}
\begin{minipage}{0.5\linewidth}
\includegraphics[height=6cm,width=9cm]{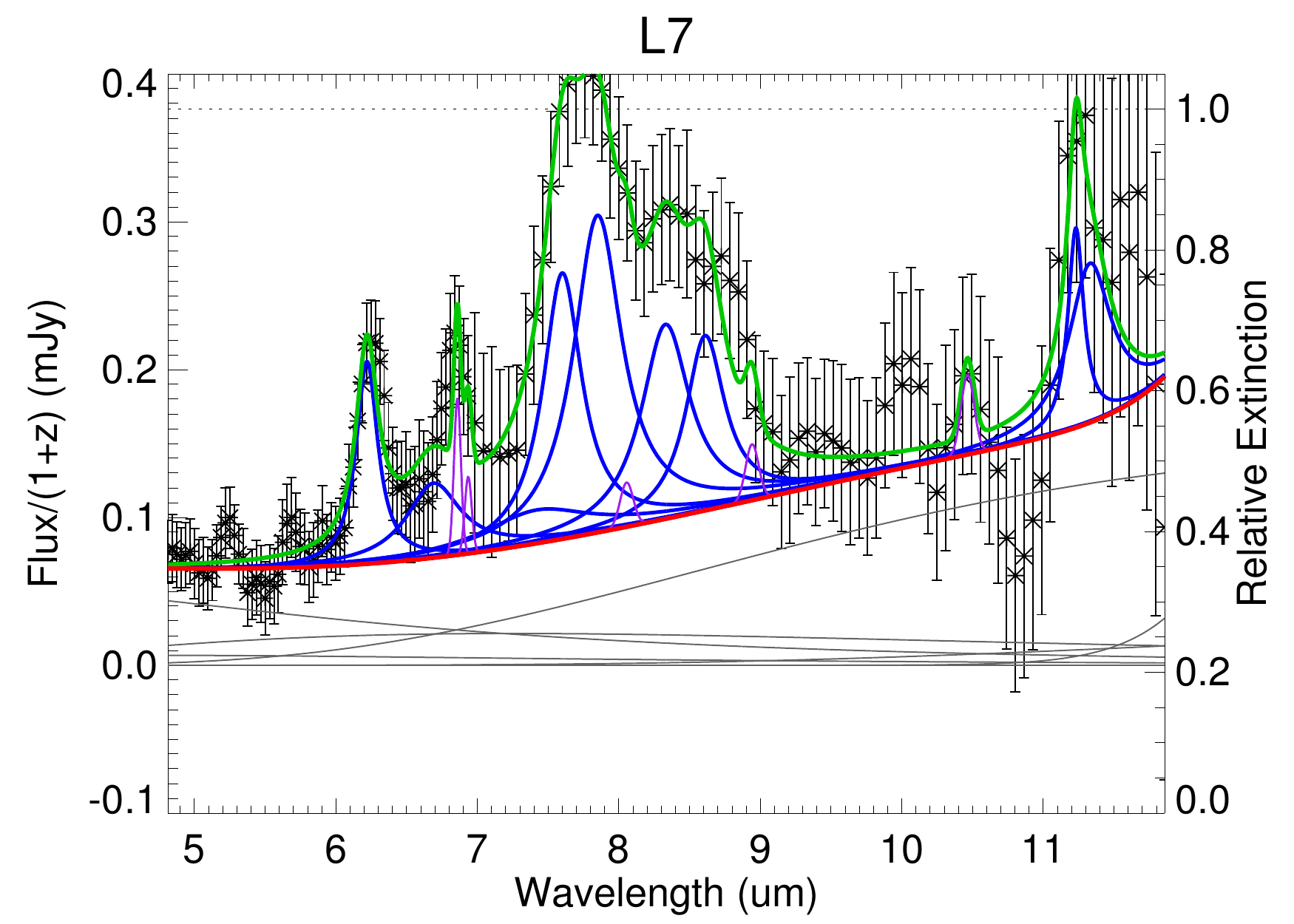}
\end{minipage}
\begin{minipage}{0.5\linewidth}
\includegraphics[height=6cm,width=9cm]{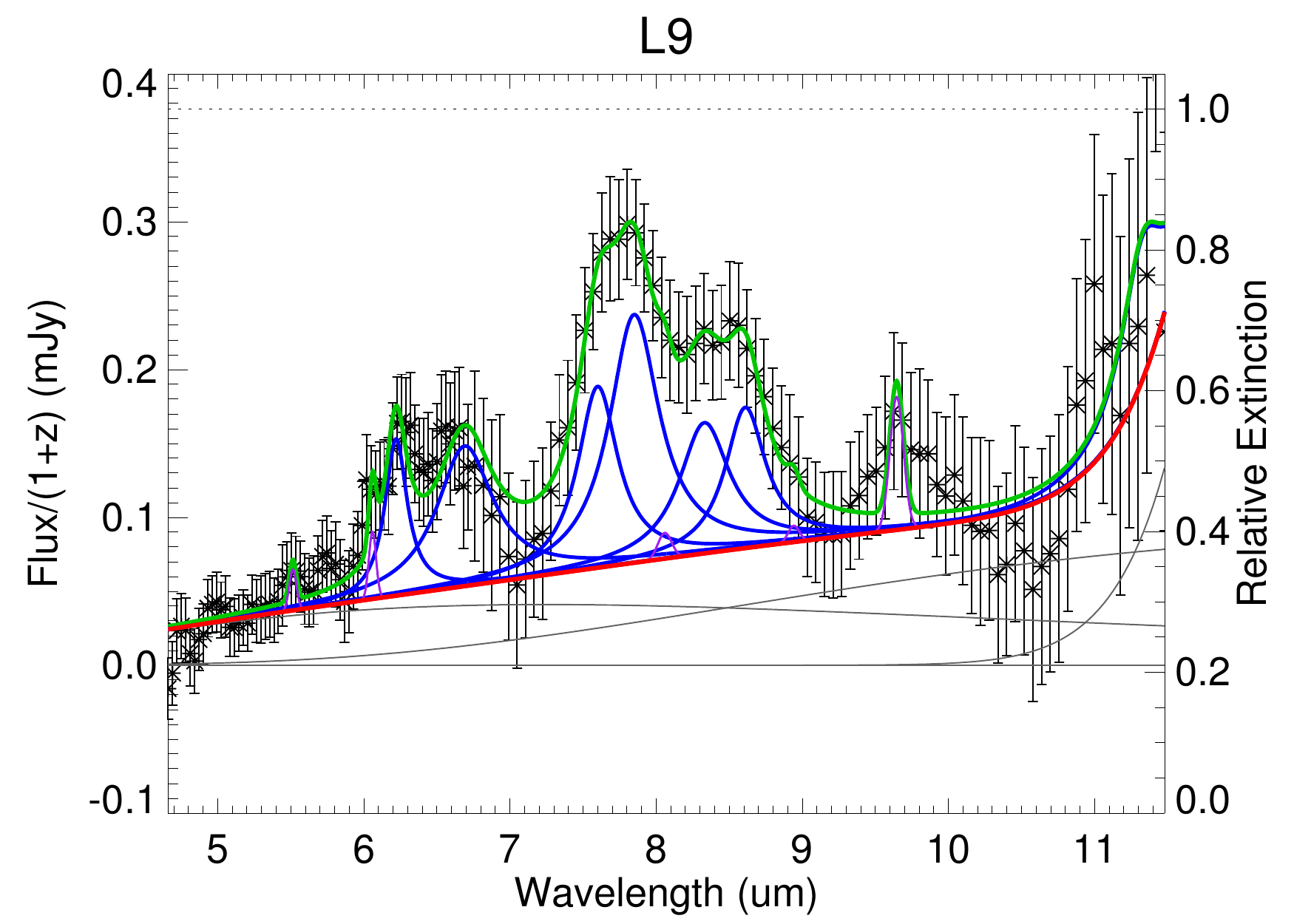}
\end{minipage}
\begin{minipage}{0.5\linewidth}
\includegraphics[height=6cm,width=9cm]{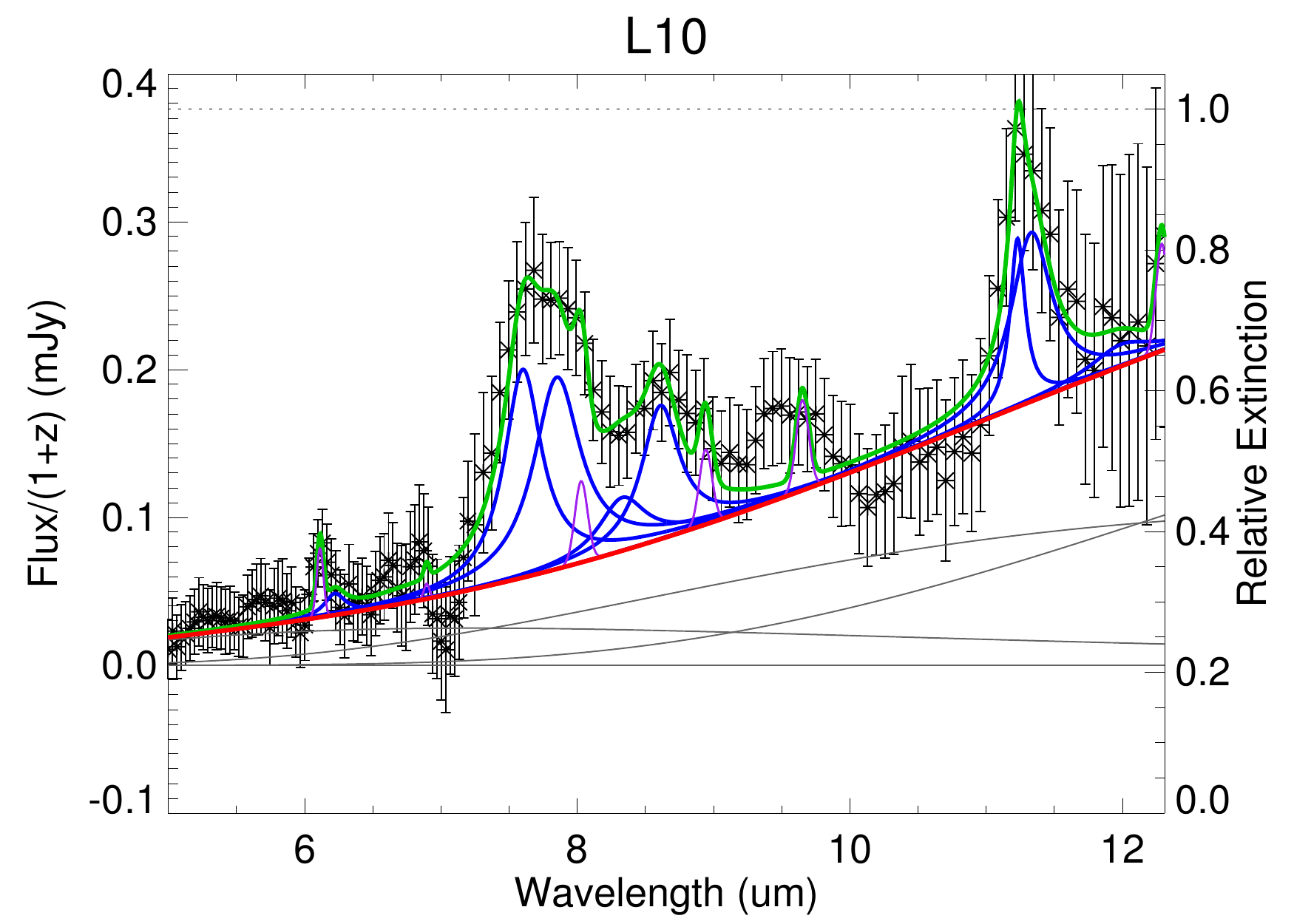}
\end{minipage}
\begin{minipage}{0.5\linewidth}
\includegraphics[height=6cm,width=9cm]{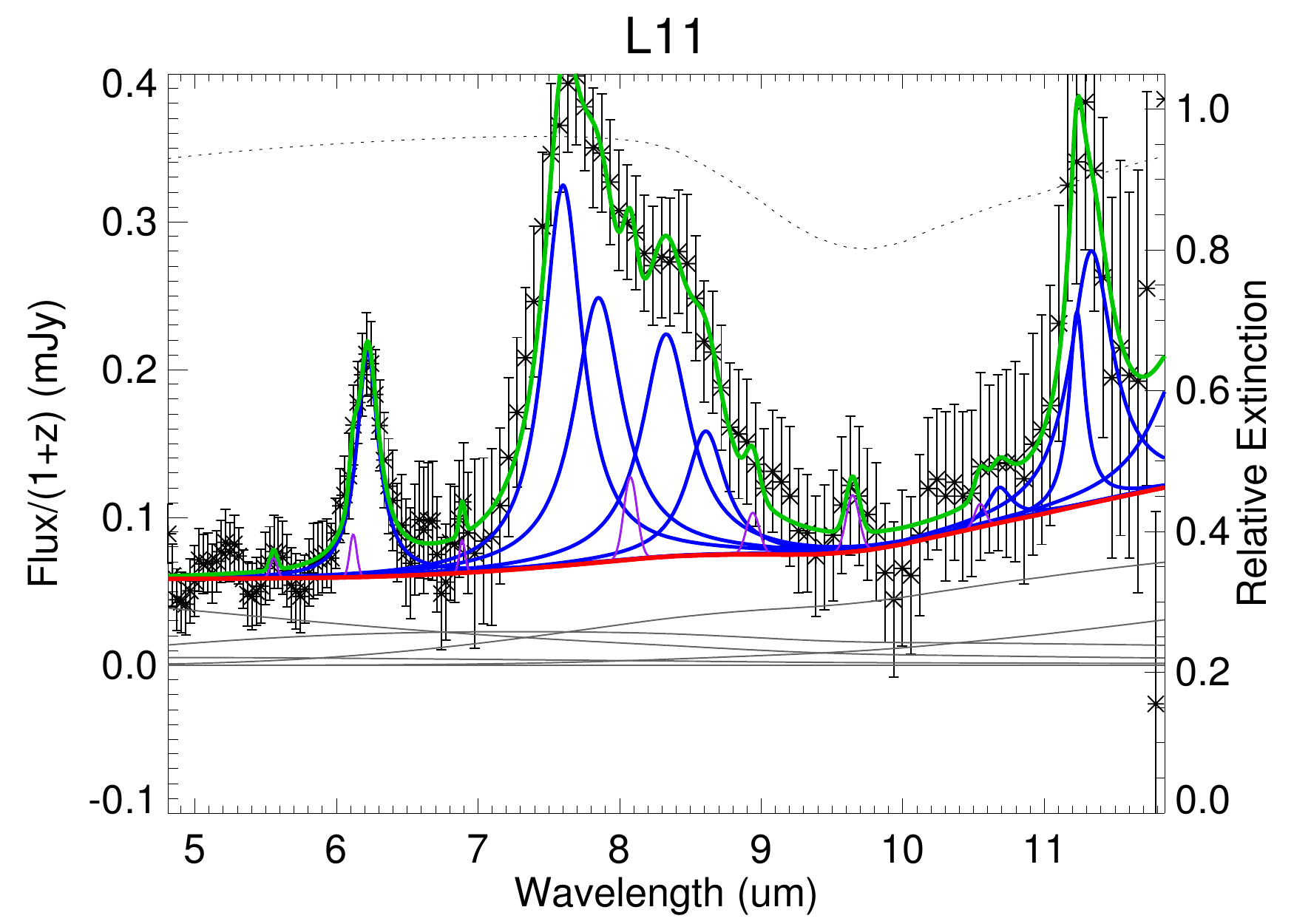}
\end{minipage}
\begin{minipage}{0.5\linewidth}
\includegraphics[height=6cm,width=9cm]{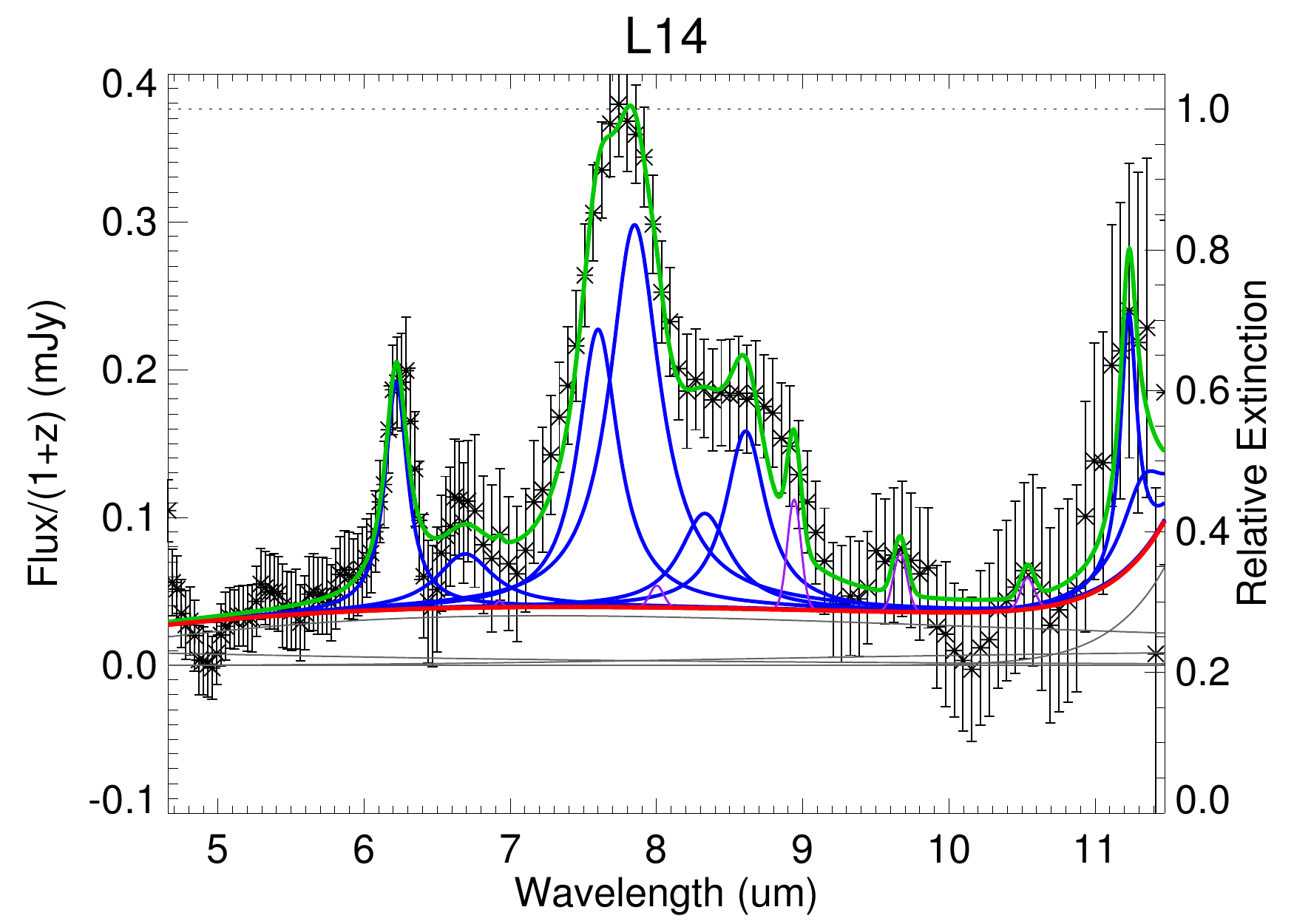}
\end{minipage}
 \caption{Decompositions of the spectra of our sources. The solid 
green line is the fitted model.  The blue lines above the continuum are the 
PAH features. The violet lines are the spectral lines. The thin grey lines 
represent the thermal dust continuum components. The thick red line shows the 
total continuum (stars+dust).  The dotted black line shows the extinction 
($e^{-\tau_{\lambda}}$, $=1$ if no extinction). The source ID is reported as 
the title of each panel.}
\end{figure*}
\addtocounter{figure}{-1}

\begin{figure*}
\begin{minipage}{0.5\linewidth}
\includegraphics[height=6cm,width=9cm]{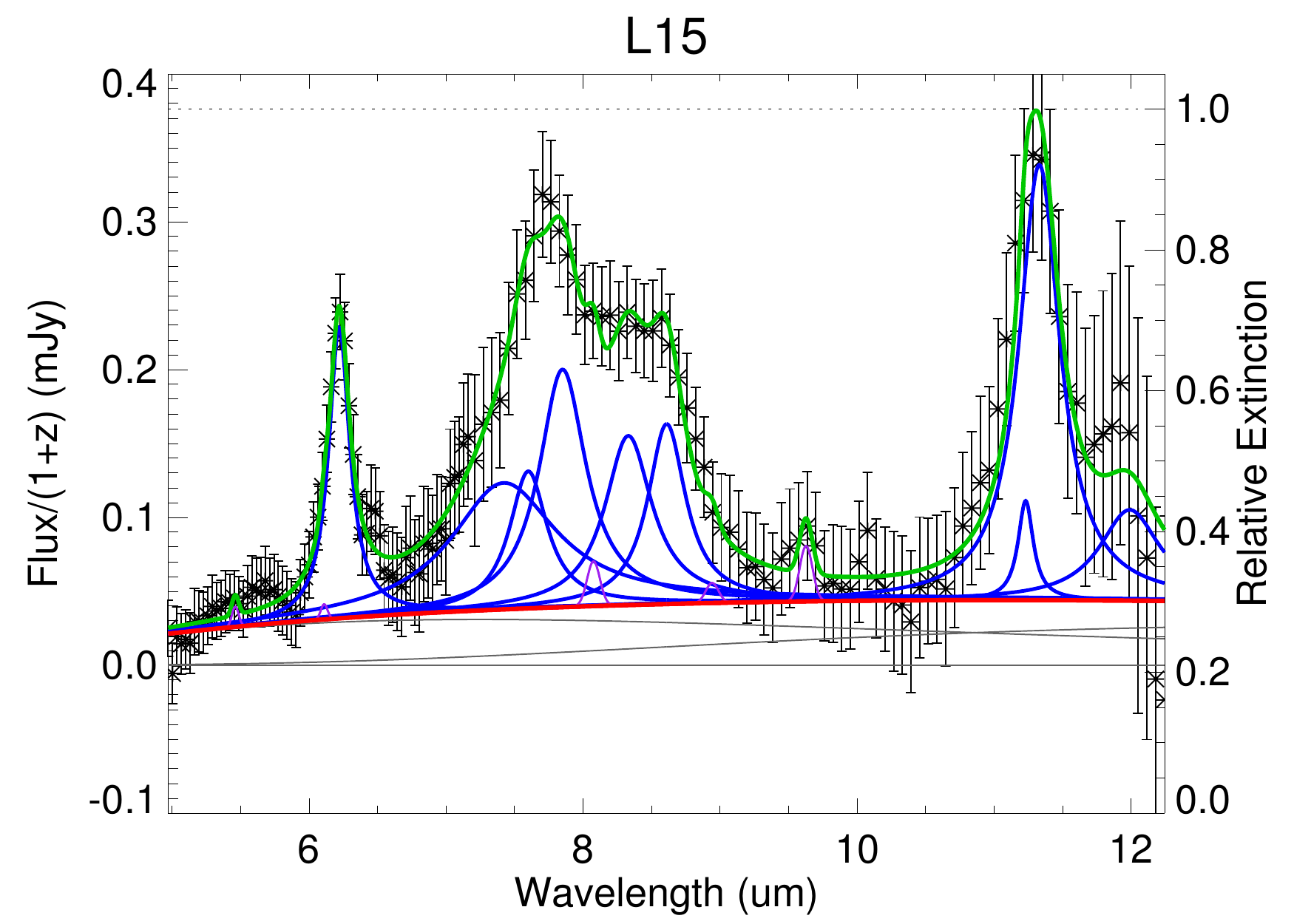}
\end{minipage}
\begin{minipage}{0.5\linewidth}
\includegraphics[height=6cm,width=9cm]{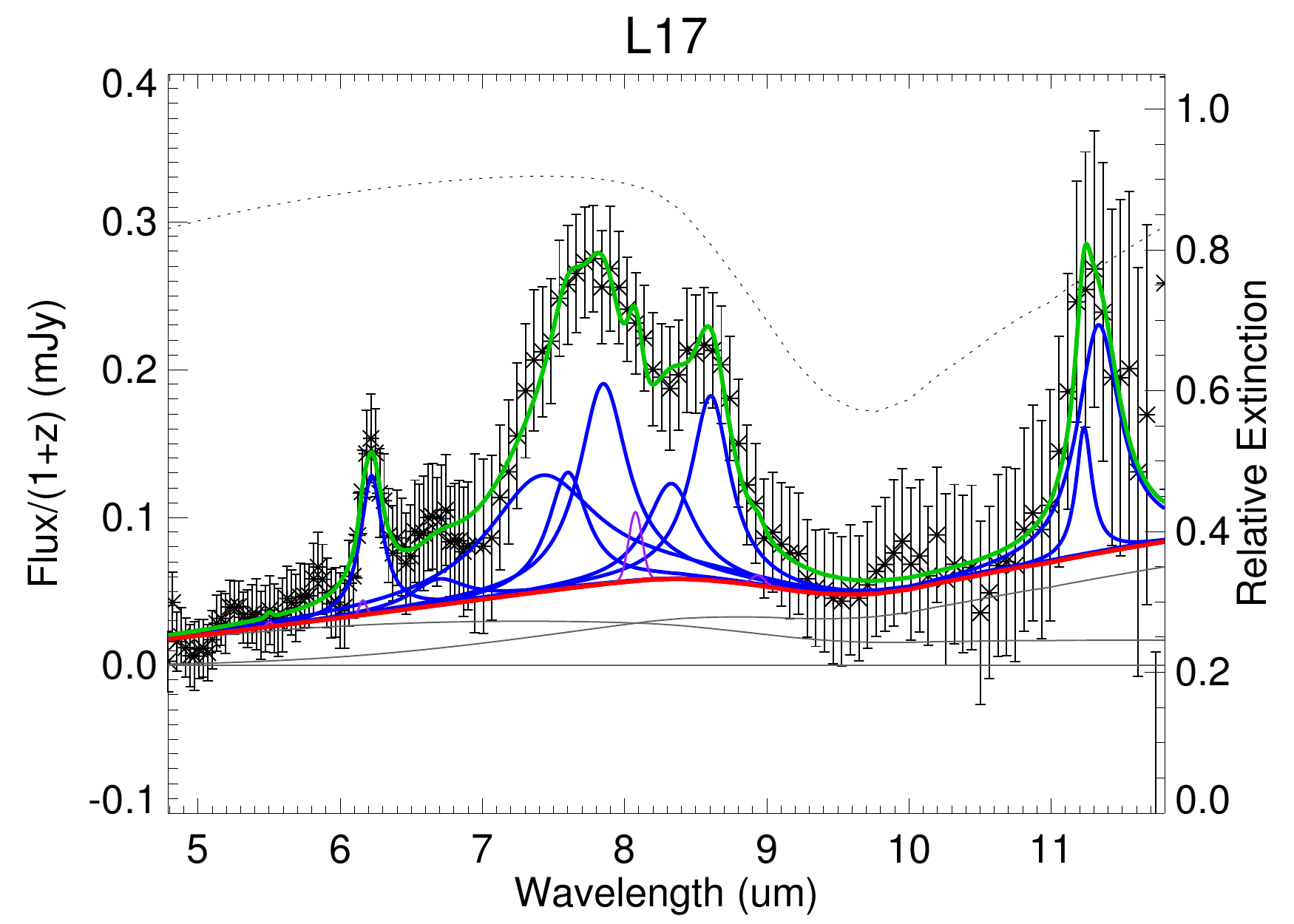}
\end{minipage}
\begin{minipage}{0.5\linewidth}
\includegraphics[height=6cm,width=9cm]{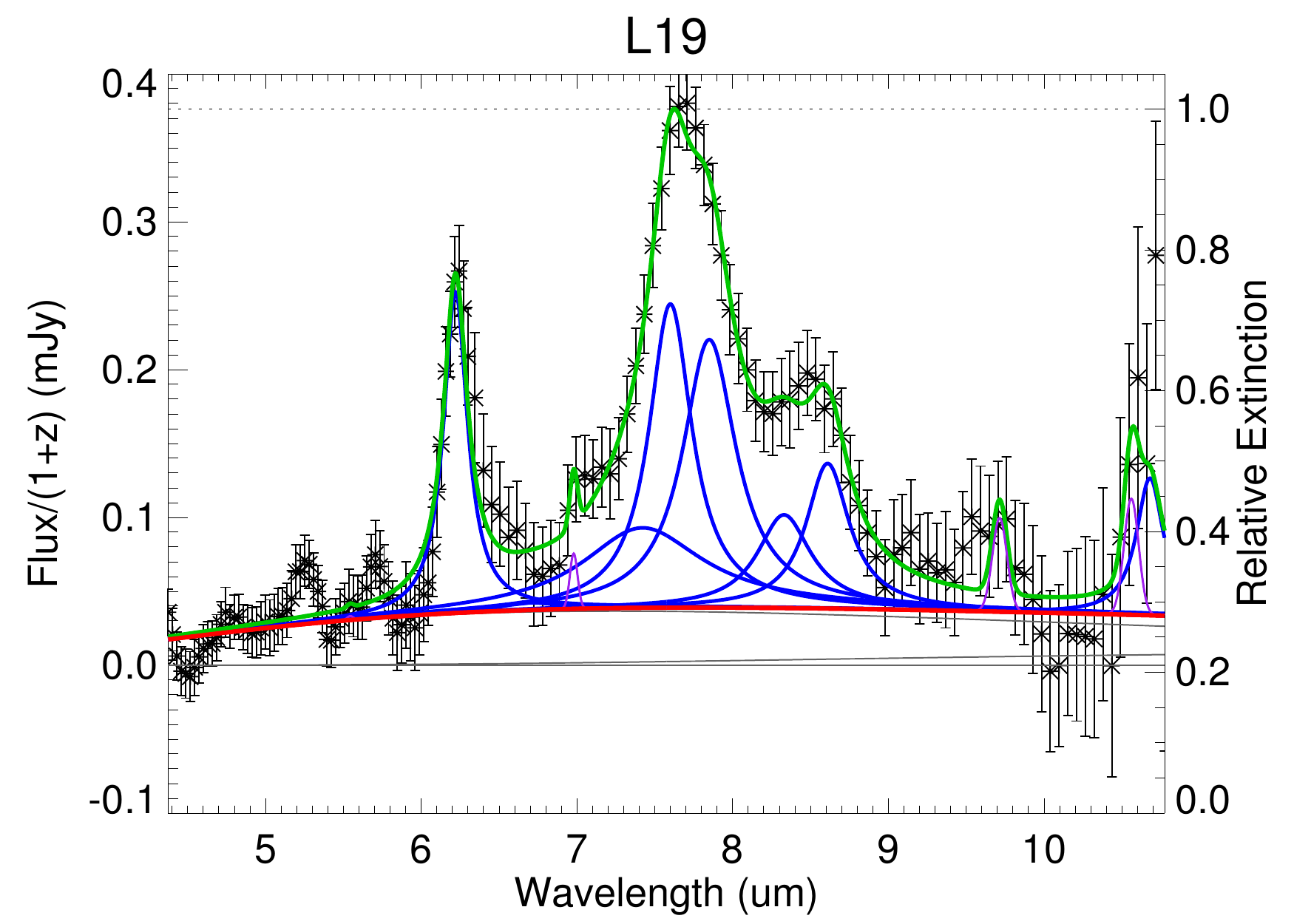}
\end{minipage}
\begin{minipage}{0.5\linewidth}
\includegraphics[height=6cm,width=9cm]{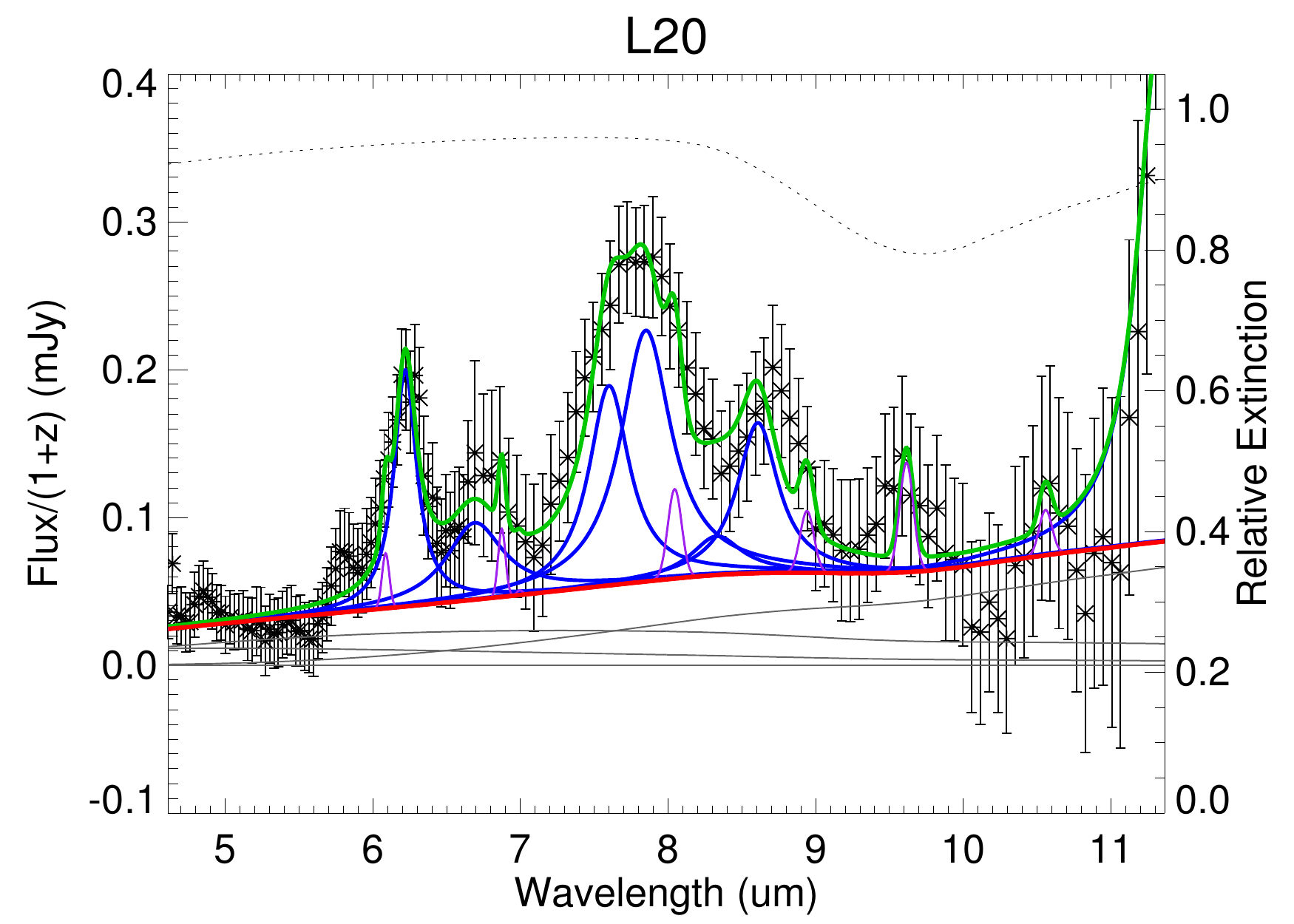}
\end{minipage}
\begin{minipage}{0.5\linewidth}
\includegraphics[height=6cm,width=9cm]{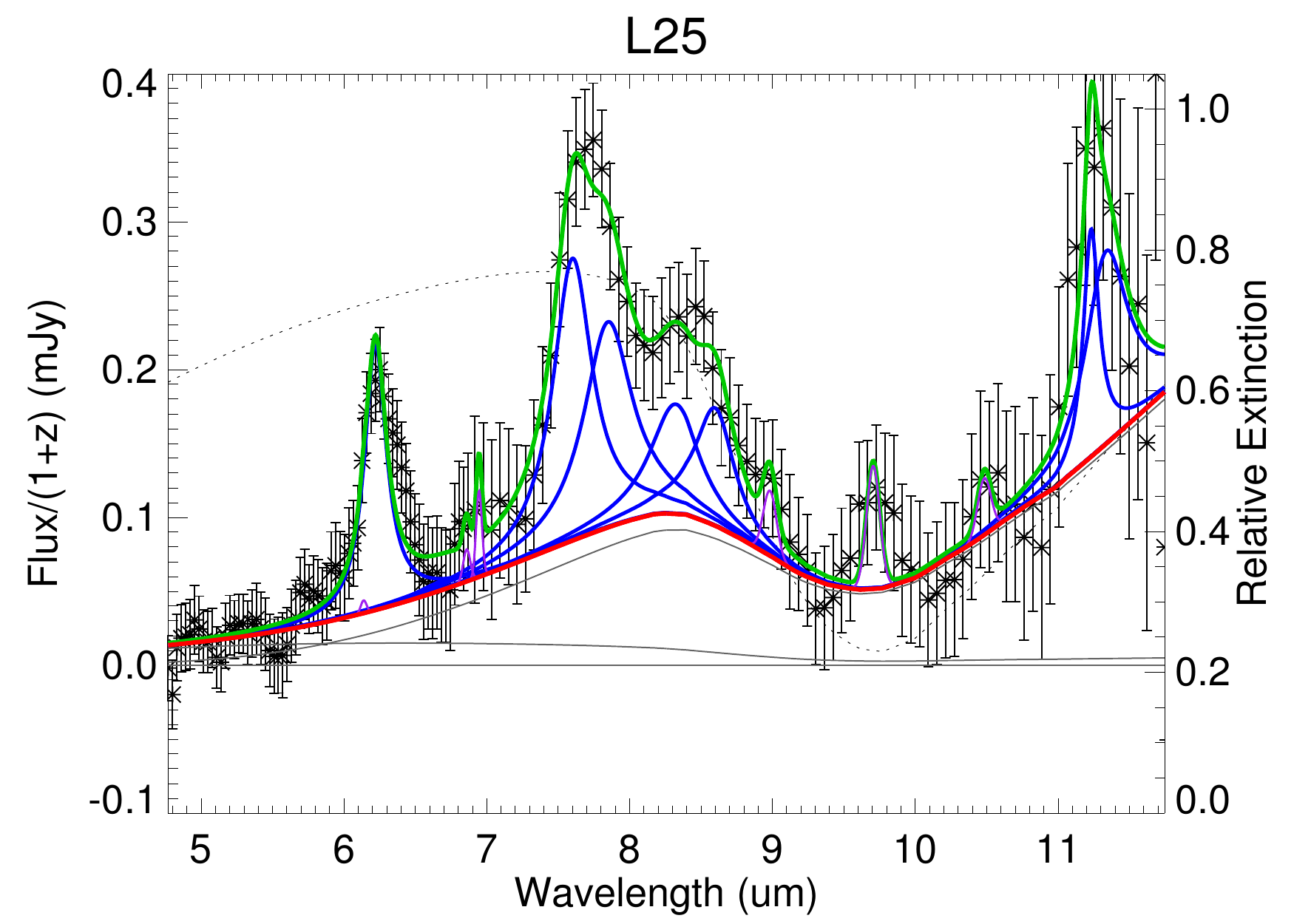}
\end{minipage}
\begin{minipage}{0.5\linewidth}
\includegraphics[height=6cm,width=9cm]{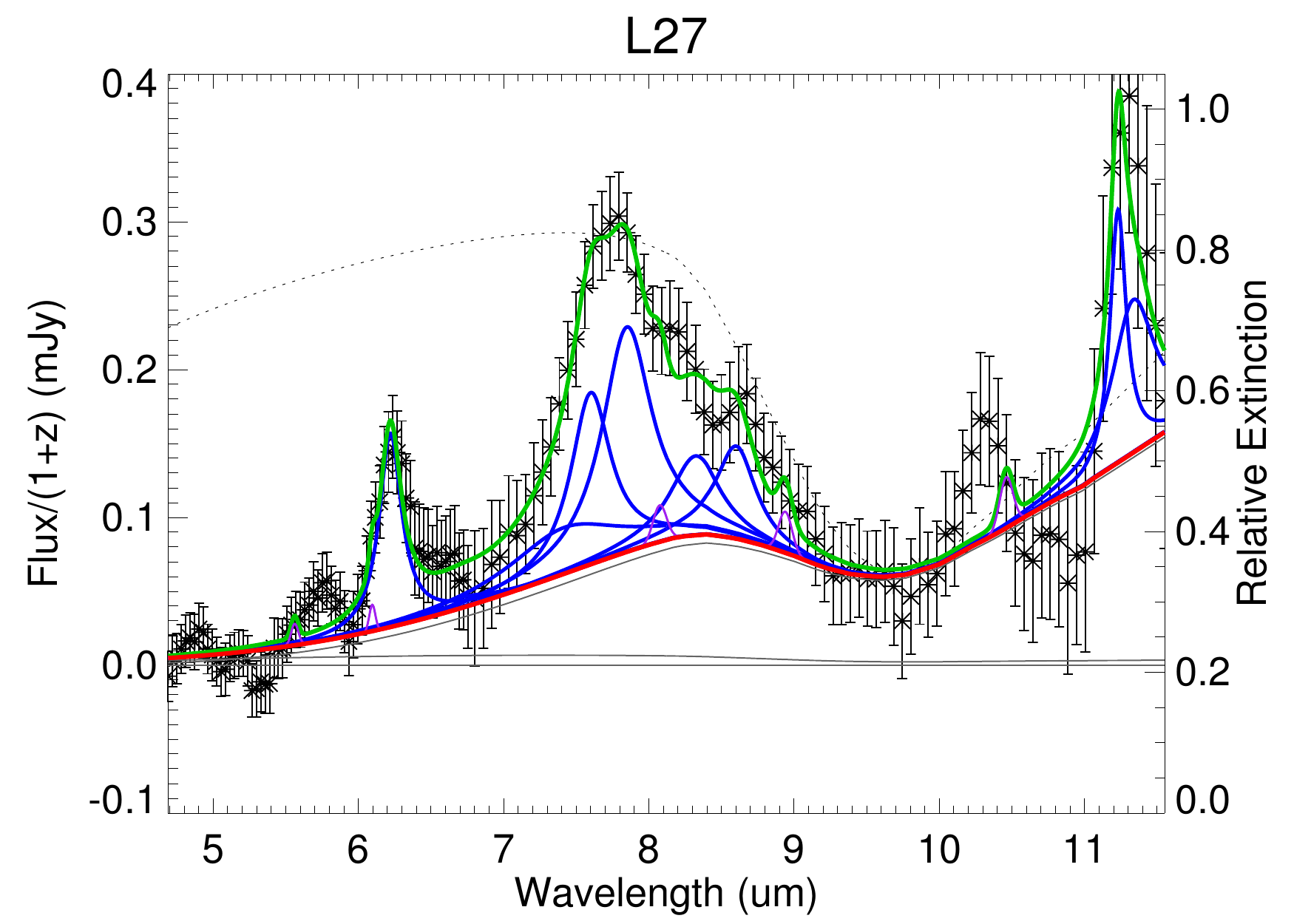}
\end{minipage}
\begin{minipage}{0.5\linewidth}
\includegraphics[height=6cm,width=9cm]{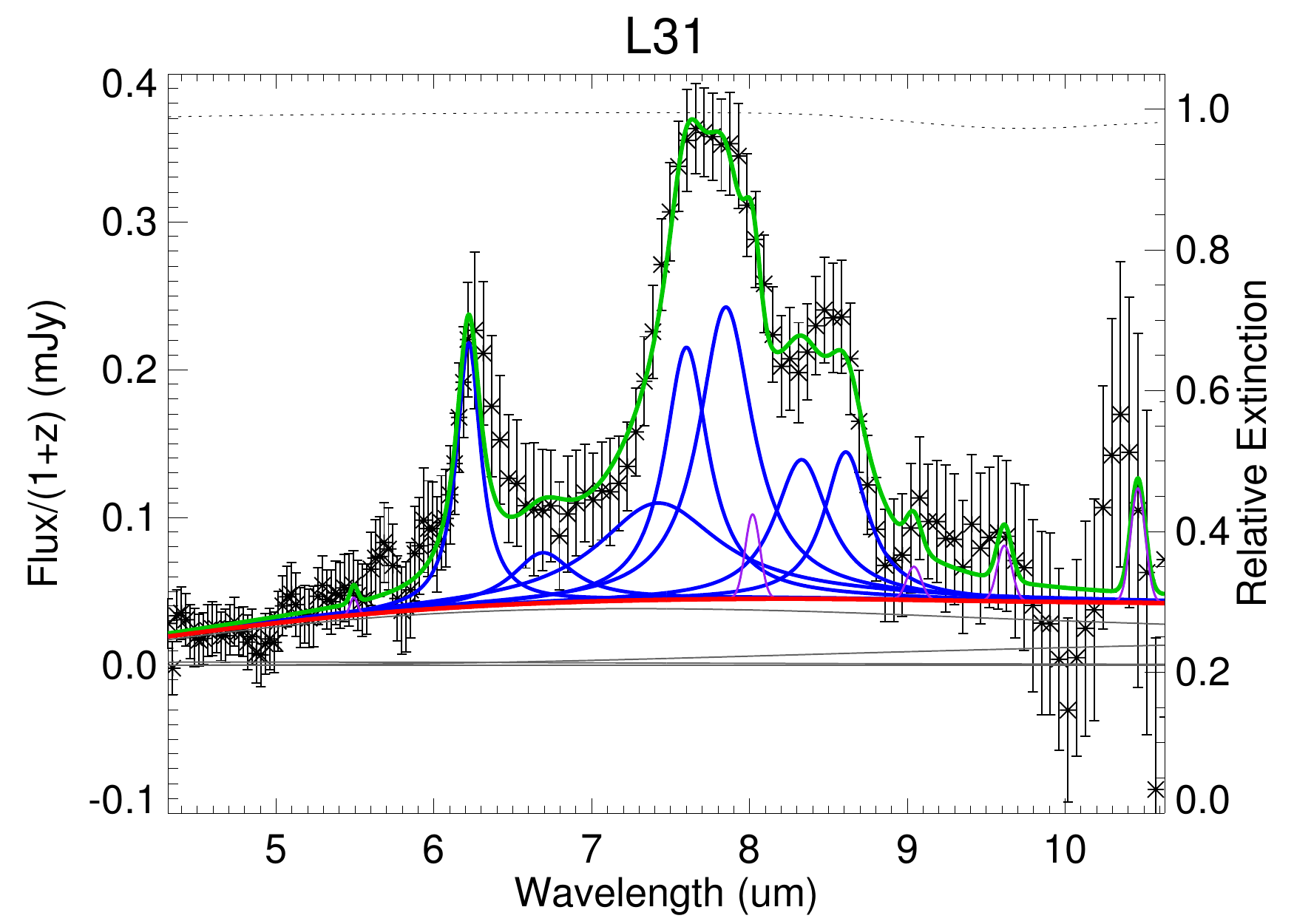}
\end{minipage}
\begin{minipage}{0.5\linewidth}
\includegraphics[height=6cm,width=9cm]{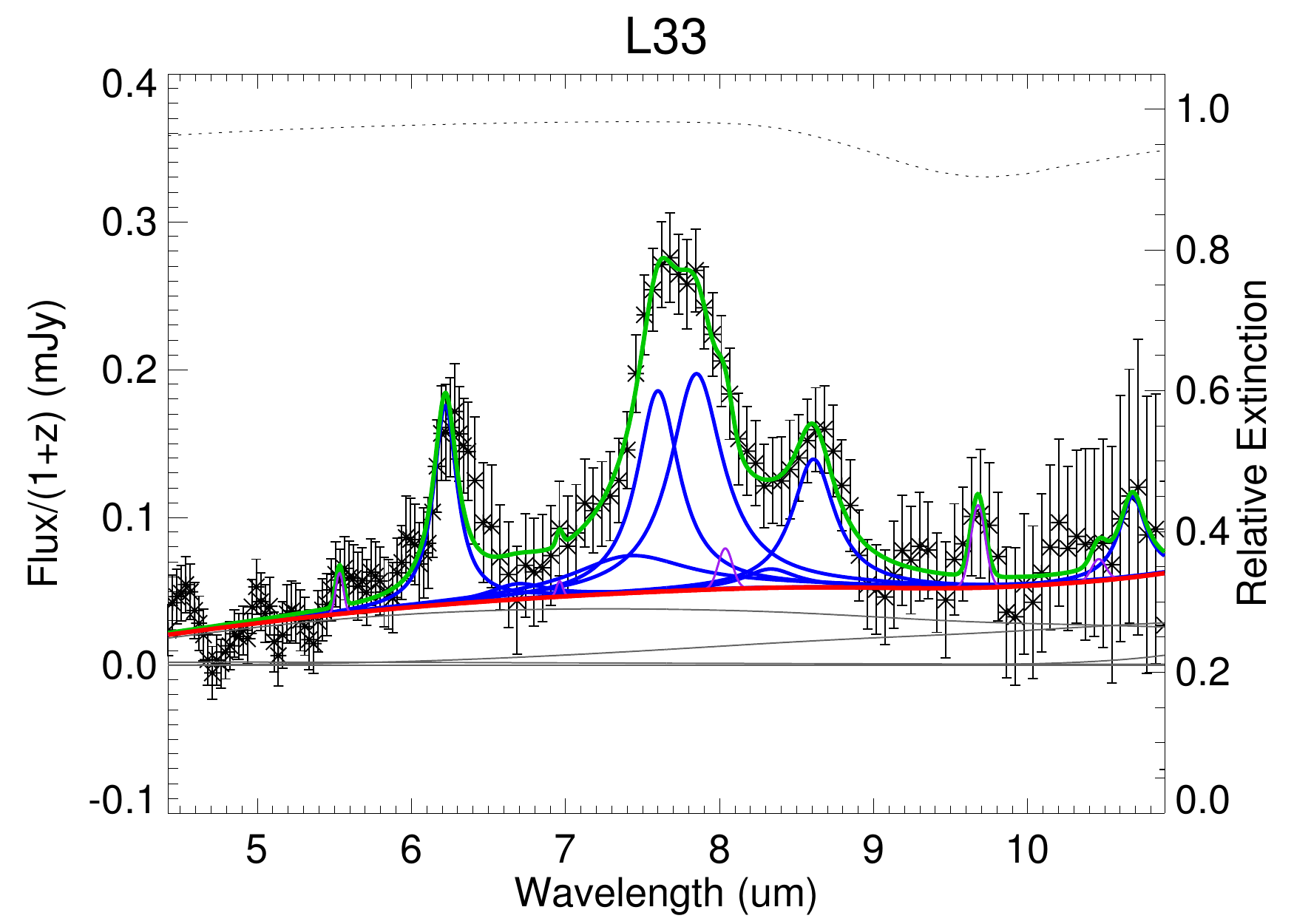}
\end{minipage}
\caption{\it{Continued}}
\end{figure*}

\end{appendix}

\onecolumn 
\centering
\begin{table}[b]
\centering

\caption{\label{new}{\it Spitzer} mid-IR, submillimeter, and radio data of 
the selected sample.}
\begin{tabular}{llcccccc}
\hline\hline{}
ID$^{a}$ & IAU name     & $S_{\rm{24\,\mu m}}$$^{b}$ & $S_{\rm24\,\mu m,spectra}$$^{c}$ & $S_{\rm{350\,\mu m}}$$^{d}$ & $S_{\rm{1.2\,mm}}$$^{a}$ & $S_{\rm{20\,cm}}$$^{e}$ & $\alpha$$^{f}$\\
\multicolumn{2}{l}{}    & \multicolumn{2}{c}{($\mu$Jy)} & \multicolumn{2}{c}{(mJy)} & \multicolumn{1}{c}{($\mu$Jy)} & \\
\hline
L1   & SWIRE4\_J104351.17+590058.1 &  664$\pm$18 & 674$\pm$38 &   30.0$\pm$10.5 &  \bf{2.95$\pm$0.66} &   76.3$\pm$9.0 & -0.84 \\
L3   & SWIRE4\_J104410.00+584056.0 &  589$\pm$12 & 495$\pm$80 & $\mathellipsis$ &  1.04$\pm$1.11 & 105.5$\pm$35.0 & +0.86 \\
L5   & SWIRE4\_J104427.52+584309.6 &  542$\pm$11 & 645$\pm$26 & $\mathellipsis$ &  \bf{2.75$\pm$0.76} &  71.0$\pm$20.0 & -1.03 \\
L7   & SWIRE4\_J104430.60+585518.6 &  825$\pm$13 & 871$\pm$105 & $\mathellipsis$ &  1.39$\pm$0.79 & 101.7$\pm$13.3 & -0.81 \\
L9   & SWIRE4\_J104440.25+585928.3 &  653$\pm$14 & 586$\pm$109 &  55.1$\pm$17.1 &  \bf{4.00$\pm$0.55} &  116.5$\pm$9.2 & -0.54 \\
L10  & SWIRE4\_J104549.77+591904.0 &  626$\pm$12 & 605$\pm$76 & $\mathellipsis$ &  1.39$\pm$0.81 &  67.2$\pm$11.7 & -0.40 \\
L11  & SWIRE4\_J104556.90+585318.8 &  661$\pm$13 & 825$\pm$55 &    38.9$\pm$5.8 &  \bf{3.08$\pm$0.58} &  314.8$\pm$4.1 & -0.45 \\
L14  & SWIRE4\_J104638.68+585612.5 &  631$\pm$13 & 661$\pm$48 &    29.7$\pm$4.9 &  \bf{2.13$\pm$0.71} &  159.5$\pm$6.0 & -0.73 \\
L15  & SWIRE4\_J104656.46+590235.5 &  579$\pm$14 & 695$\pm$41 &    25.7$\pm$5.9 &  \bf{2.36$\pm$0.62} &   68.6$\pm$3.7 & -1.03 \\
L17  & SWIRE4\_J104704.97+592332.3 &  644$\pm$14 & 629$\pm$83 &   45.3$\pm$12.4 &  \bf{2.24$\pm$0.64} & 341.0$\pm$33.0 & -0.76 \\
L19  & SWIRE4\_J104708.78+590627.2 &  556$\pm$14 & 618$\pm$42 & $\mathellipsis$ &  0.89$\pm$0.97 & 102.2$\pm$12.0 & -0.44 \\
L20  & SWIRE4\_J104717.96+590231.7 &  692$\pm$15 & 558$\pm$107 &    51.6$\pm$7.2 &  \bf{2.66$\pm$0.78} &   51.1$\pm$4.7 & -1.25 \\
L25  & SWIRE4\_J104738.32+591010.0 &  704$\pm$13 & 667$\pm$74 &    36.1$\pm$7.1 &  \bf{2.56$\pm$0.74} &   69.0$\pm$9.0 & -1.09 \\
L27  & SWIRE4\_J104744.60+591413.5 &  533$\pm$13 & 591$\pm$50 &    27.4$\pm$7.9 &  \bf{2.48$\pm$0.73} &  77.2$\pm$13.7 & -0.38 \\
L31  & SWIRE4\_J104830.70+585659.3 &  640$\pm$16 & 640$\pm$101 & $\mathellipsis$ &  1.62$\pm$0.60 & 121.0$\pm$39.0 & -0.31 \\
L33  & SWIRE4\_J104848.23+585059.3 &  536$\pm$15 & 487$\pm$61 & $\mathellipsis$ & -0.28$\pm$1.09 & 109.0$\pm$33.0 & +0.27 \\

\hline

\end{tabular}
\begin{flushleft}
$^{a}$IDs and $S_{\rm{1.2\,mm}}$  are the same as F09. The bold values are $>$3\,$\sigma$ detections\\
$^{b}$$S_{\rm{24\,\mu m}}$ comes from the internal SWIRE catalogue.\\
$^{c}$$S_{\rm24\,\mu m,spectra}$ is calculated by convolution of the 24\,$\mu$m MIPS filter and our spectra.\\
$^{d}$$S_{\rm{350\,\mu m}}$ comes from \citet{Kova10}.\\
$^{e}$$S_{\rm{20\,cm}}$ comes from \citet{Owen08}.\\
$^{f}$The radio spectral index $\alpha$ is calculated using the best power law fit, $S_{\nu}\propto \nu^{\alpha}$,  between $S_{\rm{20\,cm}}$, 
$S_{\rm{50\,cm}}$, and $S_{\rm{90\,cm}}$ as reported in F09.\\
\end{flushleft}
\end{table}

\centering
\begin{table}[b]
\centering

\caption{\label{summary} Summary of {\it Spitzer} IRS observations.}
\begin{tabular}{lcc}
\hline\hline{}
ID &  LL1 time on target & LL2 time on target \\
\multicolumn{1}{l}{} & \multicolumn{2}{c}{(sec)}\\
\hline
L7                           & 2$\times$120$\times$6 &  8$\times$120$\times$6 \\
L1, L9, L11, L17, L20, L25   & 3$\times$120$\times$6 &  8$\times$120$\times$6 \\
L3, L10, L14, L15, L31       & 4$\times$120$\times$6 &  9$\times$120$\times$6 \\
L5, L19, L27, L33            & 5$\times$120$\times$6 & 11$\times$120$\times$6 \\
\hline

\end{tabular}
\end{table}

\begin{table}
\centering

\caption{\label{redshifts} Redshifts.}
\begin{tabular}{lccc}
\hline\hline{}
ID &  $z_{\rm PAH}$$^{a}$ & $z_{\rm best}$$^{b}$ & mid-IR best matching template \\
\multicolumn{1}{l}{} & \multicolumn{2}{c}{} & from \citet{Smit07}  \\
\hline
L1   & 2.200$\pm$0.005 & 2.196 & pah2 \\
L3   & 1.921$\pm$0.005 & 1.922 & pah3 \\
L5   & 1.750$\pm$0.007 & 1.756 & pah3 \\
L7   & 1.944$\pm$0.064 & 1.886 & pah2 \\
L9   & 2.041$\pm$0.007 & 2.046 & pah5 \\
L10  & 1.836$\pm$0.005 & 1.834 & pah3 \\
L11  & 1.946$\pm$0.010 & 1.946 & pah5 \\
L14  & 2.032$\pm$0.005 & 2.034 & pah4 \\
L15  & 1.851$\pm$0.010 & 1.844 & pah4 \\
L17  & 1.959$\pm$0.004 & 1.964 & pah3 \\
L19  & 2.240$\pm$0.012 & 2.222 & pah2 \\
L20  & 2.072$\pm$0.010 & 2.066 & pah1 \\
L25  & 1.972$\pm$0.005 & 1.974 & pah3 \\
L27  & 2.022$\pm$0.008 & 2.014 & pah4 \\
L31  & 2.284$\pm$0.009 & 2.272 & pah2 \\
L33  & 2.206$\pm$0.010 & 2.200 & pah3 \\

\hline

\end{tabular}
\begin{flushleft}
$^{a}$Average PAH spectroscopic redshift.\\
$^{b}$Redshift for best-matching template.\\
\end{flushleft}
\end{table}

\begin{table}
\centering

\caption{\label{few} Fluxes and equivalent widths for PAH features from PAHFIT \citep{Smit07}.}
\begin{tabular}{lcccccccccc}
\hline\hline\\
ID  & $F_{\rm6.2\,\mu m}$ & $F_{\rm7.7\,\mu m}$ & $F_{\rm8.6\,\mu m}$ & $F_{\rm11.3\,\mu m}$ & EW$_{\rm6.2\,\mu m}$ & EW$_{\rm7.7\,\mu m}$ & EW$_{\rm8.6\,\mu m}$ & EW$_{\rm11.3\,\mu m}$ & $\tau_{9.7}$ & $S_{\rm10\,\mu m}$$^{a}$\\
\multicolumn{1}{l}{} & \multicolumn{4}{c}{(10$^{-22}$\,W.cm$^{-2})$} & \multicolumn{4}{c}{($\mu$m)} &  & (mJy)\\
\hline\\
L1  & 4.93$\pm$0.51 & 21.69$\pm$1.80 & 5.48$\pm$1.03 & $\mathellipsis$ & 2.86 &  9.42 & 2.70 & $\mathellipsis$ & 1.08 & 0.037\\
L3  & 3.44$\pm$0.51 &  8.28$\pm$0.69 & 3.22$\pm$1.30 &   2.94$\pm$0.85 & 2.90 &  8.17 & 2.67 &            1.85 & 0.00 & 0.036\\
L5  & 3.82$\pm$0.51 & 12.41$\pm$1.37 & 5.16$\pm$1.07 &   2.48$\pm$0.83 & 3.67 & 12.02 & 4.99 &            1.63 & 0.02 & 0.033\\
L7  & 3.11$\pm$0.56 & 12.44$\pm$1.82 & 6.22$\pm$2.24 &   2.13$\pm$1.26 & 0.55 &  2.82 & 1.37 &            0.52 & 0.00 & 0.133\\
L9  & 2.41$\pm$0.38 &  8.68$\pm$0.67 & 4.53$\pm$0.61 &   1.28$\pm$0.91 & 0.63 &  2.41 & 1.33 &            0.27 & 0.00 & 0.096\\
L10 & 0.29$\pm$0.50 &  7.99$\pm$1.51 & 2.83$\pm$1.33 &   2.10$\pm$0.89 & 0.10 &  2.32 & 0.77 &            0.48 & 0.00 & 0.128\\
L11 & 3.61$\pm$0.48 & 12.99$\pm$1.74 & 6.43$\pm$1.70 &   3.06$\pm$1.68 & 0.81 &  4.72 & 2.28 &            1.15 & 0.10 & 0.080\\
L14 & 3.37$\pm$0.57 & 13.17$\pm$1.10 & 4.39$\pm$1.30 &   2.50$\pm$1.87 & 1.19 &  7.54 & 2.80 &            3.17 & 0.00 & 0.034\\
L15 & 4.47$\pm$0.33 & 14.48$\pm$1.30 & 5.83$\pm$0.52 &   4.25$\pm$0.53 & 1.73 &  6.88 & 3.22 &            3.92 & 0.00 & 0.044\\
L17 & 2.38$\pm$0.51 & 14.10$\pm$1.82 & 5.39$\pm$0.75 &   3.16$\pm$0.96 & 0.84 &  5.04 & 2.21 &            1.71 & 0.54 & 0.051\\
L19 & 4.93$\pm$0.35 & 15.67$\pm$1.01 & 3.87$\pm$0.48 & $\mathellipsis$ & 1.71 &  7.52 & 2.30 & $\mathellipsis$ & 0.00 & 0.035\\
L20 & 3.85$\pm$0.63 &  9.49$\pm$1.49 & 3.11$\pm$1.77 &   5.25$\pm$2.46 & 1.31 &  4.40 & 1.30 &            2.83 & 0.18 & 0.066\\
L25 & 5.39$\pm$0.50 & 12.45$\pm$1.00 & 5.98$\pm$1.08 &   4.82$\pm$1.51 & 1.72 &  2.86 & 1.57 &            1.43 & 1.32 & 0.067\\
L27 & 3.77$\pm$0.42 & 12.77$\pm$1.54 & 3.40$\pm$0.77 &   3.54$\pm$1.18 & 1.81 &  3.43 & 1.12 &            1.01 & 1.09 & 0.069\\
L31 & 4.07$\pm$0.52 & 16.25$\pm$1.37 & 4.82$\pm$0.60 & $\mathellipsis$ & 1.26 &  6.80 & 2.44 & $\mathellipsis$ & 0.03 & 0.043\\
L33 & 3.12$\pm$0.41 & 10.57$\pm$2.11 & 2.29$\pm$0.86 & $\mathellipsis$ & 0.93 &  3.93 & 1.01 & $\mathellipsis$ & 0.10 & 0.054\\
\hline
\end{tabular}
\begin{flushleft}
$^{a}$$S_{\rm10\,\mu m}$ is the continuum flux density at 10\,$\mu$m.\\
\end{flushleft}
\end{table}

\begin{table}
\centering

\caption{\label{lum} Luminosities for different wavelengths.}
\begin{tabular}{lcccccccc}
\hline\hline\\
ID &  $\nu\,L_{\nu}$(1.6\,$\mu$m)$^a$ & $\nu\,L_{\nu}$(5.8\,$\mu$m)  & 
$L_{\rm6.2\,\mu m}$ & $L_{\rm7.7\,\mu m}$ & $L_{\rm8.6\,\mu m}$ & 
$L_{\rm11.3\,\mu m}$ & $\nu\,L_{\nu}$(1.4\,GHz)$^a$ & $L_{\rm IR}$$^{b}$\\
\multicolumn{1}{l}{} & \multicolumn{2}{c}{(10$^{11}\,L_{\odot}$)} & \multicolumn{4}{c}{(10$^{10}\,L_{\odot}$)} & (10$^{6}\,L_{\odot}$) & (10$^{12}\,L_{\odot}$)\\
\hline\\
L1  & 2.09 & 1.55 & 4.83$\pm$0.50 & 21.24$\pm$1.76 & 5.36$\pm$1.01 & $\mathellipsis$ & 8.71$\pm$1.00 & 11.00$\pm$3.80\\
L3  & 2.01 & 0.97 & 2.41$\pm$0.35 &  7.80$\pm$5.12 & 2.26$\pm$0.91 & 2.07$\pm$0.59 & 1.35$\pm$0.43 & $\mathellipsis$\\
L5  & 1.80 & 0.71 & 2.12$\pm$0.29 &  6.90$\pm$0.76 & 2.87$\pm$0.60 & 1.38$\pm$0.46 & 5.75$\pm$1.59 & $\mathellipsis$\\
L7  & 1.80 & 2.99 & 2.24$\pm$0.40 &  9.73$\pm$7.18 & 4.49$\pm$1.62 & 1.54$\pm$0.91 & 8.32$\pm$1.15 & $\mathellipsis$\\
L9  & 2.00 & 2.27 & 1.96$\pm$0.31 &  7.06$\pm$0.55 & 3.68$\pm$0.50 & 1.04$\pm$0.74 & 7.94$\pm$0.55 & 9.77$\pm$3.37\\
L10 & 1.85 & 1.06 & 0.18$\pm$0.31 &  5.01$\pm$0.95 & 1.77$\pm$0.83 & 1.32$\pm$0.56 & 3.16$\pm$0.58 & $\mathellipsis$\\
L11 & 1.55 & 2.54 & 2.61$\pm$0.34 & 10.77$\pm$4.42 & 4.65$\pm$1.23 & 2.22$\pm$1.21 & 17.78$\pm$0.41 & 8.13$\pm$2.24\\
L14 & 2.14 & 1.96 & 2.72$\pm$0.46 & 11.63$\pm$5.30 & 3.54$\pm$1.05 & 2.02$\pm$1.51 & 13.49$\pm$0.62 & 8.51$\pm$2.54\\
L15 & 2.97 & 1.50 & 2.86$\pm$0.21 &  9.25$\pm$0.83 & 3.73$\pm$0.34 & 2.72$\pm$0.34 & 6.31$\pm$0.29 & 6.31$\pm$1.89\\
L17 & 2.38 & 1.57 & 1.76$\pm$0.38 & 10.37$\pm$1.34 & 3.96$\pm$0.55 & 2.33$\pm$0.71 & 26.92$\pm$2.48 & 9.12$\pm$3.36\\
L19 & 3.63 & 2.59 & 5.05$\pm$0.35 & 16.04$\pm$1.04 & 3.96$\pm$0.49 & $\mathellipsis$ & 7.59$\pm$0.87 & $\mathellipsis$\\
L20 & 4.19 & 2.05 & 3.25$\pm$0.53 &  9.82$\pm$5.96 & 2.63$\pm$1.50 & 4.43$\pm$2.08 & 7.94$\pm$0.73 & 12.30$\pm$3.40\\
L25 & 2.36 & 1.54 & 4.03$\pm$0.37 &  9.30$\pm$0.75 & 4.47$\pm$0.81 & 3.60$\pm$1.13 & 7.94$\pm$1.10 & 7.76$\pm$2.50\\
L27 & 1.74 & 1.13 & 3.00$\pm$0.34 & 10.16$\pm$1.22 & 3.18$\pm$0.61 & 2.82$\pm$0.94 & 4.37$\pm$0.80 & 6.92$\pm$2.23\\
L31 & 1.55 & 2.97 & 4.37$\pm$0.56 & 17.44$\pm$1.47 & 5.18$\pm$0.64 & $\mathellipsis$ & 7.94$\pm$2.56 &  $\mathellipsis$\\
L33 & 2.53 & 2.43 & 3.08$\pm$0.40 & 10.42$\pm$2.08 & 2.26$\pm$0.85 & $\mathellipsis$ & 3.39$\pm$1.01 &  $\mathellipsis$\\
\hline
\end{tabular}
\begin{flushleft}
$^a$From F09. \\
$^b$From \citet{Kova10}.\\

\end{flushleft}
\end{table}

\begin{table}
\centering

\caption{\label{lumsam}Average luminosities for different samples.}
\begin{tabular}{lcccccccc}
\hline\hline\\
Sample & $\langle$z$\rangle$ & $\langle S_{\rm 24\,\mu m} \rangle$ & $L_{\rm 
6.2\,\mu m}$ & $L_{\rm 7.7\,\mu m}$ & $L_{\rm8.6\,\mu m}$ & $L_{\rm11.3\,\mu 
m}$ & $\nu\,L_{\nu}$(1.4\,GHz) & $L_{\rm IR}$\\
 & & (mJy) & \multicolumn{4}{c}{(10$^{10}\,L_{\odot}$)} & \multicolumn{1}{c}{(10$^{6}\,L_{\odot}$)} & \multicolumn{1}{c}{(10$^{12}\,L_{\odot}$)}\\
\hline\\
This work                                  & 2.017 &                 0.63 & 2.90$\pm$0.31 & 10.38$\pm$1.09 & 3.62$\pm$0.27 & 2.29$\pm$0.26 &  8.68$\pm$1.56 &  8.87$\pm$0.64 \\
This work, EW$_{7.7\mu m}$$>$6\,$\mu$m$^a$ & 2.040 &                 0.60 & 3.07$\pm$0.13 & 11.30$\pm$0.35 & 3.42$\pm$0.15 & 2.51$\pm$0.29 &  7.31$\pm$1.38 &  8.61$\pm$1.35 \\
This work, EW$_{7.7\mu m}$$<$6\,$\mu$m$^a$ & 2.000 &                 0.65 & 1.94$\pm$0.14 &  7.96$\pm$0.72 & 2.86$\pm$0.21 & 2.12$\pm$0.29 &  9.75$\pm$2.59 &  9.00$\pm$0.78 \\
\citet{Saji07}$^b$                         & 1.524 &                 1.19 & 2.66$\pm$0.46 & 13.17$\pm$3.28 & 2.87$\pm$0.67 & 1.71$\pm$0.37 & 18.33$\pm$6.50 &  6.32$\pm$1.36 \\ 
Men\'edez-D. et al. (2009)$^c$             & 2.000 &                 0.33 & 1.09$\pm$0.22 &  4.10$\pm$0.93 & 1.63$\pm$0.17 & 1.39$\pm$0.10 &  4.63$\pm$0.79 &  7.37$\pm$2.45 \\ 
\citet{Pope08a}                            & 1.910 &                 0.45 & 1.34$\pm$0.22 &  6.46$\pm$1.30 & 2.13$\pm$0.38 & 1.19$\pm$0.32 & 14.32$\pm$5.96 &  5.90$\pm$1.00 \\ 
\citet{Copp10}                             & 2.730 &                 0.51 & 3.04$\pm$0.71 & 16.28$\pm$3.52 & 2.55$\pm$0.72     & $\mathellipsis$  & $\mathellipsis$ &  10.98$\pm$1.88 \\
\citet{Farr08}                             & 1.690 &                 0.73 & 2.37$\pm$0.18 &  3.56$\pm$0.24 & $\mathellipsis$     & 1.50$\pm$0.10     & $\mathellipsis$ & $\mathellipsis$ \\
\hline\\
 & & (mJy) & \multicolumn{4}{c}{(10$^{8}\,L_{\odot}$)} & \multicolumn{1}{c}{(10$^{5}\,L_{\odot}$)} & \multicolumn{1}{c}{(10$^{11}\,L_{\odot}$)}\\
\hline\\
\citet{Bran06}                             & 0.008 & 8.52$\times$10$^{3}$ & 2.10$\pm$0.85 &  4.26$\pm$1.61 & 0.83$\pm$0.34 & 1.85$\pm$0.65 &  1.23$\pm$0.64 &  1.18$\pm$0.31 \\   
\citet{Odow09}                             & 0.092 &                 6.08 & 4.24$\pm$0.49 & 17.72$\pm$1.72 & 3.82$\pm$0.36 & 5.16$\pm$0.43 & $\mathellipsis$  & $\mathellipsis$ \\
\hline
\end{tabular}
\begin{flushleft}
$^a$The values are computed from the stacked spectra.\\
$^a$The sample of \citet{Saji07} is limited to PAH-rich sources with EW$_{\rm 7.7\mu m} > 0.8\,\mu$m. \\
$^b$Men\'edez-D. et al. (2009) designates \citet{Mene09}.\\
\end{flushleft}
\end{table}

\begin{table}
\centering

\caption{\label{ratio}Average luminosities ratios for different samples.}
\begin{tabular}{lcccccccc}
\hline\hline\\
Sample &  $\frac{L_{\rm6.2\,\mu m}}{L_{\rm7.7\,\mu m}}$ & $\frac{L_{\rm6.2\,\mu m}}{L_{\rm11.3\,\mu m}}$ & $\frac{L_{\rm7.7\,\mu m}}{L_{\rm8.6\,\mu m}}$ & $\frac{L_{\rm7.7\,\mu m}}{L_{\rm11.3\,\mu m}}$ & $\frac{L_{\rm8.6\,\mu m}}{L_{\rm11.3\,\mu m}}$ & $\frac{L_{\rm6.2\,\mu m}}{L_{\rm IR}}$ & $\frac{L_{\rm7.7\,\mu m}}{L_{\rm IR}}$ & $\frac{L_{\rm7.7\,\mu m}}{\nu L_{\nu}({\rm 1.4\,GHz})}$ \\
  & \multicolumn{5}{c}{} & \multicolumn{2}{c}{(10$^{-3}$)} & \multicolumn{1}{c}{(10$^{3}$)}  \\
\hline\\
This work                              & 0.28 & 1.27 & 2.87 & 4.53 & 1.58 & 3.27 & 11.70 & 11.96 \\
This work, EW$_{\rm 7.7\mu m} > 6\,\mu$m$^a$ & 0.27 & 1.22 & 3.30 & 4.50 & 1.36 & 3.57 & 13.12 & 15.46 \\
This work, EW$_{\rm 7.7\mu m} < 6\,\mu$m$^a$ & 0.24 & 0.92 & 2.78 & 3.75 & 1.35 & 2.16 &  8.84 &  8.16 \\
\citet{Saji07}$^b$                     & 0.20 & 1.56 & 4.59 & 7.70 & 1.68 & 4.21 & 20.84 &  7.18 \\ 
\citet{Mene09}                         & 0.27 & 0.78 & 2.52 & 2.95 & 1.17 & 1.48 &  5.56 &  8.86 \\ 
\citet{Pope08a}                        & 0.21 & 1.13 & 3.03 & 5.43 & 1.79 & 2.27 & 10.95 &  4.51 \\
\citet{Copp10}                         & 0.19 & $\mathellipsis$     & 6.38     & $\mathellipsis$     &  $\mathellipsis$    & 2.77     &  14.83  & $\mathellipsis$ \\   
\citet{Farr08}                         & 0.66 & 1.58     &  $\mathellipsis$    &  2.37    &  $\mathellipsis$ & $\mathellipsis$ &  $\mathellipsis$ & $\mathellipsis$ \\        
\citet{Bran06}                         & 0.49 & 1.14 & 5.13 & 2.30 & 0.45 & 1.77 &  3.61 &  3.46 \\   
\citet{Odow09}                         & 0.24 & 0.83 & 4.63 & 3.44 & 0.74 & $\mathellipsis$  & $\mathellipsis$ \\  
\hline
\end{tabular}
\begin{flushleft}
$^a$The values are computed from the stacked spectra.\\
$^b$The sample of \citet{Saji07} is limited to PAH-rich sources with EW$_{\rm 
7.7\mu m} > 0.8\,\mu$m. \\

\end{flushleft}
\end{table}

\end{document}